\documentclass[11pt]{article}
\usepackage{graphicx}
\usepackage{rotating}
\usepackage{bm}
\usepackage{amsmath}
\usepackage{relsize}
\usepackage{multirow}
\usepackage{natbib}

\def\be{\begin{equation}}
\def\ee{\end{equation}}
\def\deg{$^\circ$}

\def\deg{$^\circ$}

\def\be{\begin{equation}}
\def\bea{\begin{eqnarray}}
\def\ee{\end{equation}}
\def\eea{\end{eqnarray}}

\def\arcmin{$^\prime$}

\def\R0{R$_{\odot}$}

\def\pccm6{pc cm$^{-6}$}

\def\h2o{H$_2$O}
\def\bs{\boldsymbol}

\newcommand{\threevdots}{%
  \vbox{\baselineskip1ex\lineskiplimit0pt%
  \hbox{.}\hbox{.}\hbox{.}}}

\begin{document}

\begin{titlepage}
\begin{center}

\textsc{\LARGE \bf NATIONAL RADIO ASTRONOMY OBSERVATORY\\[0.4cm] Charlottesville, Virginia}\\[5cm]

\textsc{\Large \bf ELECTRONICS DIVISION INTERNAL REPORT NO. 330}\\[5cm]

{\LARGE \bfseries A Model for Phased Array Feed \\[1.0cm] }

{\Large
D. Anish Roshi$^{1}$, J. Richard Fisher$^{1}$ \\ [0.4cm]
{\small
$^{1}$ National Radio Astronomy Observatory, Charlottesville, \\
}
}

\vfill
% Bottom of the page
January 27, 2016

\end{center}
\end{titlepage}

\title{A Model for Phased Array Feed}
\author{D. Anish Roshi \& J. Richard Fisher}
\date{January 27, 2016 \\Version 2.0}
\maketitle

\section*{Abstract}
In this report we present a model for phased array feed (PAF) and compare
the model predictions with measurements. A theory for loss-less PAF is presented
first. To develop the theory we ask the question -- what is the best $T_{sys}/\eta_{ap}$ that
can be achieved when a PAF is used on a telescope to observe a source at an
angle $\theta_s, \phi_s$ from the boresight direction ? We show that
a characteristic matrix for the {\em system} (i.e. PAF+telescope+receiver)
can be constructed starting from the signal-to-noise ratio of the observations
and the best $T_{sys}/\eta_{ap}$ can be obtained from the maximum eigenvalue 
of the characteristic matrix. For constructing the characteristic 
matrix, we derive the open-circuit voltage at the output of the antenna 
elements in the PAF due to (a) radiation from source, (b) radiation 
from ground (spillover), (c) radiation from sky background and (d) noise 
due to the receiver. The characteristic matrix is then obtained from
the correlation matrices of these voltages. We then describe a modeling program
developed to implement the theory presented here. Finally the model
predictions are compared with results from test observations made toward Virgo A
with a prototype PAF (Kite array) on the GBT\citep{roshietal2015}. The main
features of the model and summary of the comparison are :
\begin{itemize}
\item
We present an {\em ab initio} model for the PAF. The model is developed from
Lorentz Reciprocity theorem (see Section~\ref{recmode}), 
which is derived from Maxwell's equation with
no further assumptions made, and also with the aid of first and second law
of thermodynamics (see Appendices). The result of the model is presented 
as a theorem (see Section~\ref{theory}).

\item
Based on the model results, we show that the receiver temperature of the PAF
can be expressed as a generalization of the equation for a single 
antenna followed by a receiver. The proof of this statement is given in 
Appendix~\ref{A6}. 

\item
The model well predicts the measured $\frac{T_{sys}}{\eta}$ vs offset angle
from boresight direction but
needs an increase in receiver temperature of about 5K to match the measured results
\footnote{In \citet{roshietal2015}, we presented our preliminary model
results, where the excess temperature needed is given as 8 K. Our improved model (see
Section~\ref{compare}) requires only 5K excess noise. Also the aperture efficiency
given in \citet{roshietal2015} is 70 \%, which has been revised to 65\% in this report.}.
The model also predicts the measured $\frac{T_{sys}}{\eta}$ vs
frequency with the additional increase in receiver temperature
mentioned above. 

\item
The aperture and spillover efficiencies obtained from the best fit 
model are 65\% and 98\% respectively between 1.3 and 1.7 GHz.

\end{itemize}

\section{Introduction}

Feeds consisting of dense, electrically small antenna arrays, referred
to as Phased Array Feed (PAF), are now of significant interest
\citep{Hotanetal2014, cortesetal2015,
grayetal2011, oosterlooetal2010}. These dense arrays sample the focal field pattern
of the telescope. Multiple beams are then formed by combining the
signals sampled by the array elements with complex weights that form
an efficient reflector illumination.  The beams
formed using a PAF can fully sample
the FOV. Additionally, a PAF can be used to improve spillover efficiency as well as the
illumination of the dish. However, mutual coupling between array 
elements are a major hurdle in designing a PAF.  Mutual coupling 
modifies the element radiation patterns.
It also couples amplifier noise. Therefore, detailed electromagnetic, noise and network
modeling is needed to design a PAF for radio astronomy applications \citep{rfisher1996}.

Several research groups have analyzed and modeled the noise performance\citep{warnicketal2009,
woestenburg2005}, electromagnetic properties of PAFs\citep{warnicketal2011, hay2010}
and signal processing aspects\citep{jeffs2008}.
In this report, we present details of a new PAF model developed at NRAO starting from
Lorentz Reciprocity theorem (see Section~\ref{recmode}), and also with the aid of 
first and second law of thermodynamics (see Appendices). The implementation of the
theory using a matlab program is discussed in Section~\ref{pafmodel}. The model
predictions are compared with measurements in Section~\ref{compare}. 

\section{Some basic Assumptions and Approximations}

The following assumptions and approximations are made during the development of the model.
\begin{itemize}
\item
The PAF is assumed to be loss-less and reciprocal device.
This assumption implies a symmetry in the impedance matrix of 
PAF, {\em viz}, $z_{ij} = z_{ji}$,
where $z_{ij}$ are the elements of the impedance matrix. 

\item
The parabolic reflector and ground are assumed to be located at the far-field of PAF. We
also neglect scattering at the edge of the reflector and in any mounting structure.

\item
The fields on the aperture of the antenna are computed using geometric optics approximations.

\item
The amplifier noise radiated from the PAF is not reflected back to the system. 
\end{itemize}

\section{Theory of loss-less Phased Array Feed}
\label{theory}

The performance of PAF is usually expressed as the ratio of the
system temperature by aperture efficiency ($T_{sys}/\eta_{ap}$)
\footnote{The directly measurable quantity is the ratio 
$\frac{T_{sys}}{\eta_{ap}\eta_{rad}}$, where $\eta_{rad}$ is the
radiation efficiency of the PAF. In this report we are considering
a loss-less PAF and hence $\eta_{rad} = 1$.} 
To develop the theory we ask the question :

\begin{description}
  \item[ ] \hfill \\ 
 ``For a PAF installed on a telescope what is 
the minimum value of $T_{sys}/\eta_{ap}$ that can be obtained when 
observing a compact source (point source) 
at angle ($\theta_s$, $\phi_s$) from the boresight direction ?'' 
\end{description}
The answer to this question is the following theorem:
\begin{description}
\item[Theorem :] Given the (spectral) impedance matrix, $\bm Z$, 
and the embedded beam patterns, $\bs{\vec{\mathcal{E}}^e}$, of the PAF,
given the amplifier noise parameters ($R_n, g_n, \rho$) and its input impedance ($Z_{in}$), 
and given the telescope geometry and source position ($\theta_s$, $\phi_s$), one can
construct a characteristics matrix $\bm M$ for the {\em system} (i.e. PAF + telescope + receiver). 
Then the best signal-to-noise ratio 
on the source is the maximum eigenvalue, $e_{max}$ of $\bm M$.
\end{description}

As described below $e_{max}$ can be converted to $T_{sys}/\eta_{ap}$ from the
knowledge of telescope aperture area and source flux density. The 
proof of the theorem is presented below.

\subsection{Schematic of PAF, Impedance Matrix and Embedded beam pattern}

\begin{figure}[t]
\begin{tabular}{cc}
%\multicolumn{2}{c}{\includegraphics[width=5.3in, height=2.5in, angle =0]{fig/pafmodel.pdf}} \\ 
%\includegraphics[width=3.3in, height=2.5in, angle =0]{fig/pafmodelb.pdf} & 
%\includegraphics[width=3.3in, height=3.5in, angle =0]{fig/pafmodelc.pdf} \\
 & \multirow{2}{*}{\includegraphics[width=3.3in, height=4.8in, angle =0]{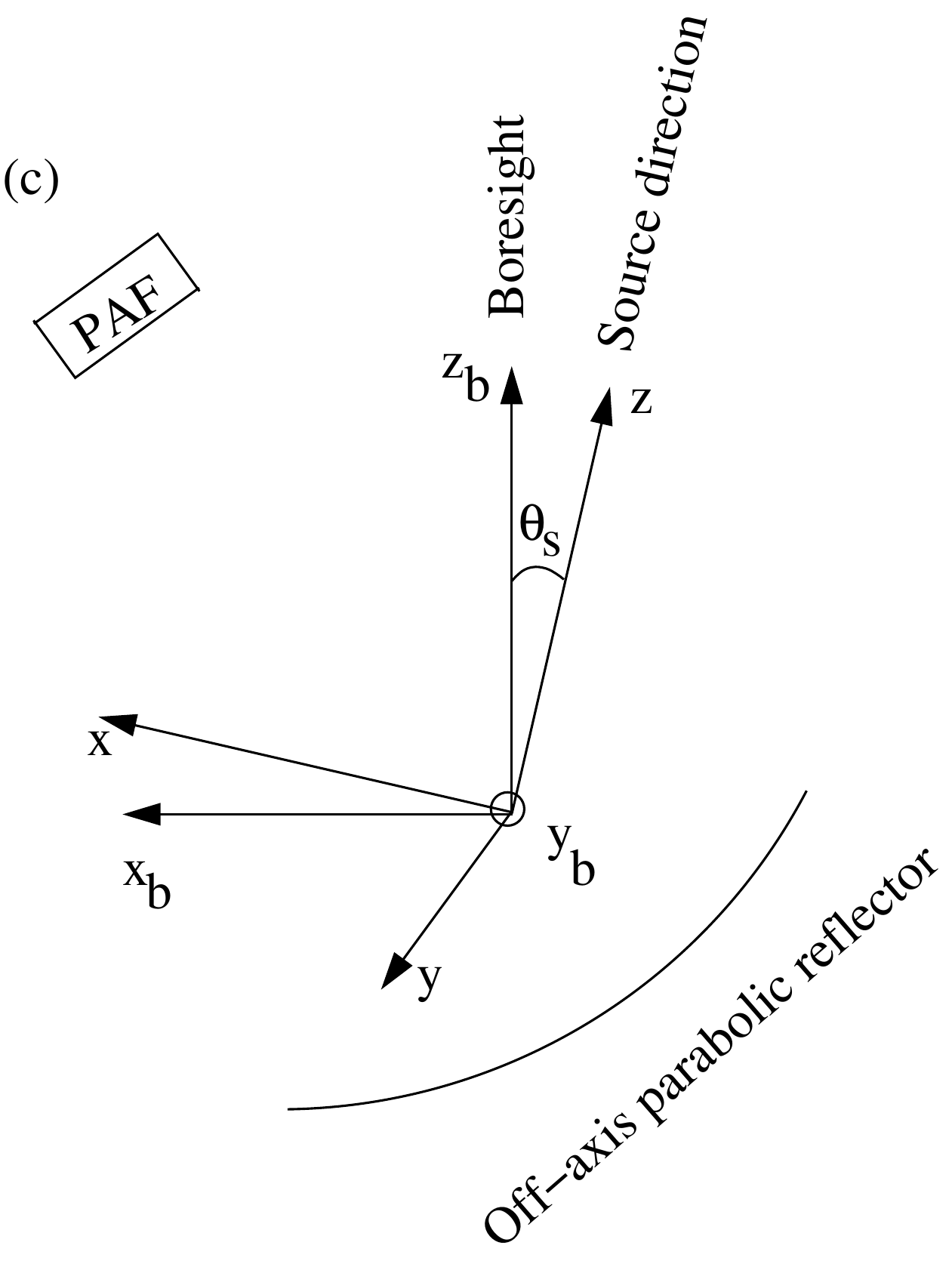}} \\ 
\includegraphics[width=3.5in, height=2.5in, angle =0]{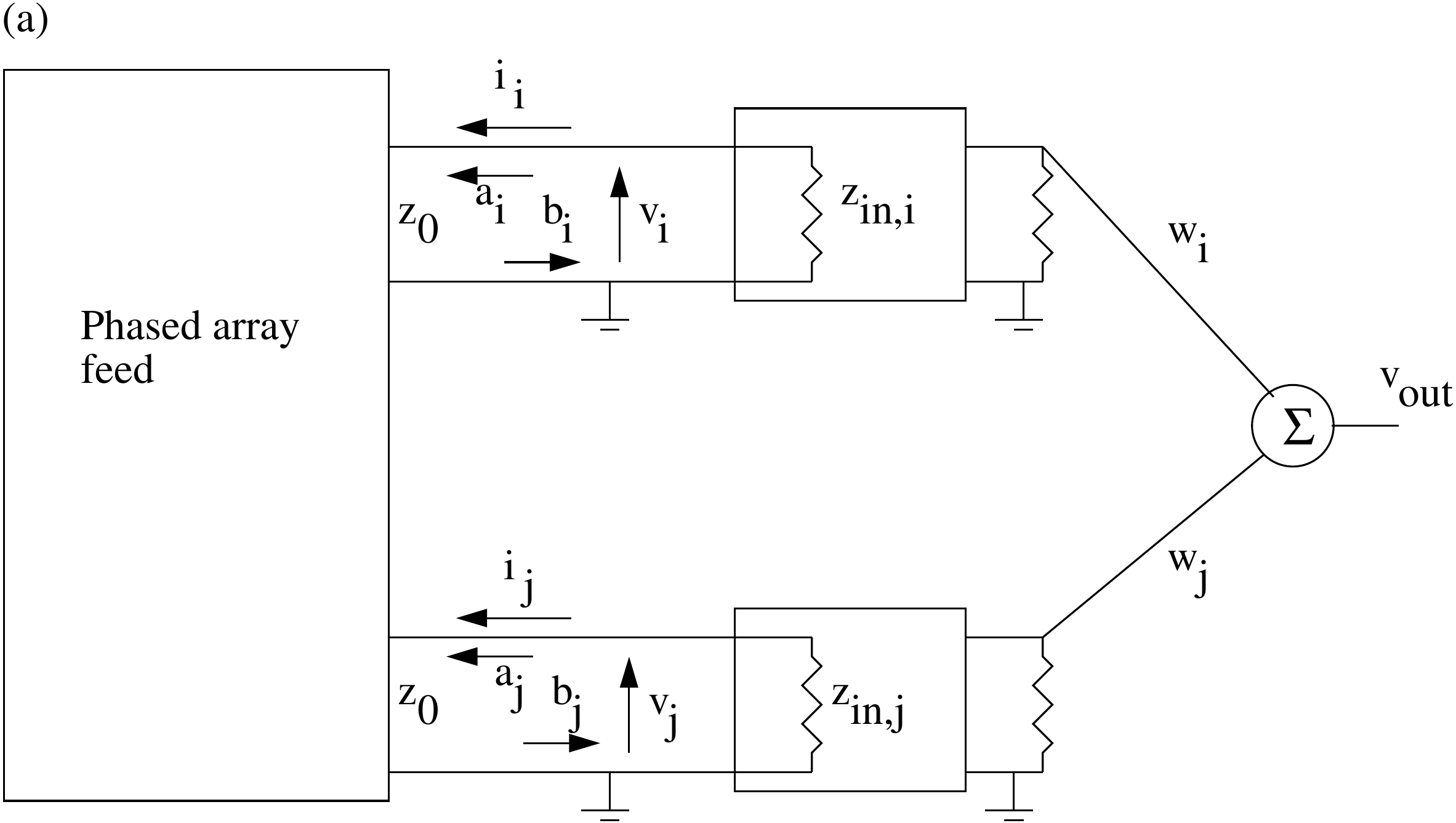} & \\  
\includegraphics[width=3.5in, height=2.5in, angle =0]{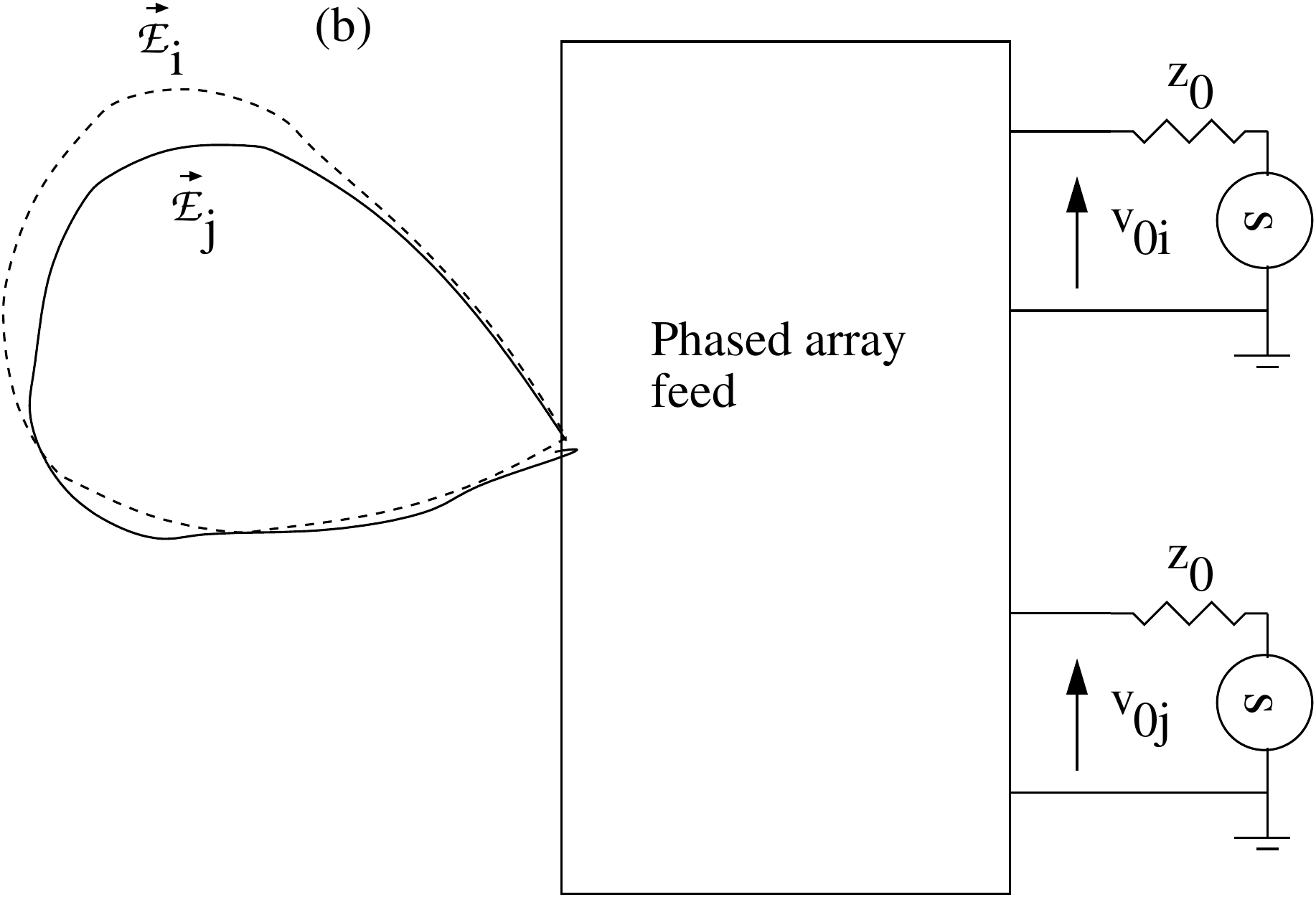} & \\
\end{tabular}
\caption{(a) Schematic showing the signal flow and processing in the PAF system. (b) 
PAF in transmitting mode. (c) PAF on an off-axis parabolic reflector. Co-ordinate
system used to specify the `boresight' direction and source direction are also
shown (see text for details).
}
\label{fig1}
\end{figure}

In this section we describe the necessary background needed on the {\em system} to prove the
theorem. A schematic of the PAF is shown in Fig.~\ref{fig1}a. The signals 
from the antenna elements are brought to the `ports' of the PAF 
through a transmission line of characteristics impedance $z_0$.
For simplicity we consider $z_0$ is equal to the reference impedance 50 $\Omega$.
Let the total number of ports of the PAF be $M$.
The network characteristics of the PAF are specified by the impedance matrix, $\bm Z$,
which relates the port voltages $v_i$ and currents $i_i$. Another method
used to specify the network characteristics is the Scattering matrix, $\bm S$,
which relates the forward and reverse traveling waves at the port. The amplitude
of these waves are $a_i$ and $b_i$ respectively for the forward and reverse waves.   
All voltages, currents and wave amplitudes are harmonic quantities (i.e. they
are quantities per unit frequency interval) and their values
are specified as {\em peak} values. For simplicity, we omit adding the term
$e^{j\omega t}$.  The relationship between the network 
parameters is discussed in Appendix~\ref{A1}.

Fig.~\ref{fig1}a also shows schematically the signal processing done in the PAF system. 
Each antenna element of 
the PAF is followed by a low-noise amplifier (LNA). The input impedance of 
the $i^{th}$ LNA is $Z_{in_i}$. The signals are further amplified and 
scaled with a complex weight vector $\bm w^T = [w_1, w_2, ....]$. We
consider that the weights are normalized such that $\bm w^H \bm w = 1$. The
scaled quantities are then added together to get the signal from a beam. The
operation is also referred to as beamforming. Multiple beams
are formed by adding the signals after scaling them with different weight vectors.

In addition to the network characteristics we need to specify the radiation pattern
of the PAF. In Section~\ref{recmode} we show that it is convenient to express the
radiation pattern in terms of the {\em embedded beam
patterns}. The $j^{th}$ embedded beam pattern, $\vec{\mathcal{E}}^e_j$ is defined 
as the beam pattern of the PAF when $j^{th}$ port is excited with
1 V (i.e. $v_{0_j} = 1$ V) and all other ports are short circuited 
(i.e. $v_{0_i} = 0$ V for $i \neq j$; see Fig.~\ref{fig1}b).
The source impedance for excitation is considered to be equal to $z_0$.
Thus there are $M$ embedded beam patterns, which are represented conveniently
as a vector $\bs{\vec{\mathcal{E}}^e}$,
\be
\bs{\vec{\mathcal{E}}^e}^T = \left[\vec{\mathcal{E}}^e_1, \vec{\mathcal{E}}^e_2, ... \right]
\ee 
These beam patterns are function of the position vector $\vec{r}$, the origin of the
coordinate system is located at the center of the PAF (see Appendix~\ref{A8a} for further
details on the coordinate system). The beam patterns are specified at the far-field 
ie $|\vec{r}| >> \frac{2 D_{array}^2}{\lambda}$, where $D_{array}$ is the maximum
physical size of the PAF and $\lambda$ is the wavelength of operation of the PAF.  
As described below, the embedded beam patterns are scaled by the
port voltage and summed up to get the resultant beam pattern of the PAF for an arbitrary
set of port voltages. Hence the dimension 
of the embedded beam pattern is m$^{-1}$. 
At far-field, the beam pattern can be described by an outgoing spherical wave,
\be
\vec{\mathcal{E}}^e_i(\vec{r}) =  \vec{E}^e_i(\theta, \phi) \; \frac{e^{j\vec{k}.\vec{r}}}{r}
\label{farf}
\ee
where $\vec{\mathcal{E}}^e_i$ is the $i^{th}$ embedded beam pattern,
$r$ and $\hat{r}$ are the magnitude and the unit vector in the direction 
of $\vec{r}$ respectively, $\vec{k} = \frac{2\pi}{\lambda} \hat{r}$ is 
the propagation vector. The vector function $\vec{E}^e_i$ depends only 
on the coordinates $\theta, \phi$. It should be noted that the 
geometric phase due to the location
of elements (or in other words the excitation current distribution) away
from the co-ordinate center is included in $\vec{E}^e_i$. From the 
definition of embedded pattern it follows that $\vec{E}^e_i$ is dimensionless. 
As in the case of network parameters,
the field are harmonic quantities and for simplicity we omit the term $e^{j\omega t}$.
Since the beam patterns are specified at free space the angular frequency $\omega$
and $\lambda$ are related through
\be
c = \frac{\omega}{2\pi} \lambda,
\ee
where $c$ is the velocity of light in free space.
Moreover the embedded magnetic field patterns are given by
\be
\vec{\mathcal{H}}^e_i = \frac{\hat{k}}{z_f} \times \vec{\mathcal{E}}^e_i
\ee
where $\hat{k}$ is the unit vector in the direction of $\vec{k}$ and
$z_f$ is the free space impedance. The relationship 
between the network parameters and embedded beam patterns are discussed 
in Appendix~\ref{A1}.

The radiation pattern of the PAF when excited by a set of arbitrary
port voltages can be expressed in terms of the embedded beam
patterns as (see Fig.~\ref{fig1}b and also Section~\ref{recmode})
\bea 
\vec{\mathcal{E}}(\vec{r}) & =  & \sum_{i=1,M} v_{0_i} \vec{\mathcal{E}}^e_i(\vec{r}) \nonumber \\ 
                           & =  & \bm V_0^T \bs{\vec{\mathcal{E}}^e}  
\eea
where $\bm V_0$ is the vector of port voltages $v_{0_i}$. The 
radiation pattern $\vec{\mathcal{E}}$ has the units V/m. At far-field,
the ($\theta, \phi$) dependence of the radiation pattern can be written
in a similar fashion,
\be
\vec{E}(\theta, \phi) = \bm V_0^T \bs{\vec{E}^e}.
\ee
The units of $\vec{E}$ is V. In the report, we refer to $\bs{\vec{\mathcal{E}}^e}$
and $\bs{\vec{E}^e}$ as embedded beam patterns -- the $\vec{r}$ dependence is implied
in the usage of calligraphic symbol. 

For our application, the PAF is placed at the prime focus of a 
(off-axis) parabolic antenna. A schematic showing the geometry of the
configuration along with the observing source direction is 
shown in Fig.~\ref{fig1}c. To specify the source direction, we
consider a co-ordinate system with $z_b$-axis pointing toward
the `boresight' direction of the telescope and the $x_b-y_b$ plane
coinciding with the `aperture plane'. The source direction
is specified by the polar angle ($\theta_s, \phi_s$) 
with respect to this co-ordinate system. A second co-ordinate
system with the $z$-axis pointing towards the source and the
$x-y$ plane coinciding with the `projected
aperture plane' in the direction of the source is used
to calculate the PAF field pattern on the projected aperture 
plane (see Section~\ref{recmode}). The unit vectors in the
direction of $x,y,z$ axes 
are referred to as $\hat{u}_x, \hat{u}_y$
and $\hat{u}_z$ respectively. 

\subsection{Signal to Noise Ratio}

We start the proof of the theorem by considering the
signal to noise ratio (SNR) at the output of the PAF. For a typical
observation with the PAF, on-source and off-source measurements 
are made. The SNR can then be written as
\bea
\textrm{SNR}  =  \frac{P_{on-source}}{P_{off-source}} 
\eea
where $P_{on-source}$ is the increase in power spectral
density at the output of a beam due to the source relative to the
off-source spectral density, $P_{off-source}$. Below, we compute
this SNR using the signal processing done in the PAF.  

The output voltage per unit frequency interval 
after beamforming can be written as 
\be
v_{out} = \bm w^T \bs{\tilde{V}}
\ee
where $\bs{\tilde{V}}$ is the voltage vector at the input of the combiner
(see Fig.~\ref{fig1}a).
This voltage vector can be written as a sum of at least four components;
\be
\bs{\tilde{V}} = \bs{\tilde{V}}_{signal} + \bs{\tilde{V}}_{spill} + 
                \bs{\tilde{V}}_{rec} + \bs{\tilde{V}}_{sky} 
\ee
where $\bs{\tilde{V}}_{signal}, \bs{\tilde{V}}_{spill}, \bs{\tilde{V}}_{rec}$ 
and $\bs{\tilde{V}}_{sky}$ are the 
contributions due to source, spillover noise, receiver noise and sky
background noise. For a typical application using the PAF, the output signal power or
power spectral density is measured. This power spectral density is proportional
to $\langle v_{out} v_{out}^{*} \rangle$, which can be written as
\bea 
\langle v_{out} v_{out}^{*} \rangle & = & \bm w^H \langle \bs{\tilde{V}} \bs{\tilde{V}}^H \rangle \bm w \nonumber \\  
 & = & \bm w^H \bs{\tilde{R}}\bm w \nonumber \\
 & = & \bm w^H \bs{\tilde{R}}_{signal} \bm w + \bm w^H \bs{\tilde{R}}_{spill} \bm w +
       \bm w^H \bs{\tilde{R}}_{rec} \bm w + \bm w^H \bs{\tilde{R}}_{sky} \bm w 
\label{vout}
\eea
where $\bs{\tilde{R}}$ is the matrix of voltage correlation per unit frequency interval.  
Since the different voltage components are 
uncorrelated, $\bs{\tilde{R}}$ can be written as the sum of the correlations of
its components, {\em viz}, $\bs{\tilde{R}}_{signal}, \bs{\tilde{R}}_{spill}, \bs{\tilde{R}}_{rec}$
and $\bs{\tilde{R}}_{sky}$ which correspond to the voltage correlations due to source,
spillover noise, receiver noise and sky background noise respectively.
The SNR on the source is then 
\bea
\textrm{SNR} & = & \frac{P_{on-source}}{P_{off-source}}  \nonumber \\
    & \approx & \frac{\bm w^H \bs{\tilde{R}}_{signal} \bm w}{\bm w^H \bs{\tilde{R}}_{spill} \bm w +
       \bm w^H \bs{\tilde{R}}_{rec} \bm w + \bm w^H \bs{\tilde{R}}_{sky} \bm w}.
\label{snr}
\eea
The approximate sign in Eq.~\ref{snr} is introduced to
indicate (a) that the noise power in on-source measurement is dominated by components other than
the source and (b) that the sky and spillover components are approximately same for 
the on-source and off-source positions. (We ignore the approximate
sign below for convenience.) 

We wish to write Eq.~\ref{snr} in terms of open circuit voltages at the output of 
the PAF antenna elements.
It follows from network analysis that the output voltage vector 
\be
\bs{\tilde{V}}  =  \bm G \bm A \bm V_{oc} 
\ee
where
\be
\bm  A = -\bm  Z_{in}(\bm  Z + \bm  Z_{in})^{-1},
\ee
$\bm Z_{in}$ is the input impedance matrix, which is a 
diagonal matrix with elements $Z_{in_i}$, $\bm G$ is the overall system
gain matrix, which again is a diagonal matrix and $\bm V_{oc}$
is the open circuit voltage vector at the output of the PAF.
In terms of the open circuit voltage Eq.~\ref{snr} becomes
\bea 
\textrm{SNR} & = & \frac{\bm w_1^H  \bm R_{signal} \bm w_1}
        {\bm w_1^H \bm R_{spill} \bm w_1 +
       \bm w_1^H \bm R_{rec} \bm w_1 + \bm w_1^H \bm R_{sky} \bm w_1} \nonumber \\
        & = & \frac{\bm w_1^H \bm R_{signal} \bm w_1}{\bm w_1^H \bm N \bm w_1}
\label{snr1}
\eea 
where 
\be
\bm w_1  =  \bm A^H \bm G^H \bm w,
\label{w1}
\ee
is the modified weights and
$\bm R_{signal}, \bm R_{spill}, \bm R_{rec}$ and $\bm R_{sky}$ are the
open circuit voltage correlations corresponding to source, spillover
noise, receiver noise and sky background noise respectively. 
We refer to $\bm R_{signal}$ as the signal matrix, $\bm R_{spill}$
as the spillover matrix, $\bm R_{rec}$ as the receiver matrix and
$\bm R_{sky}$ as the sky matrix. Further, we define  
\be 
\bm N  \equiv \bm R_{spill} + \bm R_{rec} + \bm R_{sky},
\label{nmatrix}
\ee
and refer to as the noise matrix. Since $\bm N$ is a correlation matrix, 
it is Hermitian. Therefore it can be uniquely decomposed into 
products of two matrices (i.e. $\bm N = \bs{\mathcal{N}} \bs{\mathcal{N}}$) using the 
eigen structure of $\bm N$ \citep{hudson1981}. Thus Eq.~\ref{snr1} can be written as
\be
\textrm{SNR} = \frac{\bm w_2^H \bs{\mathcal{N}^{-1}} \bm R_{signal}  \bs{\mathcal{N}^{-1}} \bm w_2}{\bm w_2^H \bm w_2} 
\label{snr2}
\ee
where 
\be
\bm w_2 = \bs{\mathcal{N}} \bm w_1.
\label{w2}
\ee
We define the system characteristics matrix as
\be
\bm M \equiv \bs{\mathcal{N}^{-1}} \bm R_{signal}  \bs{\mathcal{N}^{-1}}.  
\label{charmat}
\ee
From Eq.~\ref{snr2} the maximum SNR is given by the maximum eigenvalue, $e_{max}$, 
of $\bm M$. Using Eq.~\ref{w1} and \ref{w2}, the weight vector 
$\bm w_{max}$ that needs to be applied at the output to
get the maximum SNR is
\be
\bm w_{max} = \left(\bm G^H\right)^{-1} \left(\bm A^H\right)^{-1} \bs{\mathcal{N}^{-1}} \bm w_{2max} 
\label{wtvec}
\ee
where $\bm w_{2max}$ is the eigenvector corresponding to $e_{max}$.

We now convert $e_{max}$ to the ratio $\frac{T_{sys}}{\eta}$. 
The power spectral density due to the source is proportional 
to the antenna temperature $T_A$,
which can be written as
\be
T_A = \frac{1}{2} \frac{S_{source}}{k_B} A_{ap} \eta_{ap}
\ee
where $S_{source}$ is the flux density of the source used for the measurement,
$A_{ap}$ is the aperture area of the antenna, $k_B$ is the Boltzmann constant
and $\eta_{ap} = \eta_{tap} \eta_{spill}$
is the aperture efficiency of the system. $\eta_{tap}$ and $\eta_{spill}$ are the
taper and spillover efficiencies respectively. The off-source power spectral
density is
proportional to the system temperature $T_{sys}$. Then the SNR can be
written as
\be
\textrm{SNR} = \frac{T_A}{T_{sys}}.
\label{snr3}
\ee
Using Eq.~\ref{snr2} and \ref{snr3}, the best performance of PAF is 
\be
\frac{T_{sys}}{\eta_{ap}} = \frac{S_{source} A_{ap}}{2 k_B e_{max}}.
\label{tsyseta}
\ee

The computation of the signal, receiver, 
spillover, and sky correlation matrices in order to construct $\bm M$
is discussed below. We first develop
the method used to compute the open circuit voltage at the output of 
the array due to incident radiation, which is presented in
Section~\ref{recmode}.

\subsection{PAF in Receiving mode}
\label{recmode}

\begin{figure}[t]
\begin{tabular}{cc}
\includegraphics[width=3.0in, height=3.3in, angle =0]{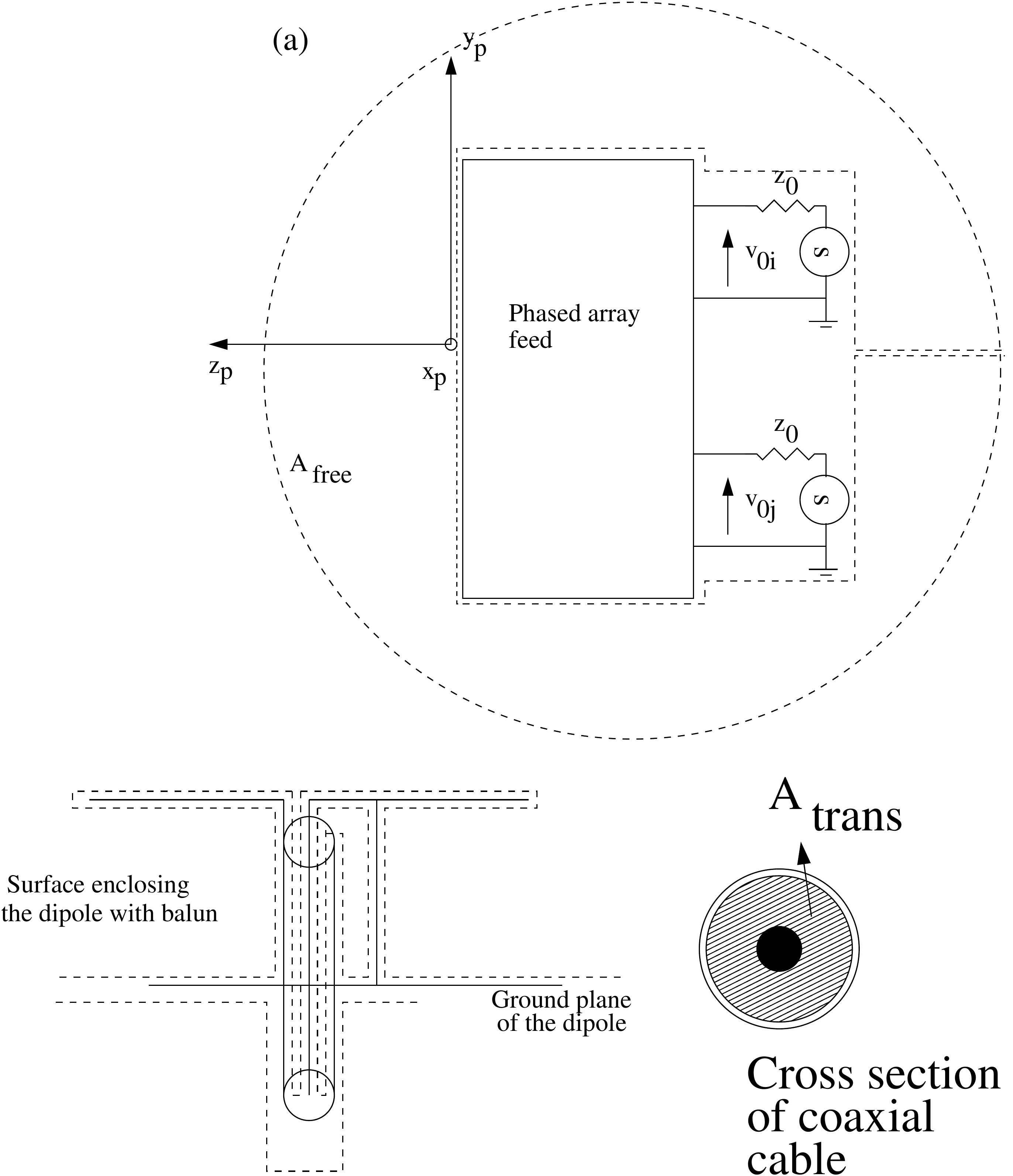} &   
\includegraphics[width=3.0in, height=3.3in, angle =0]{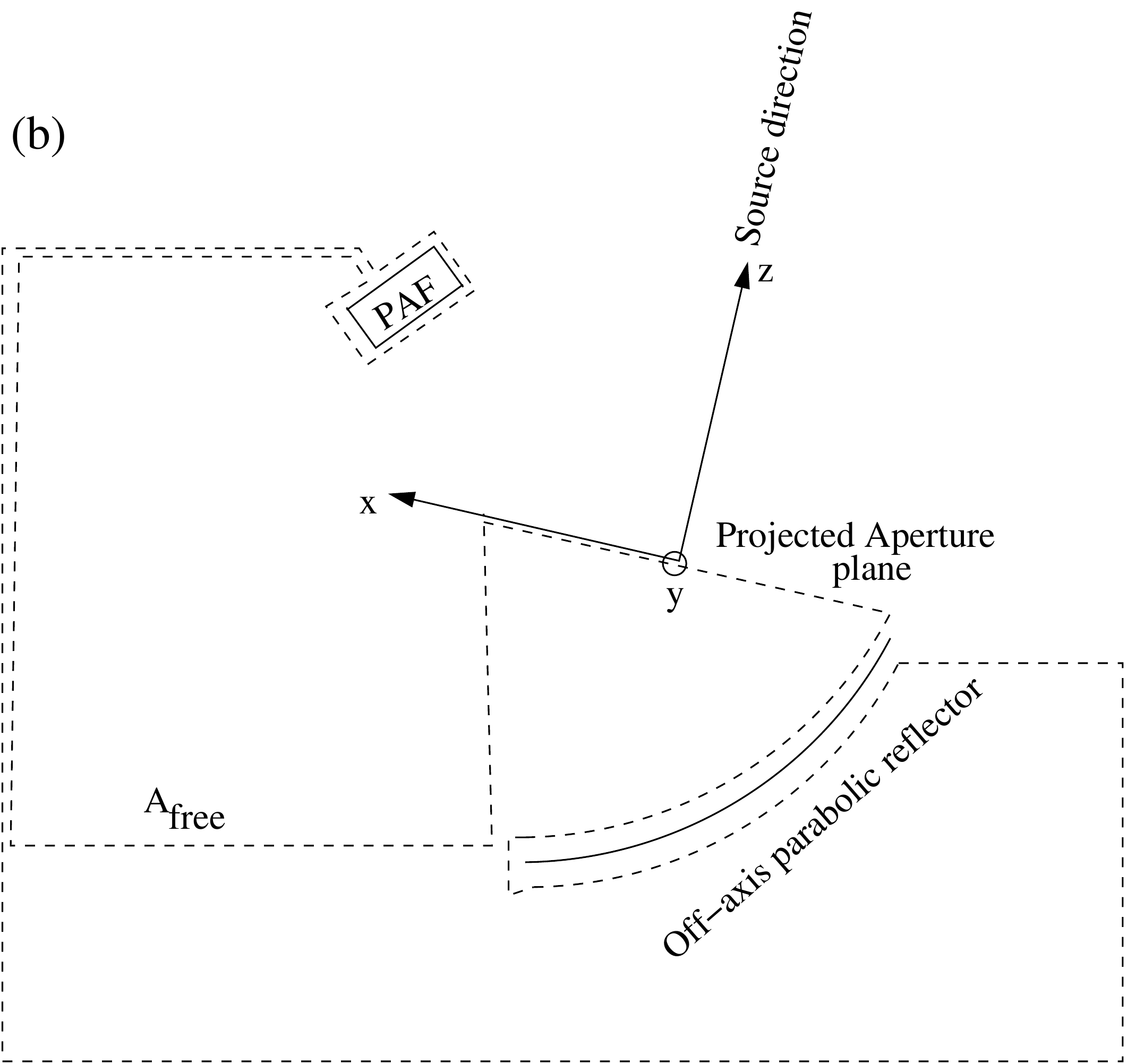} \\
\end{tabular}
\caption{(a) Schematic showing the enclosed source free volume for the application
of Lorentz theorem (see Section~\ref{recmode}). Details of the source free
surface enclosing the dipole and balun and inside the transmission line is also
shown. (b) PAF on an off-axis parabolic reflector and the corresponding
enclosed source free volume used to get the open circuit voltage due to a source. 
}
\label{fig2}
\end{figure}

To compute the open circuit voltage at the output of the PAF
in receiving mode, we start with the Lorentz reciprocity theorem. 
This theorem is derived from Maxwell's equation with no further 
assumptions made. The particular form of the theorem we use is
\be
\oint_A \vec{\mathcal{L}} \cdot \hat{n}\; \textrm{d}A  =  0
\label{lorentz}
\ee
where
\be 
\vec{\mathcal{L}}  \equiv  \vec{\mathcal{E}} \times \vec{\mathcal{H}}_r -
         \vec{\mathcal{E}}_r \times \vec{\mathcal{H}}.
\ee
The fields $\vec{\mathcal{E}}, \vec{\mathcal{H}}, \vec{\mathcal{E}_r}$
and $\vec{\mathcal{H}_r}$ are two solutions to Maxwell's equation within
a source free volume and on the enclosed surface $A$, $\hat{n}$ is the
unit vector inward to the surface normal to the elementary area $\textrm{d}A$
\citep{clarkebrown1980}. 
We consider $\vec{\mathcal{E}}, \vec{\mathcal{H}}$ as the 
electric and magnetic fields, respectively, when the PAF is in transmitting mode,
$\vec{\mathcal{E}_r}, \vec{\mathcal{H}_r}$
as the electric and magnetic fields, respectively, incident on the PAF
when the array is in receiving mode. By appropriately choosing the surface, the integration can be
broken up into two parts -- (a) integration over the surface area, $A_{trans}$ 
inside the transmission
line and (b) the integration over the surface, $A_{free}$ outside the PAF
(see Fig.~\ref{fig2}). The integral in
all other parts of the surface close to the metal surface is zero since the
tangential component of the electric field vanishes. We consider the ports
of the PAF are connected to the signal sources with source impedance $z_0$ during
transmission, and during reception, all ports are terminated with the impedance $z_0$.
As shown in Fig.~\ref{fig2}, the $x_p, y_p, z_p$ coordinate system is 
located on the PAF. The axis $z_p$ and the unit vectors $\hat{u}^p_z$
along its direction are parallel to the transmission line.
The field in the wave guide for the transmission and reception cases can be written as
\bea
\vec{\mathcal{E}} & = & \sqrt{z_0} (a_i + b_i) \vec{e}(x_p,y_p) e^{-j \beta z_p} \label{tranf1} \\
\vec{\mathcal{H}} & = & \frac{1}{z_0} \hat{u}^p_z \times \vec{\mathcal{E}} \label{tranf2}\\
\vec{\mathcal{E}_r} & = & \sqrt{z_0} b_{r_i} \vec{e}(x_p,y_p) e^{j \beta z_p} \\
\vec{\mathcal{H}_r} & = & -\frac{1}{z_0} \hat{u}^p_z \times \vec{\mathcal{E}_r}
\eea
where $a_i$, $b_i$ are the traveling wave amplitudes in the transmission case,
$b_{r_i}$ is the traveling wave amplitude in the receiving case, $\vec{e}(x_p,y_p)$ is
the normalized electric field inside the transmission line and $\beta$ is the
propagation constant in the transmission line. $\vec{e}(x_p,y_p)$
is a function of the geometric parameters of the transmission line \citep{pozar2005}.
The integral over $A_{trans}$ can then be written as
\be
\int_{A_{trans}} \vec{\mathcal{L}} \cdot \hat{n}\; \textrm{d}A  =
-2 \sum_{i=1,M} b_{r_i} (a_i + b_i) \int_{A_{trans}} |\vec{e}(x_p,y_p)|^2 \textrm{d}A \\
\label{inttrans}
\ee
The transmitted power in {\em each} transmission line or port can be written as
\be
\textrm{Re}\frac{1}{2} \int_{A_{trans}} \vec{\mathcal{E}} \times \vec{\mathcal{H}}^* \cdot 
\hat{n} \; \textrm{d}A  =  \frac{1}{2} |a_i|^2 + \frac{1}{2} |b_i|^2
\label{tranpwr1}
\ee
Using Eq.~\ref{tranf1} and \ref{tranf2}, Eq.~\ref{tranpwr1} can be written as
\be 
(a_i + b_i) (a_i + b_i)^* \int_{A_{trans}} |\vec{e}(x_p,y_p)|^2 \textrm{d}A  = 
            |a_i|^2 + |b_i|^2.
\label{tranpwr}
\ee
Substituting Eq.~\ref{tranpwr} in Eq.~\ref{inttrans}, the integration over $A_{trans}$ becomes
\bea
\int_{A_{trans}} \vec{\mathcal{L}} \cdot \hat{n}\; \textrm{d}A & = & 
                     -2 \sum_{i=1,M} b_{r_i} (a_i - b_i) \nonumber \\
          & = & - \sum_{i=1,M} v_{oc_i} i_{0_i}  \nonumber \\
          & = & - \bm V_{oc}^T \bm I_0 
\label{parta}
\eea
where $\bm V_{oc}$ is the open circuit (peak) voltage vector 
with elements $v_{oc_i}$, induced by the receiving field and $\bm I_0$ is the
excitation (peak) current vector with elements $i_{0_i}$,
which produces the transmission beam pattern (see Appendix~\ref{A1} for
the relationship between traveling wave amplitude and voltages and currents). 
Substituting Eq.~\ref{parta} in Eq.~\ref{lorentz} we get
\bea
\bm V_{oc}^T \bm I_0 & = & \int_{A_{free}} \vec{\mathcal{L}} \cdot \hat{n}\; \textrm{d}A \nonumber \\
        & = & \int_{A_{free}} \left(\vec{\mathcal{E}} \times \vec{\mathcal{H}_r} -
                 \vec{\mathcal{E}_r} \times \vec{\mathcal{H}}\right) \cdot \hat{n}\; \textrm{d}A
\label{lorenz1}
\eea
The right hand side of Eq.~\ref{lorenz1} can be written as a summation
analogues to the left hand side, i.e.
\bea
\vec{\mathcal{E}} & = & \sum_{i=1,M} v_{0_i} \vec{\mathcal{E}}^e_i \nonumber \\
                   & = & \bm V_0^T \bs{\vec{\mathcal{E}}^e}
\label{embed}
\eea
where the elements of the vector $\bs{\vec{\mathcal{E}}^e}$ are $\vec{\mathcal{E}}^e_i$, which
are the radiation patterns of the PAF for $v_{0_i} = 1V$ for $i = j$ and 
0 V for $i \neq j$. We define the elements of the vector $\bs{\vec{\mathcal{E}}^e}$ as the
{\em embedded beam patterns}. Using Eq.~\ref{embed} we can write Eq.~\ref{lorenz1} 
as 
\be
\bm V_{oc}^T \bm I_0  =  \bm V_0^T \int_{A_{free}} 
      \left(\bs{\vec{\mathcal{E}}^e}^T \times \bs{\mathcal{I}} \vec{\mathcal{H}_r} -
      \bs{\mathcal{I}} \vec{\mathcal{E}_r} \times \bs{\vec{\mathcal{H}}^e} \right) 
      \cdot \hat{n}\; \textrm{d}A,
\label{lorenz2}
\ee
where $\bs{\mathcal{I}}$ is the identity matrix, $\bs{\vec{\mathcal{H}}^e}$ is the embedded magnetic field
patterns (see Appendix~\ref{A9} for an explanation of the notation used in Eq.~\ref{lorenz2}). 
Since Eq.~\ref{lorenz2} is true for arbitrary excitation voltages it follows that
\be
\bm V_{oc} = \bm Z \int_{A_{free}} \left(\bs{\vec{\mathcal{E}}^e}^T 
             \times \bs{\mathcal{I}} \vec{\mathcal{H}_r} -
             \bs{\mathcal{I}} \vec{\mathcal{E}_r} \times \bs{\vec{\mathcal{H}}^e}\right) 
             \cdot \hat{n}\; \textrm{d}A,
\label{ocvolt}
\ee
where we have made use of the relation $\bm V_0 = \bm Z \bm I_0$ between the excitation
voltage and current (see Appendix~\ref{A1}). We show in Appendix~\ref{A3} 
that Eq.~\ref{ocvolt} gives the voltage correlation expected from
the first and second law of thermodynamics when the PAF is in thermal
equilibrium with a black body field. Below we evaluate the right 
hand side of Eq.~\ref{ocvolt} for two cases
of interest here, {\em viz} (a) incident plane wave radiation field and (b) incident radiation 
from a (compact) source when the PAF is on the telescope.

\subsubsection{Plane Wave}

Let the PAF be incident by a plane wave of field $\vec{E}_{inc} = E_{inc}\hat{p}$,
where $\hat{p}$ denotes a vector parallel to the polarization of the electric
field. The incident direction $\vec{r}$ is parallel to the propagation vector
$\vec{k}_{inc}$. The surface considered for the integration is shown in
Fig.~\ref{fig2}a. Using stationary-phase algorithm,  the integral over $A_{free}$
can be written as \citep{clarkebrown1980} 
\be
\int_{A_{free}} \vec{\mathcal{L}} \cdot \hat{n}\; \textrm{d}A
= \frac{jk\lambda^2r e^{jkr}}{\pi z_f} E_{inc} \hat{p} \cdot \vec{\mathcal{E}}
\label{pwave}
\ee
where $\vec{\mathcal{E}}$ is the beam pattern of the PAF for an arbitrary excitation, 
$r = |\vec{r}|$, $k = |\vec{k}_{inc}|$,
$z_f$ is the free space impedance. This equation is valid for excitation that
produces the embedded beam pattern as well. Thus by writing Eq.~\ref{pwave} 
in terms of the embedded beam pattern and substituting in Eq.~\ref{ocvolt}
we get (see Appendix~\ref{A9} for an explanation of notation used in Eq.~\ref{ocvplane}) 
\be
\bm V_{oc} = \frac{j 4\pi r e^{jkr}}{kz_f} E_{inc} \bm Z \bs{\mathcal{I}} \hat{p} \cdot \bs{\vec{\mathcal{E}^e}}.
\label{ocvplane}
\ee
 
\subsubsection{PAF on the telescope}

We now consider the case when the PAF is placed at the prime focus of a telescope
with a parabolic reflector. An appropriate closed surface enclosing a source free region for
the integration is shown in Fig.~\ref{fig2}b for a (off-axis) parabolic antenna (eg. 
Green Bank Telescope; GBT). The source is located at an angle $(\theta_s, \phi_s)$ from 
the boresight of the telescope.
The major contribution to the integral over the surface $A_{free}$ comes from the 
projected aperture plane (see Fig.~\ref{fig2}b). 
Although the surface geometry of $A_{free}$ has changed
compared to Fig.~\ref{fig2}a, the part (a) of the surface integral 
discussed in Section~\ref{recmode} remain the same. The part of
the surface close to the reflector will not contribute to the integral because
the tangential component of the field should be zero on the surface of the metal.
The phase difference between the incoming waves and the embedded beam patterns
changes rapidly in all other parts of the surface $A_{free}$ 
so that the net contribution to the integral will be zero. Thus
\be
\int_{A_{free}} \vec{\mathcal{L}} \cdot \hat{n}\; \textrm{d}A =
\int_{A_{pap}} \vec{\mathcal{L}} \cdot \hat{n}\; \textrm{d}A
\ee
where $A_{pap}$ denotes integration over the projected aperture plane.
Using Eq.~\ref{ocvolt}, the open circuit voltage due to the source is 
\be
\bm V_{oc} = \bm Z \int_{A_{pap}} 
     \left(\bs{\vec{\mathcal{E}}^e_{pap}}^T \times \bs{\mathcal{I}} \vec{\mathcal{H}_r} -
           \bs{\mathcal{I}} \vec{\mathcal{E}_r} \times \bs{\vec{\mathcal{H}}^e_{pap}} \right)
           \cdot \hat{n}\; \textrm{d}A.
\label{ocvpap}
\ee
Here $\vec{\mathcal{H}}_r$ and $\vec{\mathcal{E}}_r$ are fields on the projected
aperture plane due to the source, $\bs{\vec{\mathcal{E}}^e_{pap}}$ and $\bs{\vec{\mathcal{H}}^e_{pap}}$ 
are embedded beam patterns propagated to the projected aperture plane after reflection
from the parabolic surface. By representing the incident field due to the source  
on the projected aperture plane
as $\vec{E}_{inc} = E_{inc}\hat{p}$ and the embedded magnetic
field as $\bs{\vec{\mathcal{H}}^e_{pap}} = - \frac{1}{z_f} \bs{\mathcal{I}} \hat{n} \times 
\bs{\vec{\mathcal{E}}^e_{pap}}$, Eq.~\ref{ocvpap} becomes 
\be
\bm V_{oc} = 2 \frac{\bm Z}{z_f} \int_{A_{pap}} E_{inc} 
                      \bs{\mathcal{I}} \hat{p} \cdot \bs{\vec{\mathcal{E}}^e_{pap}} \textrm{d} A.
\label{ocvpap1}
\ee
Note that $E_{inc}$ and $\hat{p}$ are in general functions of positions in the projected
aperture plane.

For the case of an astronomical source the incident
field can be broken up into two orthogonal polarizations, the
directions are conveniently taken along the x and y axis, i.e.,
\be
\vec{E}_{inc} = E_{inc,x}\hat{u}_x + E_{inc,y}\hat{u}_y.
\ee 
Here $E_{inc,x}$ and $E_{inc,y}$ are random variables and
for simplicity we assume that they are uncorrelated, i.e., the
source is unpolarized. Eq.~\ref{ocvpap1} can then be written as
\be
\bm V_{oc} = 2 \frac{\bm Z}{z_f}  
        \int_{A_{pap}} \left( \bs{\mathcal{I}} E_{inc,x} \bs{\mathcal{E}^e_{pap,x}} 
             + \bs{\mathcal{I}} E_{inc,y} \bs{\mathcal{E}^e_{pap,y}} \right) \textrm{d} A
\label{ocvsou}
\ee
where $\bs{\mathcal{E}^e_{pap,x}}$ and $\bs{\mathcal{E}^e_{pap,y}}$ are vectors
of the x and y components of the embedded beam patterns, 
$\bs{\mathcal{E}^e_{pap}}$, respectively. We show in Appendix~\ref{A5}
that this equation gives the well known relationship between source antenna
temperature and aperture efficiency.

\subsection{The Spillover matrix}
\label{spillmat}

The PAF placed at the prime focus of the telescope
picks up thermal radiation from the ground. These radiation
are received by the PAF from the solid angles
larger than the solid angle subtended by the parabolic reflector.
We refer to the total solid angle from which the ground radiation
is picked up as $\Omega_{spill}$. We consider the ground radiation field 
as a set of plane waves incident at different directions. The open circuit
voltage due to a plane wave in a direction $\vec{r}$ is given by Eq.~\ref{ocvplane}.
The correlation of this voltage is
\bea
\bm R_1 & = & \langle \bm V_{oc1} \bm V_{oc1}^H \rangle \nonumber \\
        & = & \frac{(4\pi)^2 r^2} {k^2 z_f^2}
    \left< \left(E_{inc} \bm Z \bs{\mathcal{I}} \hat{p} \cdot \bs{\vec{\mathcal{E}^e}} \right)
    \left(\bs{\vec{\mathcal{E}^e}}^H \cdot \bs{\mathcal{I}} \hat{p} \bm Z^H E_{inc}^* \right) \right>
\label{spill1}
\eea
where $\bm R_1$ is the correlation of open circuit voltage $\bm V_{oc1}$ due to a plane wave. 
For thermal radiation, $E_{inc}$ is a stationary random variable and $\hat{p}$ has random 
orientation. Therefore, without loss of generality, $E_{inc}\hat{p}$ can be decomposed 
into two orthogonal directions of the field $\vec{\mathcal{E}^e_i}$. 
The two polarization components of $E_{inc}\hat{p}$ are uncorrelated for the
thermal radiation field. 
Keeping in mind that a port of PAF is sensitive to one of the polarizations, Eq.~\ref{spill1} 
can be written as
\be
\bm R_1  =  \frac{(4\pi)^2 r^2}{k^2 z_f^2} \frac{\langle E_{inc}E_{inc}^* \rangle}{2} 
            \bm Z \bs{\vec{\mathcal{E}^e}} \cdot \bs{\vec{\mathcal{E}^e}}^H \bm Z^H.
\label{spill2}
\ee
In the Rayleigh-Jeans approximation the thermal field is related to the brightness temperature 
\be
\frac{1}{z_f} \langle E_{inc} E_{inc}^* \rangle = \frac{2 k_B T_g}{\lambda^2} \textrm{d}\Omega,
\label{spill3}
\ee
where $T_g$ is the physical temperature of the ground. 
Using Eq.~\ref{spill3}, $\bm R_1$ can be written as
\be
\bm R_1  =  \frac{4 k_B T_g r^2}{z_f} 
        \bm Z \bs{\vec{\mathcal{E}^e}} \cdot \bs{\vec{\mathcal{E}^e}}^H \bm Z^H \textrm{d} \Omega.
\ee
Since the spillover noise is due to thermal radiation from the ground, the voltages due to 
plane waves from different directions are uncorrelated. The resultant
voltage correlation due to spillover noise is then 
\bea 
\bm R_{spill} & = & \int_{\Omega_{spill}} \bm R_1 \text{d}\Omega, \nonumber \\ 
              & = & \frac{4 k_B T_g}{z_f} \bm Z \left( 
        \int_{\Omega_{spill}} \bs{\vec{\mathcal{E}^e}}\cdot\bs{\vec{\mathcal{E}^e}}^H r^2 \textrm{d}\Omega 
                 \right) \bm Z^H, \nonumber \\
& = & \frac{4 k_B T_g}{z_f} \bm Z \left(
        \int_{\Omega_{spill}} \bs{\vec{E^e}} \cdot \bs{\vec{E^e}}^H \textrm{d}\Omega
                 \right) \bm Z^H, \label{spill4} \\
& = & \frac{4 k_B T_g}{z_f} \bm Z \bm C_{ce1} \bm Z^H. 
\label{spill5}
\eea
where $\Omega_{spill}$ is the solid angle over which the PAF embedded beam patterns 
are receiving radiation from ground. We also made use of Eq.~\ref{farf} to introduce 
the embedded beam patterns $\bs{\vec{E^e}}$ in Eq.~\ref{spill4}. We define
\be
\bm C_{ce1} \equiv \int_{\Omega_{spill}} \bs{\vec{E^e}}\cdot \bs{\vec{E^e}}^H \textrm{d}\Omega,
\label{spill6}
\ee
which is the correlation matrix of the embedded beam patterns integrated over $\Omega_{spill}$.
In Appendix~\ref{A4} we use Eq.~\ref{spill5} to expressed the spillover noise
in terms of a physical temperature, referred to as spillover temperature. We also
show that the derived spillover temperature gives the well known relationship
between the ground temperature and the spillover efficiency when applied to
a single port antenna.

\subsection{The Signal Matrix}

The open circuit voltage at the ports of the PAF when observing an
astronomical source is given by Eq.~\ref{ocvsou}. The correlation
of the voltage due to source, $\bm V_{oc_s}$ can be written as
\bea
\bm R_{signal} & = & \langle \bm V_{oc_s} \bm V_{oc_s}^H \rangle \nonumber \\
               & = & \frac{4}{z_f^2} \bm Z \Biggl< \left( 
      \int_{A_{pap}} \bs{\mathcal{I}} E_{inc,x} \bs{\mathcal{E}^e_{pap,x}} 
             + \bs{\mathcal{I}} E_{inc,y} \bs{\mathcal{E}^e_{pap,y}} \textrm{d} A \right) \nonumber \\
              &   &   \left(
      \int_{A_{pap}}\bs{\mathcal{E}^e_{pap,x}}^H \bs{\mathcal{I}} E_{inc,x}^* 
             + \bs{\mathcal{E}^e_{pap,y}}^H \bs{\mathcal{I}} E_{inc,y}^* \textrm{d} A \right) 
      \Biggr> \bm Z^H \label{signal0} \\
             &  = & \frac{4}{z_f} \frac{S_{source}}{2} \bm Z \left( 
              \int_{A_{pap}} \bs{\vec{\mathcal{E}}^e_{pap}} \textrm{d} A \right) \cdot  \left(
              \int_{A_{pap}} \bs{\vec{\mathcal{E}}^e_{pap}} \textrm{d} A \right)^H \bm Z^H \nonumber \\
            &  = & \frac{2 S_{source}}{z_f} \bm Z \bm C_{Ie} \bm Z^H.
\label{signal1}
\eea
where we define 
\be
\bm C_{Ie} \equiv \left(\int_{A_{pap}} \bs{\vec{\mathcal{E}}^e_{pap}} \textrm{d} A \right) \cdot
                  \left(\int_{A_{pap}} \bs{\vec{\mathcal{E}}^e_{pap}} \textrm{d} A \right)^H
\label{forsig}
\ee
which is the correlation matrix of the {\em net} aperture fields due to embedded beam patterns. 
To arrive at Eq.~\ref{signal1}, we used the relationship 
$\frac{1}{z_f}\langle E_{inc,x}E_{inc,x}^* \rangle = 
\frac{1}{z_f}\langle E_{inc,y}E_{inc,y}^* \rangle = \frac{S_{source}}{2}$ and, for simplicity, 
$\langle E_{inc,x}E_{inc,y}^* \rangle = \langle E_{inc,y}E_{inc,x}^* \rangle = 0$
for an unpolarized source of flux density $S_{source}$. 
Note that $\bm C_{Ie}$ is a function of source position $\theta_s, \phi_s$.
In Appendix~\ref{A5}
we use Eq.~\ref{signal1} to derive an expression for the source antenna
temperature. We also show that the expression for
antenna temperature reduces to the well known 
relationship between source flux density, physical area of the telescope aperture, 
and aperture efficiency for a single port antenna.

\subsection{The Receiver Matrix}

\begin{figure}[!t]
%\centering
\includegraphics[width=6.5in, height=2.5in, angle=0]{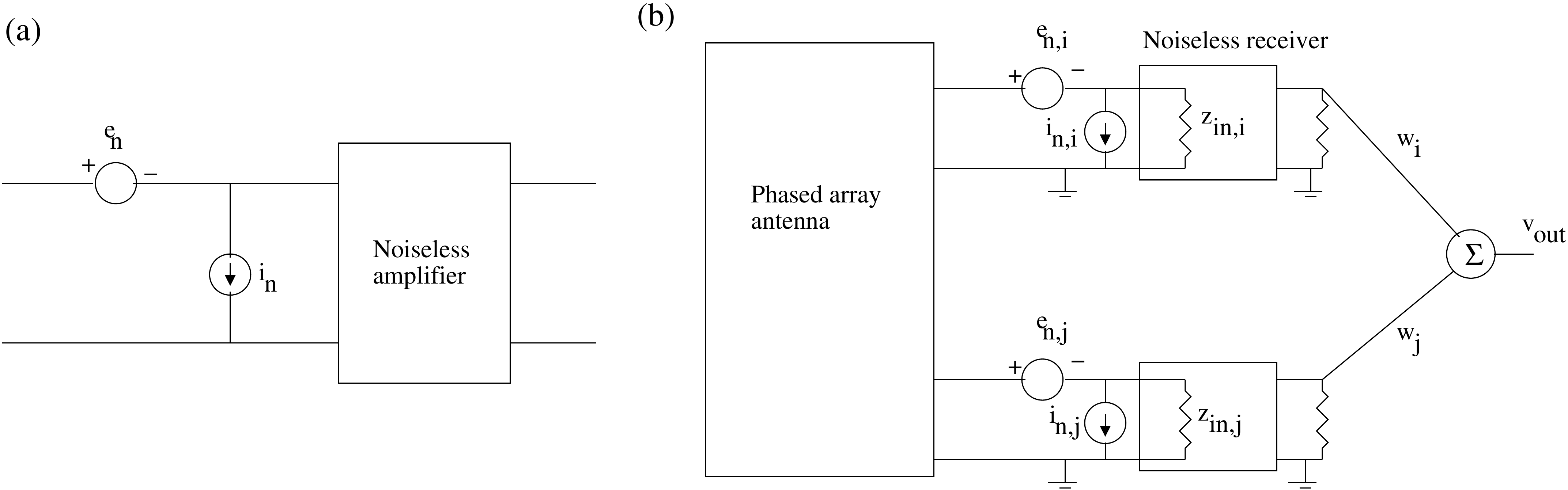}
\caption{(a)Schematic showing the noise voltage and current generators
to model the noise properties of a low-noise amplifier (LNA).
(b) The network model of the PAF system along with the noise voltage
and current generators of the LNA. $z_{in,i}$, $z_{in,j}$ etc are the 
input impedance of the amplifier
}
\label{fig3}
\end{figure}

To derive the receiver matrix, we need the noise parameters, $R_n$, $g_n$ and
$\rho$ (defined below), of the amplifier in addition to the impedance matrix of the PAF.
It is well known that the spectral (or spot) noise factor of noisy two-ports
is equivalent to noise-free two-ports plus two (in general) partially correlated
noise generators\citep{hausetal1960}. A convenient representation is to place
a noise voltage generator ($e_n$) in series and 
a noise current generator ($i_n$) in parallel to the noise-free two-ports (see Fig~\ref{fig3}a).
The noise voltage and current fluctuations per unit frequency interval, $\langle e_n^2\rangle$ and
$\langle i_n^2\rangle$ respectively, and their complex correlation coefficient
per unit frequency interval $\rho$ are expressed in terms
of the noise parameters as
\bea
\langle e_n^2\rangle & = & 4k_BT_0R_n \label{nf1} \\
\langle i_n^2\rangle & = & 4k_BT_0g_n  \label{nf2}\\
\rho    & = & \frac{\langle e_n^*i_n\rangle}{\sqrt{\langle e_n^2\rangle \langle i_n^2\rangle}}
\label{nf3}
\eea
where $k_B$ is the Boltzmann constant, $T_0 = 290$ K,  $R_n$ is the noise resistance
and $g_n$ is the noise conductance.

Fig.~\ref{fig4}b shows the network model of PAF along with the noise generators
of the LNA. For simplicity, we assume the noise contribution from the  receiver stages
following the LNA can be neglected. From network analysis, the open circuit voltage 
vector, $\bm V_{oc_r}$, due to LNA noise is
\be
\bm V_{oc_r}  =  \bm E_n + \bm Z \bm I_n, 
\ee
where $\bm E_n$ is a diagonal matrix with elements $e_{n,i}$, the noise voltage
generators of the LNA, $\bm I_n$ is a diagonal matrix with elements $i_{n,i}$,
the noise current generators of the LNA and $\bm Z$ is the impedance matrix of
the PAF. The open circuit voltage correlation is then
\bea
\bm R_{rec} & =  & \langle \bm V_{oc_r} \bm V_{oc_r}^H \rangle \nonumber   \\
      & =  & \langle \bm E_n\bm E_n^H\rangle + \bm Z\; \langle \bm I_n \bm E_n^H \rangle + \nonumber   \\
      &    & \langle \bm E_n \bm I_n^H\rangle \; \bm Z^H + 
             \bm Z \; \langle \bm I_n \bm I_n^H\rangle \; \bm Z^H. 
\label{ww1}
\eea
Here $\langle \bm E_n \bm E_n^H\rangle$, $\langle \bm I_n \bm I_n^H\rangle$ and 
$\langle \bm I_n \bm E_n^H\rangle$
are diagonal matrices of noise fluctuations and their correlations.
For identical LNAs connected to the PAF, using Eq.~\ref{nf1} to \ref{nf3}, we get
\be
\bm R_{rec}  =  4 k_B T_0 \; \Big(R_n \bs{\mathcal{I}} + \sqrt{R_n g_n} \;\big(\rho \bm Z + \rho^* \bm Z^H\big) + g_n \bm Z\bm Z^H \Big).
\label{tn}
\ee

In Appendix~\ref{A6}, we use Eq.~\ref{tn} to get the receiver temperature for the PAF.
We also show that the receiver temperature of PAF can be 
written as a generalization of the 
equation used to express the receiver temperature of a single antenna followed by 
an LNA. 

\subsection{The Sky Matrix}

The sky background radiation has components due to Cosmic microwave background (CMB), galactic
and extragalactic radiation. The open circuit voltage correlation due to these components
adds up since the voltages due to the different components are uncorrelated. The correlation
due to CMB can be derived by considering the PAF + telescope in a 
black-body radiation field with temperature $T_{cmb} = 2.7$ K. The voltage correlation
$\bm R_{cmb}$ in this case is (see Appendix~\ref{A2})
\be
\bm R_{cmb} = 2 k_B T_{cmb} (\bm Z + \bm Z^H).
\label{rcmb}
\ee
The voltage correlation due to other sky background components can be calculated
from a knowledge of their visibility function.  In this report, we take
\be
\bm R_{sky} \approx 2 k_B T_{sky} (\bm Z + \bm Z^H)
\label{sky}
\ee
where 
\be
T_{sky} = T_{cmb} + T_{bg,\nu_0} \left(\frac{\nu}{\nu_0}\right)^{-2.7}
\ee
is the temperature of the sky background at the observed off-source
position. Here $T_{bg,\nu_0}$ is the galactic background radiation
temperature at $\nu_0$ and $\nu$ is the frequency at which $\bm R_{sky}$
is computed.

\section{PAF model}
\label{pafmodel}

%\begin{figure}[t]
%\begin{tabular}{cc}
%\includegraphics[width=2.5in, height=3.3in, angle =0]{fig/pafpic.pdf} & 
%\includegraphics[width=4.3in, height=3.3in, angle =0]{fig/PAFmapping2.pdf}  
%\end{tabular}
%\caption{19-element Kite Array and the Kite dipole are shown on the left.
%The physical dimensions of the PAF are given on the right.
%}
%\label{fig4}
%\end{figure}

\begin{figure}[t]
\begin{tabular}{c}
\includegraphics[width=5.3in, height=2.3in, angle =0]{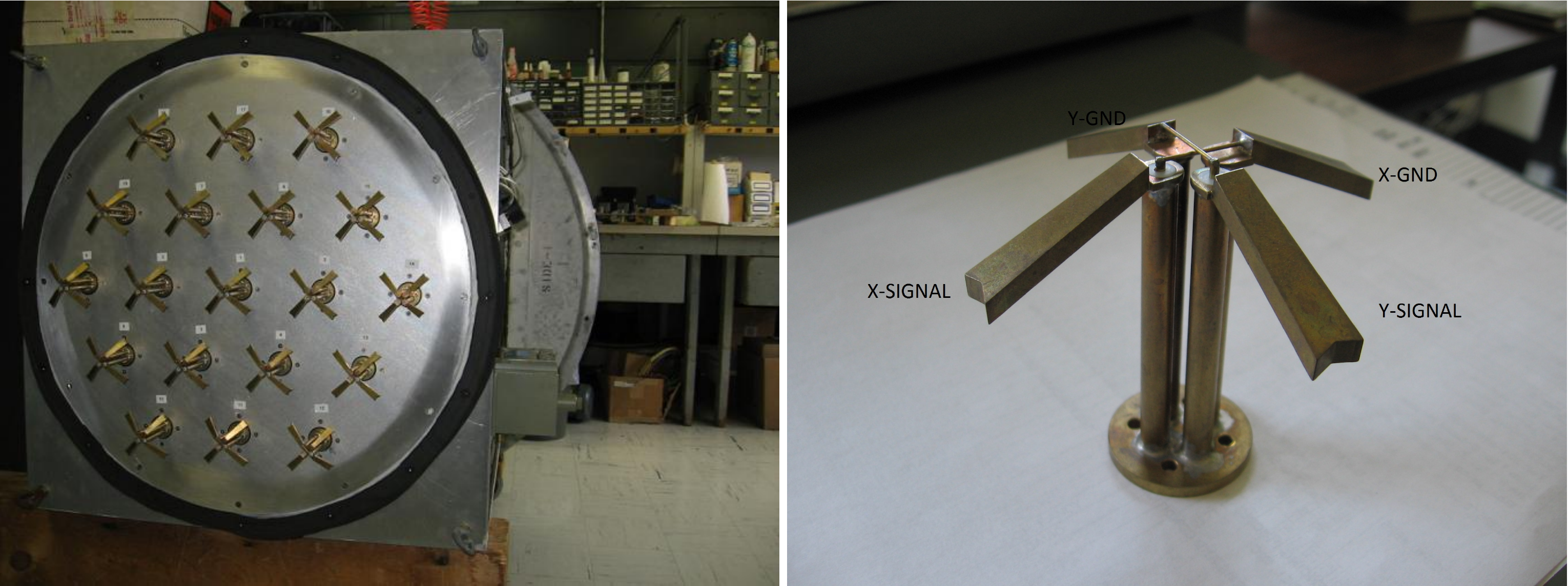} \\ 
\includegraphics[width=6.9in, height=6.0in, angle =0]{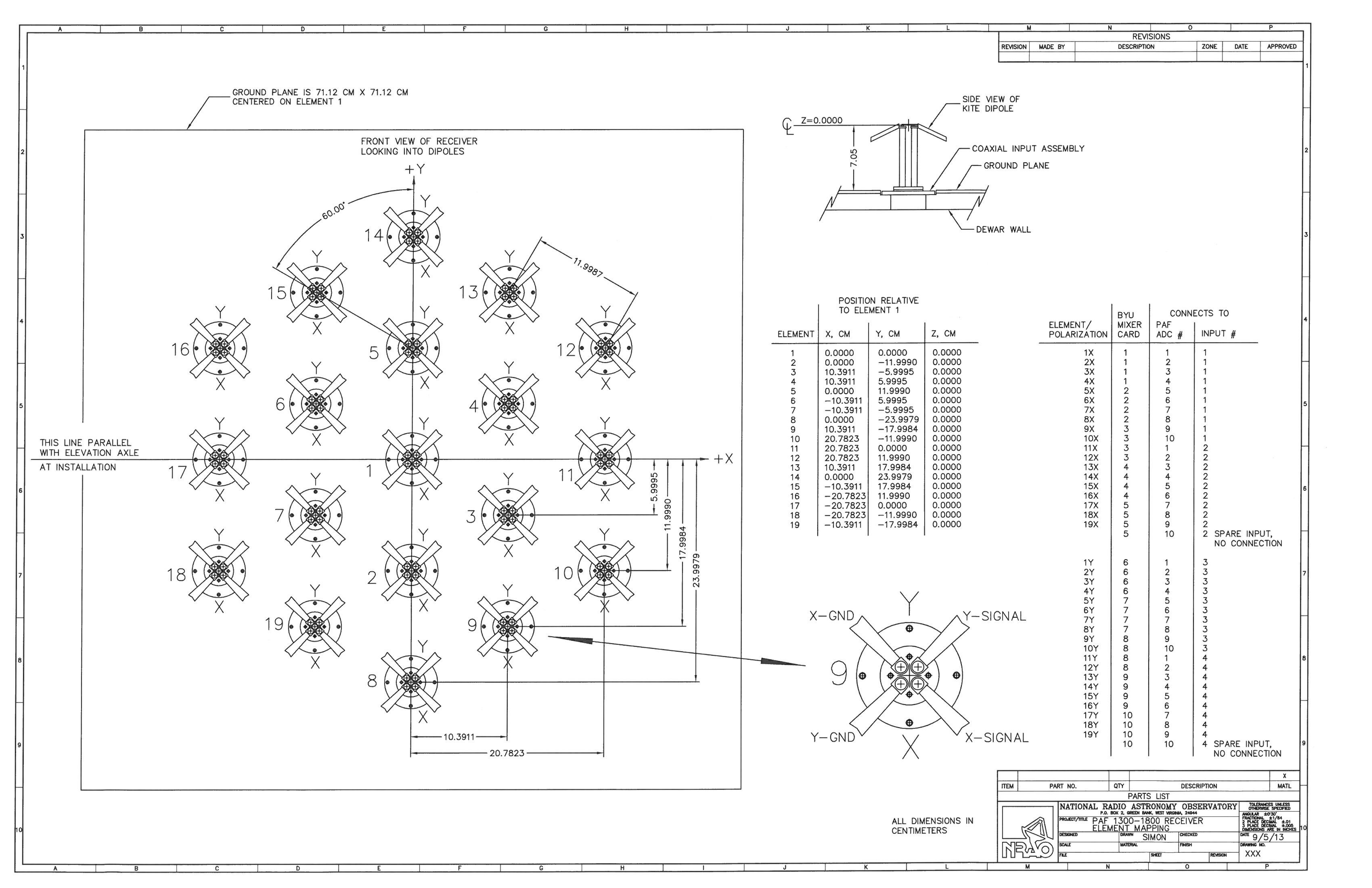}  
\end{tabular}
\caption{19-element Kite Array and the Kite dipole are shown on the top.
The physical dimensions of the PAF are given on the bottom.
}
\label{fig4}
\end{figure}

As described in Section~\ref{theory}, we need (1) the impedance matrix
of PAF, (2) the embedded beam pattern of the PAF, (3) the amplifier
noise parameters, (4) the telescope geometry and (5) the ground and
sky background temperature for modeling. The impedance matrix
and embedded beam patterns are obtained from simulation
using a commercial software package (CST). CST package is a ``software for the simulation 
of electromagnetic fields in arbitrary three-dimensional structures'' (
see \verb! https://en.wikipedia.org/wiki/Computer_Simulation_Technology!).
The mutual coupling effects between elements in the array are included in
the CST simulation.
To describe the steps involved in the modeling process we consider a 
19 dual polarized dipole PAF, shown in Fig.~\ref{fig4}. 
The dipoles are referred to as kite dipoles (see Fig.~\ref{fig4}). 
The dipoles are kept in front of a ground plane. The transmission 
lines at the end of the dipoles are 50 $\Omega$ co-axial cable. The
mechanical dimensions of the PAF are shown in Fig.~\ref{fig4}.

\subsection{CST simulation}
\label{cstpart}

\begin{figure}[t]
\includegraphics[width=6.8in, height=5.3in, angle =0]{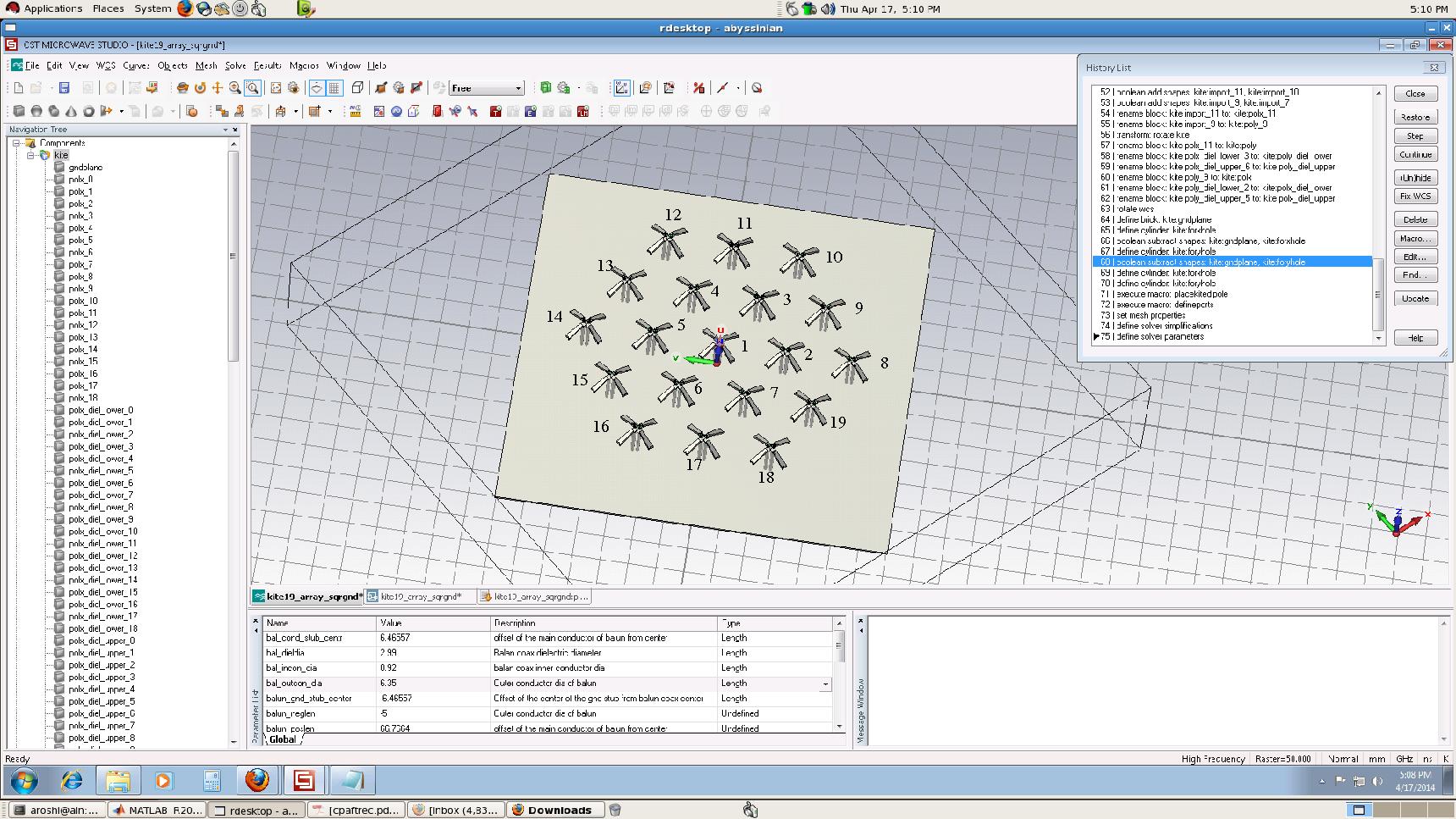}  
\caption{19-element Kite Array Mode in CST. The far-field patterns are
computed with respect to the x-y-z co-ordinate system. The origin of this
system is located at the intersection of ground plane and dipole 1.
}
\label{fig5}
\end{figure}

\begin{figure}[t]
\includegraphics[width=5.3in, height=5.3in, angle =0]{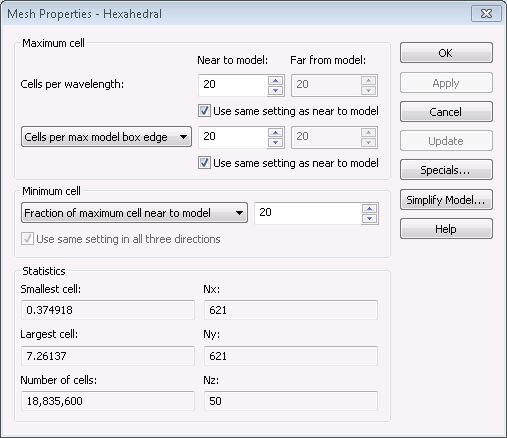}
\caption{Mesh parameters used for the simulation in CST.
}
\label{fig6}
\end{figure}

A mechanical model of the PAF is created in CST first (see Fig.~\ref{fig5}).
The model is created by importing the Autodesk Inventor 
(see \verb! https://en.wikipedia.org/wiki/Autodesk_Inventor!) CAD model of a
dipole pair. We wrote a Visual Basic program in CST to place copies
of the dipole at the 19 locations. The ground plane is created
in CST. The ground plane and dipoles are assigned with Perfect-electric-conductor (PEC)
in CST. The dielectric material in the coaxial cable are assigned
with (loss-less) teflon material. Waveguide ports
are connected to the 19$\times$2 coaxial input of the PAF for excitation. 
A hexahedral mesh size of $\lambda/20$ is used for running the Transient
solver. Other details on meshing are given in Fig.~\ref{fig6}. The solution
accuracy in the solver is set to -40 dB and the frequency range is set
to 1 -- 2 GHz. The solver takes about 200 hrs to
provide the results. The results of interest for the PAF modeling are (a)
Scattering matrix as a function of frequency and (b) the beam patterns
as a function of frequency. The CST computes the beam patterns when one 
port is excited with a 50 $\Omega$ source and all other ports are 
terminated with 50 $\Omega$. This computation is repeated for all the
19 $\times$ 2 ports. We export the Scattering matrix in TOUCHSTONE format
which are normalized to 50 $\Omega$ reference impedance. The beam patterns
are exported in `source format' and are sampled at 5\deg\ interval in
$\theta, \phi$. The coordinate system (x-y-z) used to obtain the beam patterns
is shown in Fig.~\ref{fig5}. The origin of this coordinate system is
located at the intersection of the ground plane with dipole 1. 
Some basic sanity checks on the CST products are done
using MATLAB programs as described in Appendix~\ref{A7} before further
proceeding with the modeling.   

\subsection{MATLAB simulation}

The MATLAB simulation starts with the computation of (a) the impedance matrix
from the Scattering matrix as described in Appendix~\ref{A1}, and (b) the 
embedded beam patterns as described in Appendix~\ref{A8}. The embedded
beam patterns are used to compute the beam correlation matrices
$C_{Ce1}$ (Eq.~\ref{spill6}) and $C_{Ie}$ (Eq.~\ref{forsig}). These
are functions of position angle of the source as well as the geometry
of the telescope. The correlation matrices are computed for 10 offset
angles away from boresight. The telescope geometry used is that of GBT.
The embedded beam patterns are propagated to the projected aperture plane using
Geometric optics approximation for each offset angle. To compute $\bm R_{rec}$ (Eq.~\ref{tn}),
we need the amplifier noise parameters.  These 
parameters are taken from S. Weinreb's design \citep{weinrebetal2009}. The
values used for the noise parameters are $R_n = 0.7\;\; \Omega$, 
$g_n = 1.3 \times 10^{-4}\;\; \Omega^{-1}$, $\rho = 0.1622 + j0.2040$,
$Z_{opt} = 72 + j15\;\; \Omega$ 
and that for the input impedance is 50 $\Omega$. The noise parameters
are measured at 1.6 GHz and assumed to be constant over 1 to 2 GHz. 
The values used
for other quantities needed to construct the system characteristics matrix 
are $T_g = 300$ K, $T_{cmb} = 2.7$ K, $T_{bg,\nu_0} = 0.7$ K, $\nu_0 = 1.42$ GHz
and a model for Virgo A flux density. The background temperature
is the temperature at the observed off-source position obtained
from \citep{reich1986}. The spectral index of
sky background temperature is taken as $-2.7$. The model used for
Virgo A flux density is 285 Jy at 1 GHz with a spectral index of $-0.856$.
The characteristics matrix $\bm M$ is then constructed using Eq.~\ref{charmat}.
The maximum eigenvalue of $\bm M$ is determined, which is used to
get the best $\frac{T_{sys}}{\eta_{ap}}$ (Eq.~\ref{tsyseta}). The weight
vector corresponding to the best $\frac{T_{sys}}{\eta_{ap}}$ are obtained
using Eq.~\ref{wtvec}. The gain matrix is taken as identity matrix to
get the weights. The modeling program provides $\frac{T_{sys}}{\eta_{ap}}$
and weight vector as a function of frequency and offset positions ($\theta_s$,
$\phi_s = 0$ ) from boresight. 

\subsection{Excess noise temperature}

The model presented here is for loss-less PAF and thus will not take into
account of antenna loss. Any increase in receiver temperature due to
change in amplifier noise parameters compared to the value used in the
model (for example replaced amplifiers) and transmission line loss 
ahead of amplifier is not accounted in the model. Some of these 
noise contributions are measured
in the laboratory. When comparing the model results with measurements,
the open circuit voltage correlation due to these noise contributions are
assumed to be of the form given in Eq.~\ref{rcmb} with $T_{cmb}$
replaced by the measured (frequency independent) excess noise temperature.
This noise correlation is added to the noise matrix, defined in Eq.~\ref{nmatrix},
while obtaining $\frac{T_{sys}}{\eta_{ap}}$ from the model. 
 
\section{Comparison with measurement}
\label{compare}

A set of measurements using the Kite array on the GBT was made in January 2015.
The details of the measurements are given in \citet{roshietal2015}.
Comparison of results from a previous version of the model with these measurements 
are also presented in \citet{roshietal2015}. The two major improvements in 
the model presented here are : (1) improvement in the computation of embedded
beam pattern and (2) the orientation of the Kite array on the GBT is correctly 
accounted for as described in Appendix~\ref{A8a}. With these improvements,
the unaccounted noise temperature needed to match the measured 
$\frac{T_{sys}}{\eta_{ap}}$
toward Virgo A with the model
is 5 K compared to 8 K noted in \citet{roshietal2015}. Also the aperture
efficiency predicted by the best fit model is 65\% compared to 70\%
noted in \citet{roshietal2015}. Below we provide a detail comparison
of January 2015 measurements and model results. 

\subsection{Boresight beam $\frac{T_{sys}}{\eta_{ap}}$ measured on Virgo A}
\label{bsm}

\begin{table}
\caption{Excess noise contribution to the receiver temperature}
\begin{center}
\begin{tabular}{|l|r|} \hline
\multicolumn{2}{|c|}{Noise contribution included in the model} \\\hline
Loss in thermal transition & 4 K \\
Excess noise due to replacement transistor & 5 K \\
Unaccounted excess noise  & 5 K \\ \hline
\multicolumn{2}{|c|}{Possible origin of unaccounted noise} \\ \hline
Antenna Loss   & ? K \\
Excess noise in the down converter$^1$   & $\sim$ 4 K \\
Excess spillover noise due to scattering from feed support   & ? K  \\ \hline
\end{tabular}
\end{center}
$^1$ \verb! https://safe.nrao.edu/wiki/pub/Main/PafDevelop/! \\
\verb! Test_of_PAF_Backend_Amplitude_and_Noise_Linearity.docx!     
\label{tab1}
\end{table}

\begin{figure}[t]
\includegraphics[width=6.0in, height=5.0in, angle =0]
{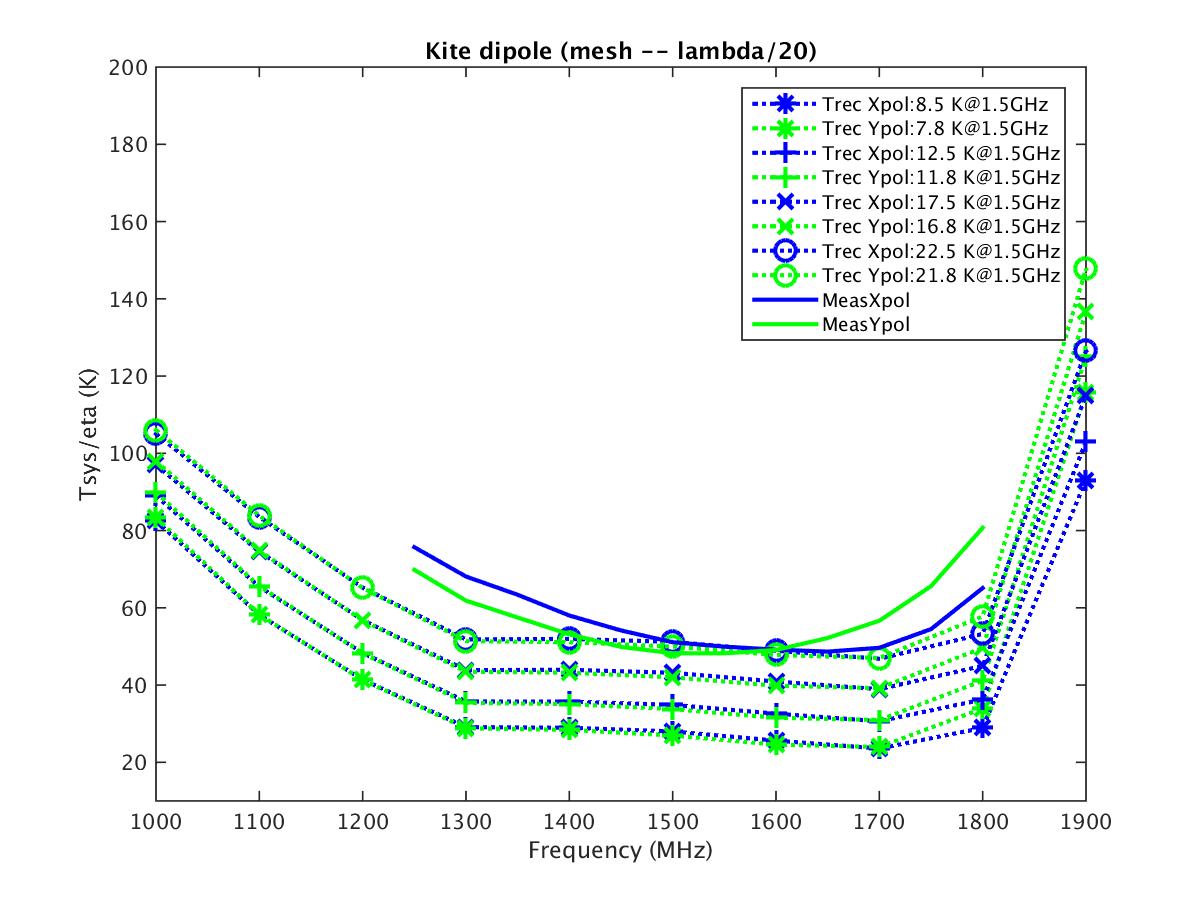} 
\caption{
$\frac{T_{sys}}{\eta}$ vs frequency from Virgo A observations (solid line)
and PAF model (dotted line). The model $\frac{T_{sys}}{\eta}$ is calculated
for different receiver temperature (see text for details). The receiver temperature
of each set of models at 1.5 GHz is marked on the plot. 
}
\label{comfig1}
\end{figure}

\begin{figure}[t]
\begin{tabular}{cc}
\includegraphics[width=3.0in, height=3.5in, angle =0]{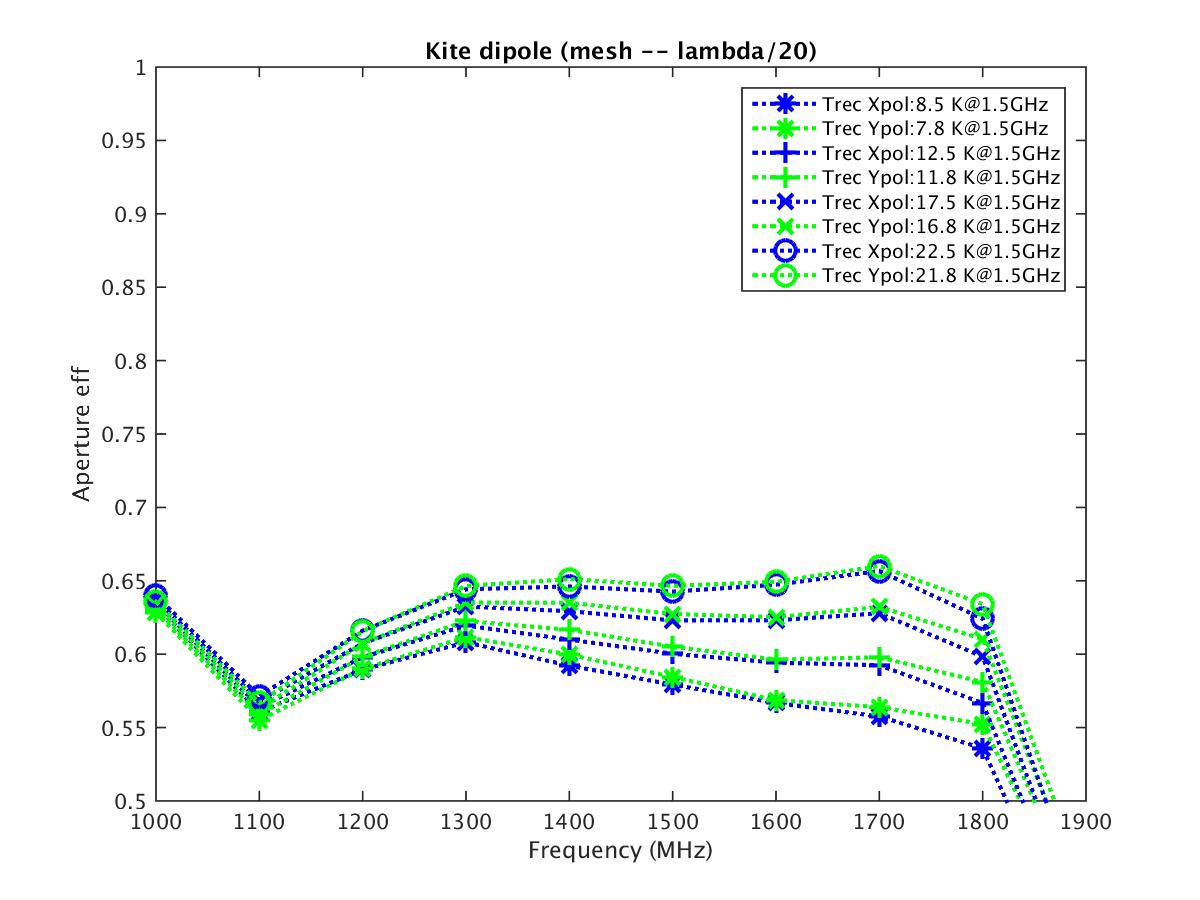} &
\includegraphics[width=3.0in, height=3.5in, angle =0]{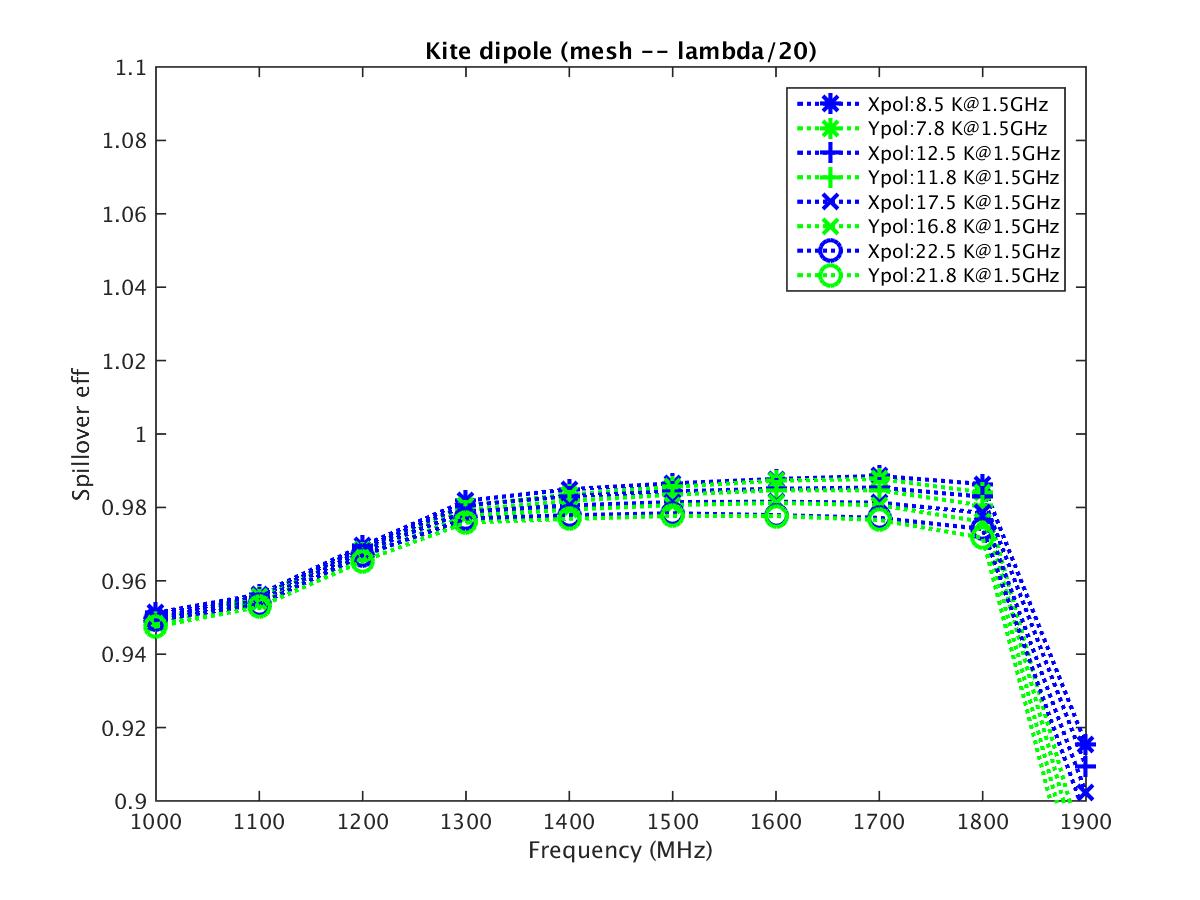} \\
\includegraphics[width=3.0in, height=3.5in, angle =0]{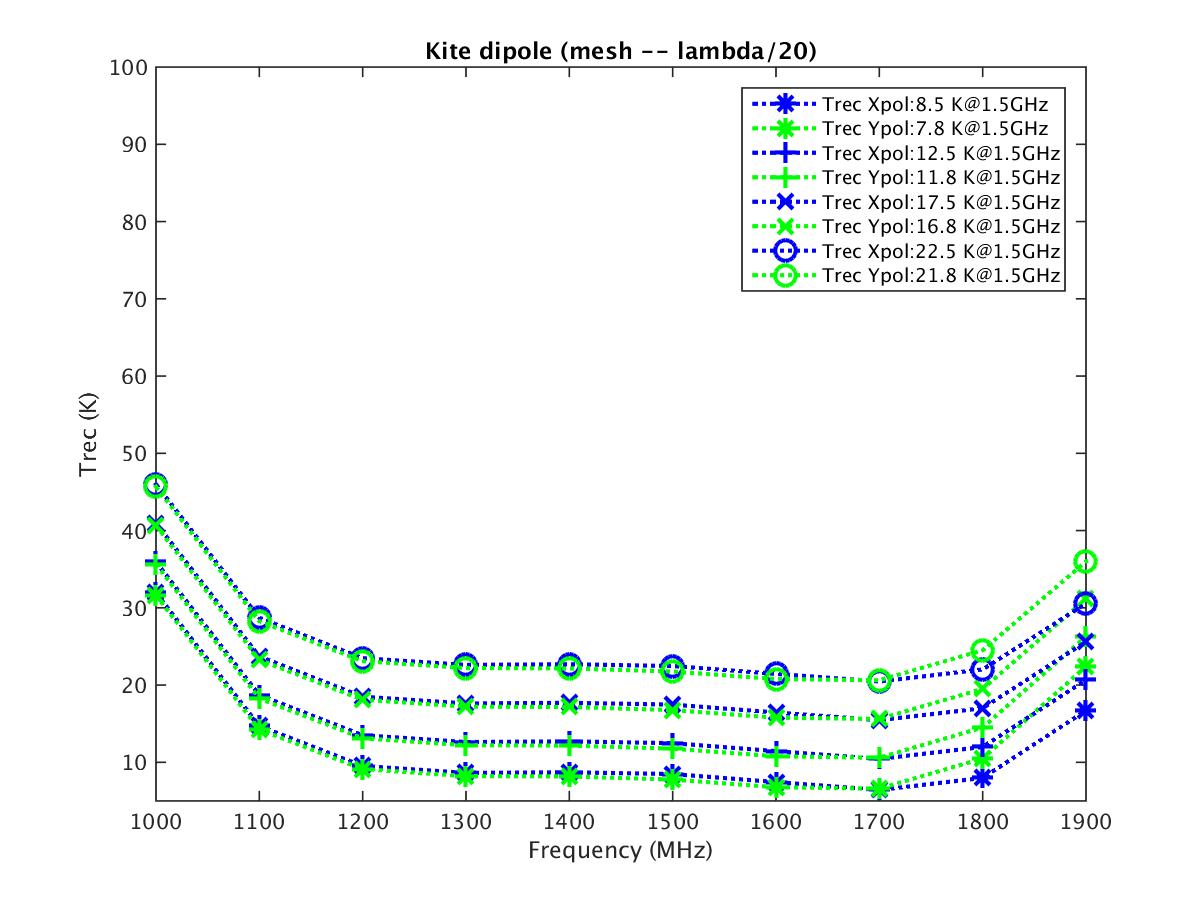} &
\includegraphics[width=3.0in, height=3.5in, angle =0]{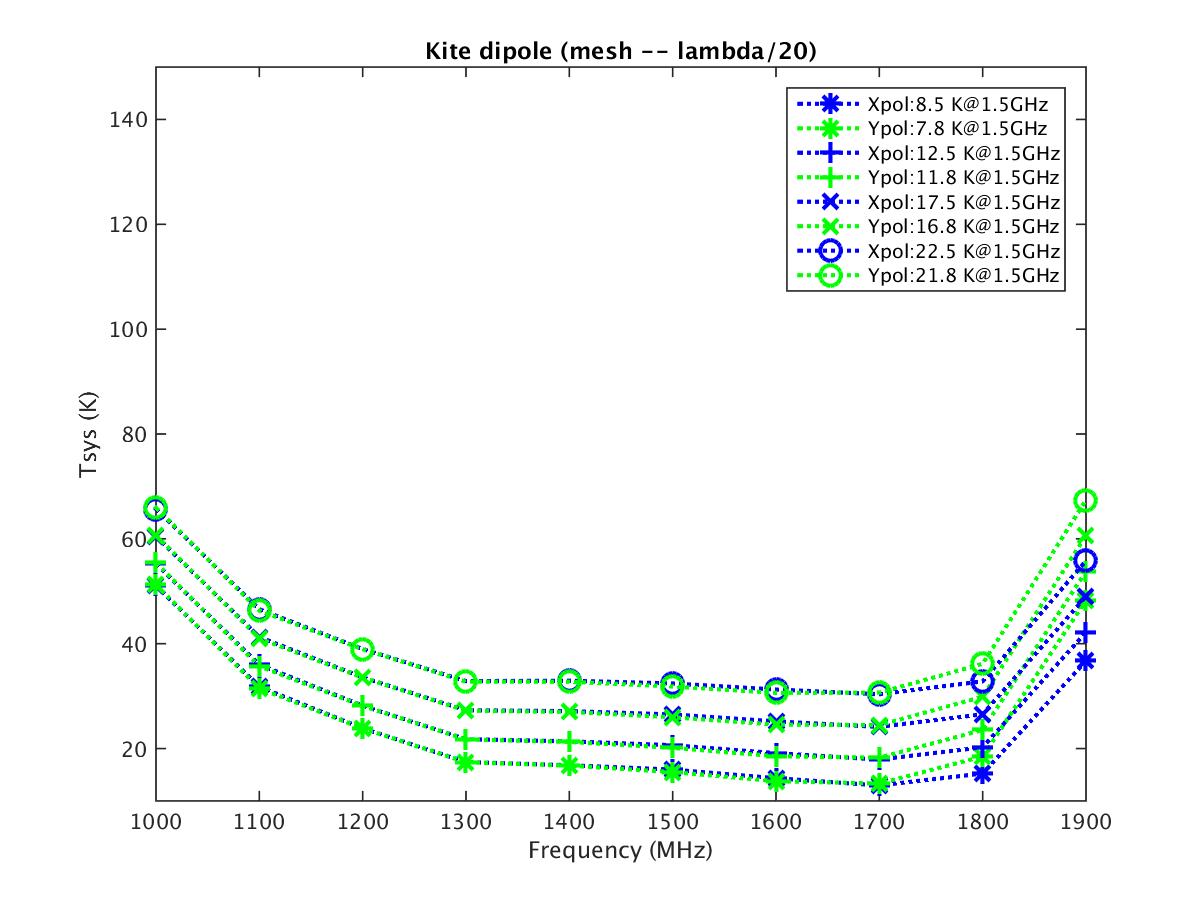} \\
\end{tabular}
\caption{
Model aperture efficiency (top right), spillover efficiency (top left),
receiver temperature (bottom right) and system temperature (bottom left) 
vs frequency. The receiver temperature
of each set of models at 1.5 GHz is marked on the plot. 
}
\label{comfig2}
\end{figure}

We compared $\frac{T_{sys}}{\eta_{ap}}$ vs frequency obtained from 
Virgo A measurements made on 25 January 2015 (TGBT14B\_913\_02, 
scans 25 to 48; see \citet{roshietal2015}) with the model predictions.
In Fig.~\ref{comfig1}, we show model $\frac{T_{sys}}{\eta_{ap}}$ for 
4 receiver temperatures along with measured values. 
The 4 receiver temperatures at 1.5 GHz are (1) those with amplifier 
noise alone which is $\sim$ 8 K, (2) with 
addition of losses in the thermal transition which gives $\sim$ 12 K, 
(3) with the excess noise due to replacement transistor which gives $\sim$ 17 K
and (4) with an excess noise of 5 K which give $\sim$ 22 K.
Table~\ref{tab1} summarizes the additional contributions to the receiver
temperature over that due to the amplifier noise alone 
(see also \citet{roshietal2015}).
A 5 K excess noise of unknown  origin need to be added to the
receiver temperature (i.e. $\sim 22$ K)
to match model prediction with the measured value (see Fig~\ref{comfig1}). 
In Table~\ref{tab1}, we list possible origin for this excess noise temperature.

Fig~\ref{comfig2} shows the model aperture efficiency, spillover efficiency, receiver
temperature and system temperature vs frequency. The aperture and spillover
efficiencies between 1.3 and 1.7 GHz are about 65 \% and 98 \% respectively. 

\subsection{Off-boresight beam $\frac{T_{sys}}{\eta_{ap}}$ at 1.7 GHz}

\begin{figure}[t]
\includegraphics[width=6.0in, height=5in, angle =0]
{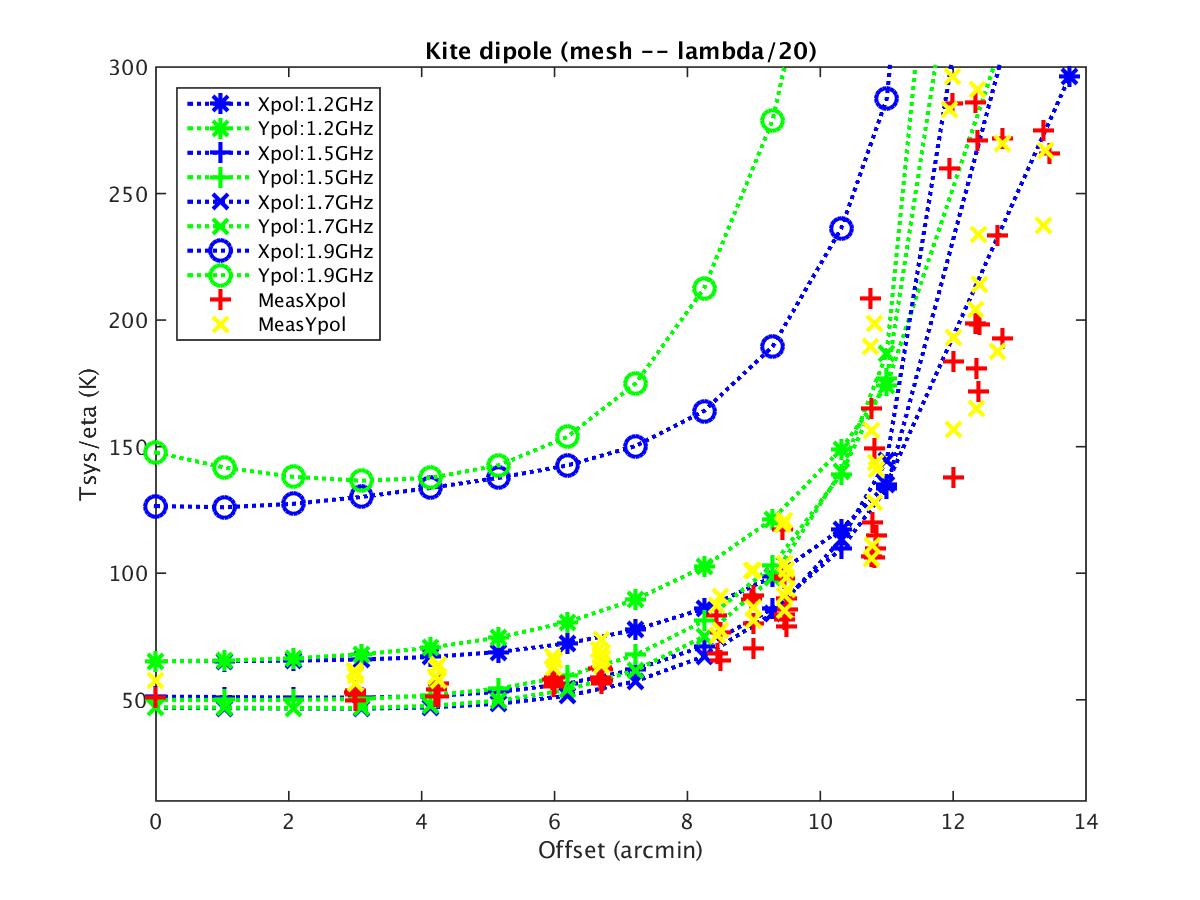} 
\caption{
$\frac{T_{sys}}{\eta_{ap}}$ vs offset from boresight direction from Virgo A `grid' 
observations (+ \& x points) and PAF model (dotted line). 
The model $\frac{T_{sys}}{\eta_{ap}}$ is calculated
for receiver temperature $\sim$ 22 K at 1.5 GHz. The model values are plotted for different
frequencies as marked on the plot. The measurements are made at 1.7 GHz.
}
\label{comfig3}
\end{figure}

\begin{figure}[t]
\begin{tabular}{cc}
\includegraphics[width=3.0in, height=3.5in, angle =0]{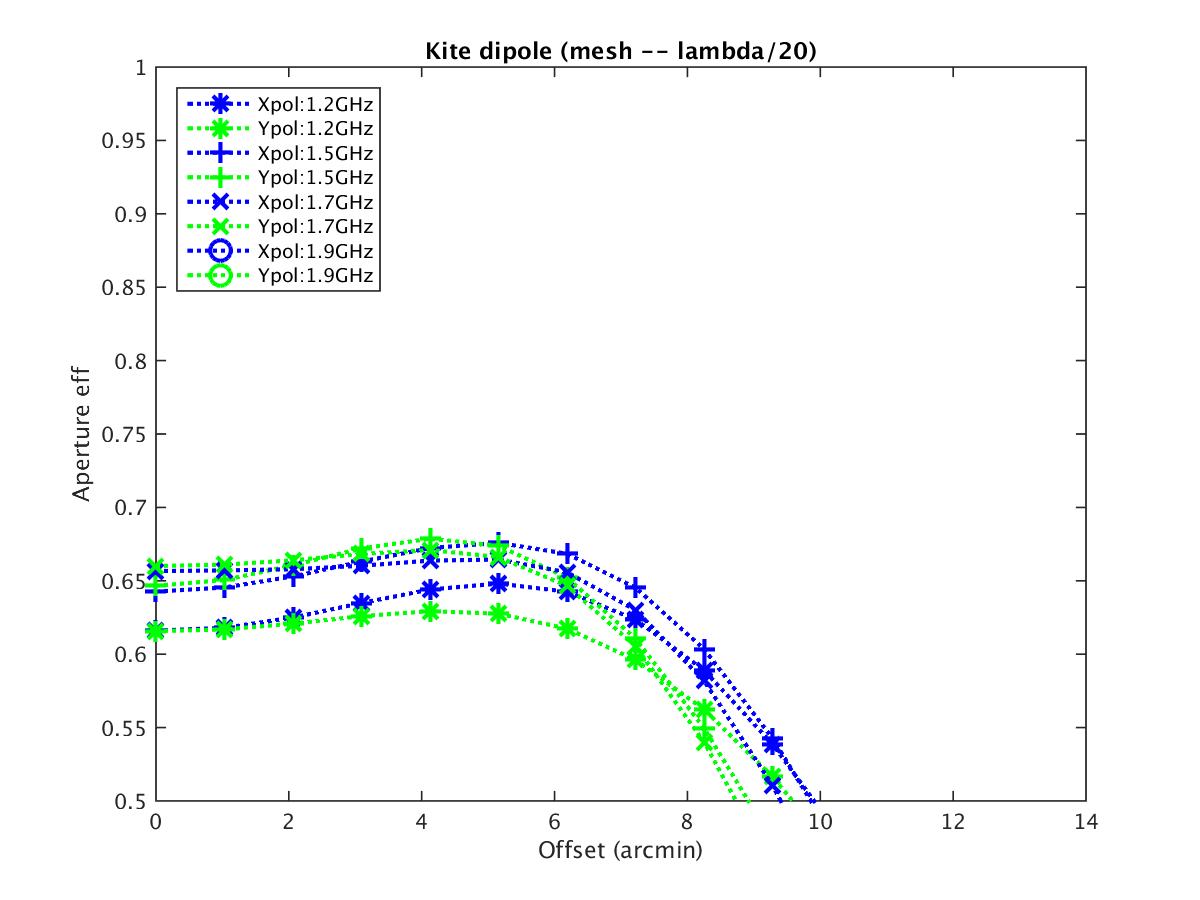} &
\includegraphics[width=3.0in, height=3.5in, angle =0]{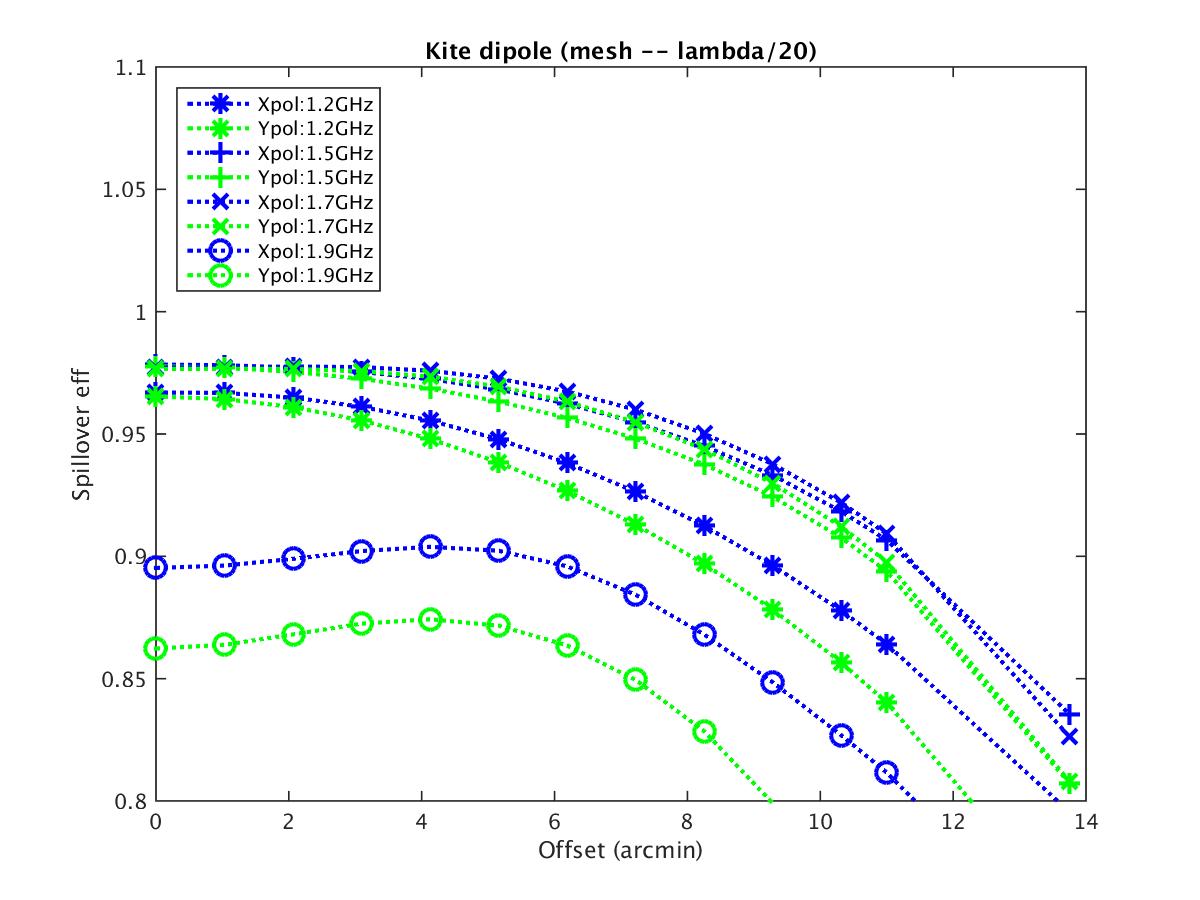} \\
\includegraphics[width=3.0in, height=3.5in, angle =0]{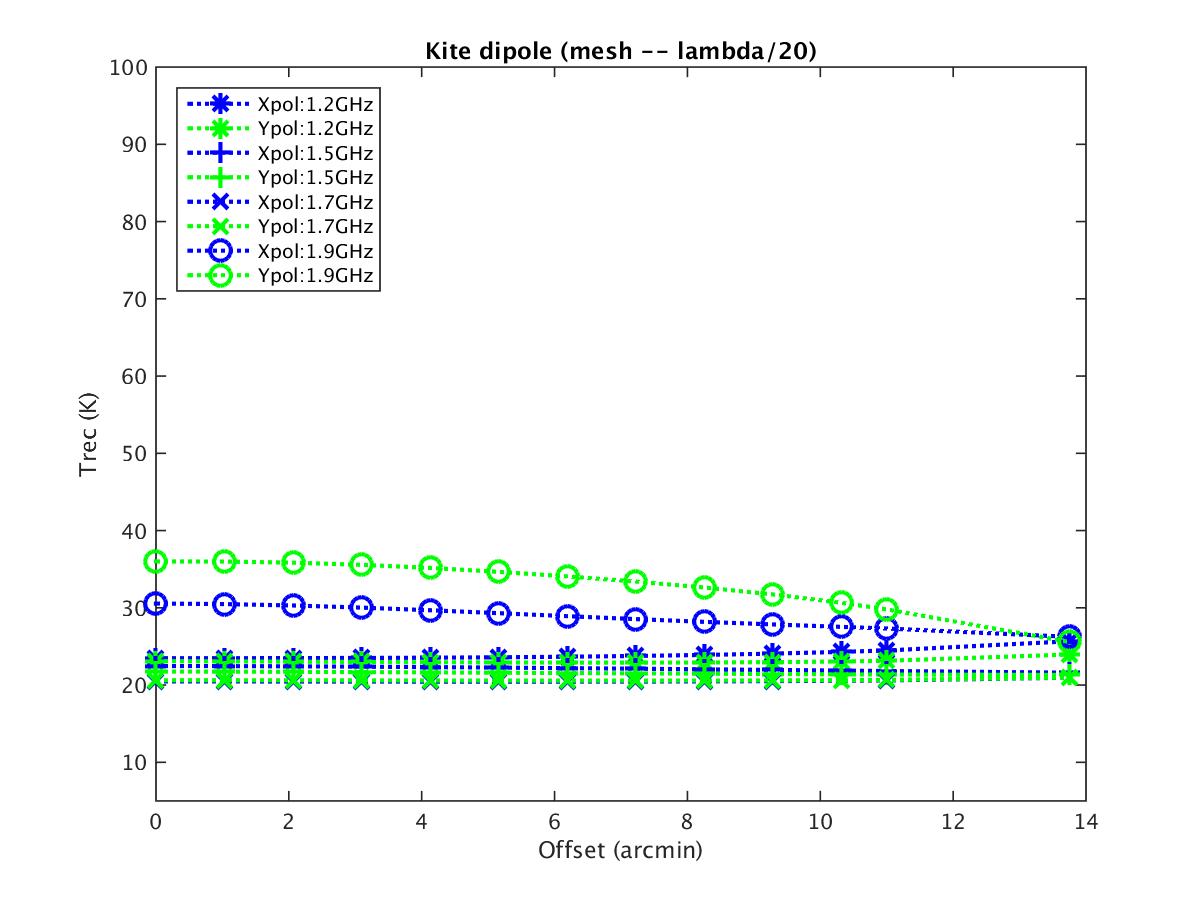} &
\includegraphics[width=3.0in, height=3.5in, angle =0]{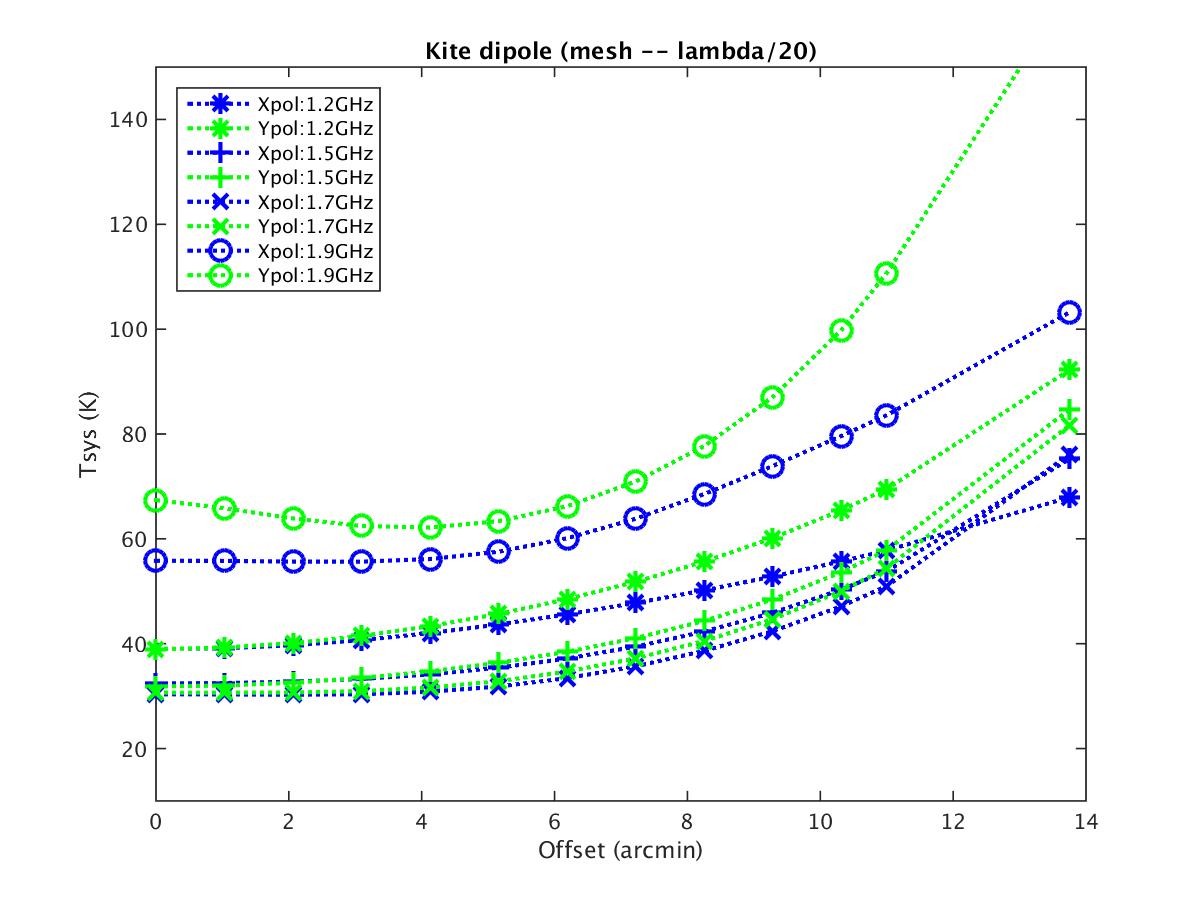} \\
\end{tabular}
\caption{
Model aperture efficiency (top right), spillover efficiency (top left),
receiver temperature (bottom right) and system temperature (bottom left) 
vs offset from boresight direction. The receiver temperature
for the models is $\sim$ 22 K at 1.5 GHz. Model values for different
frequencies are plotted as marked on the plots 
}
\label{comfig4}
\end{figure}

The off-boresight beam $\frac{T_{sys}}{\eta_{ap}}$ measurements were
obtained from `grid' observations toward Virgo A at 1.7 GHz 
(observation made on 25 January 2015 TGBT14B\_913\_02,
scans 49 to 489; see \citet{roshietal2015}).
Fig.~\ref{comfig3} shows the model $\frac{T_{sys}}{\eta_{ap}}$ at 1.7 GHz 
vs offset angle along with the measurements. The model values
are obtained for receiver temperature $\sim$ 22 K at 1.5 GHz
(see Section~\ref{bsm}). As seen in Fig.~\ref{comfig3}, the predicted
values match the measured values very well for frequency 1.7 GHz. 

Fig~\ref{comfig4} shows the model aperture efficiency, spillover efficiency, receiver
temperature and system temperature vs offset angle from boresight direction. 
The aperture efficiency drops significantly
below 65 \% beyond 6\arcmin offset angle, which, as shown in \citet{roshietal2015},
is due to the finite size of the PAF. 

\subsection{Comparison of measured noise and signal correlations with the model}

\begin{figure}[t]
\begin{tabular}{cc}
\includegraphics[width=3.0in, height=3.5in, angle =0]{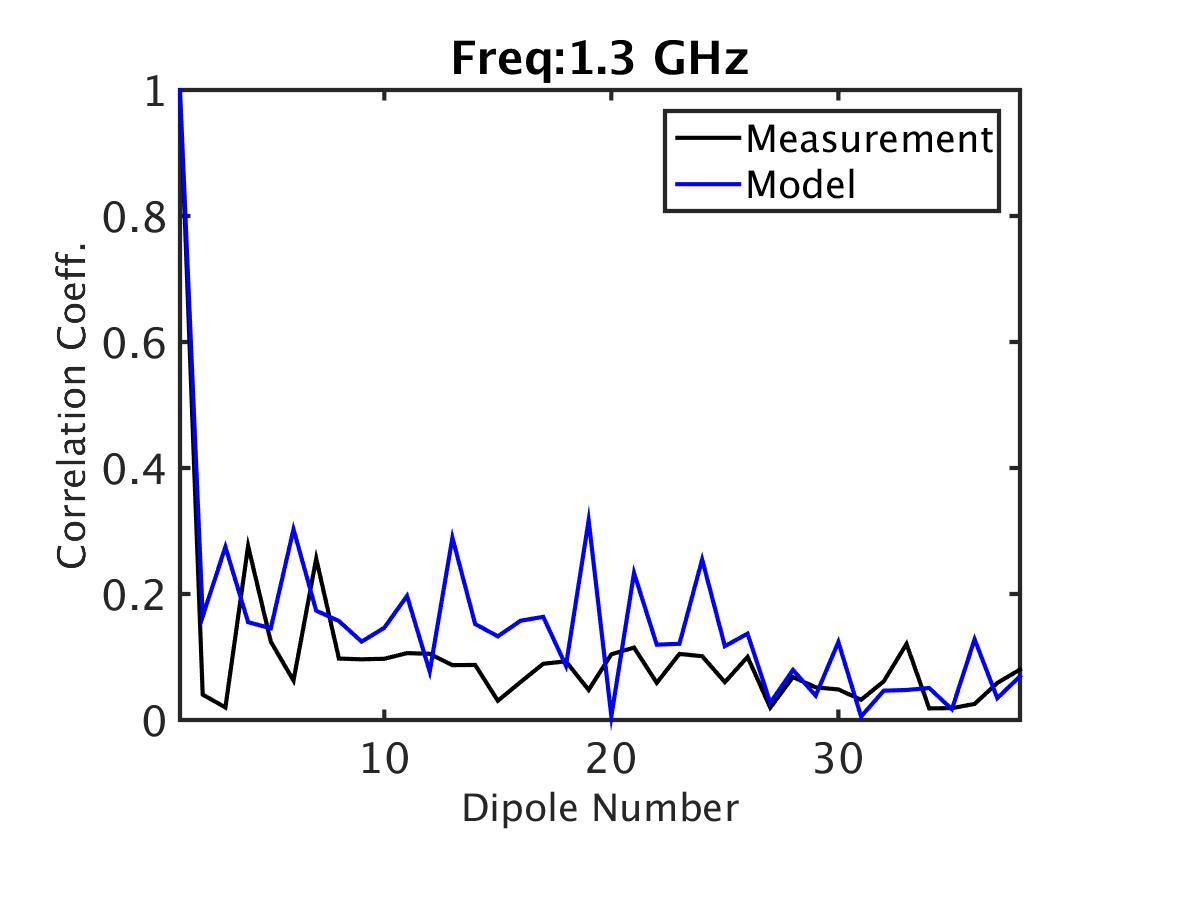} &
\includegraphics[width=3.0in, height=3.5in, angle =0]{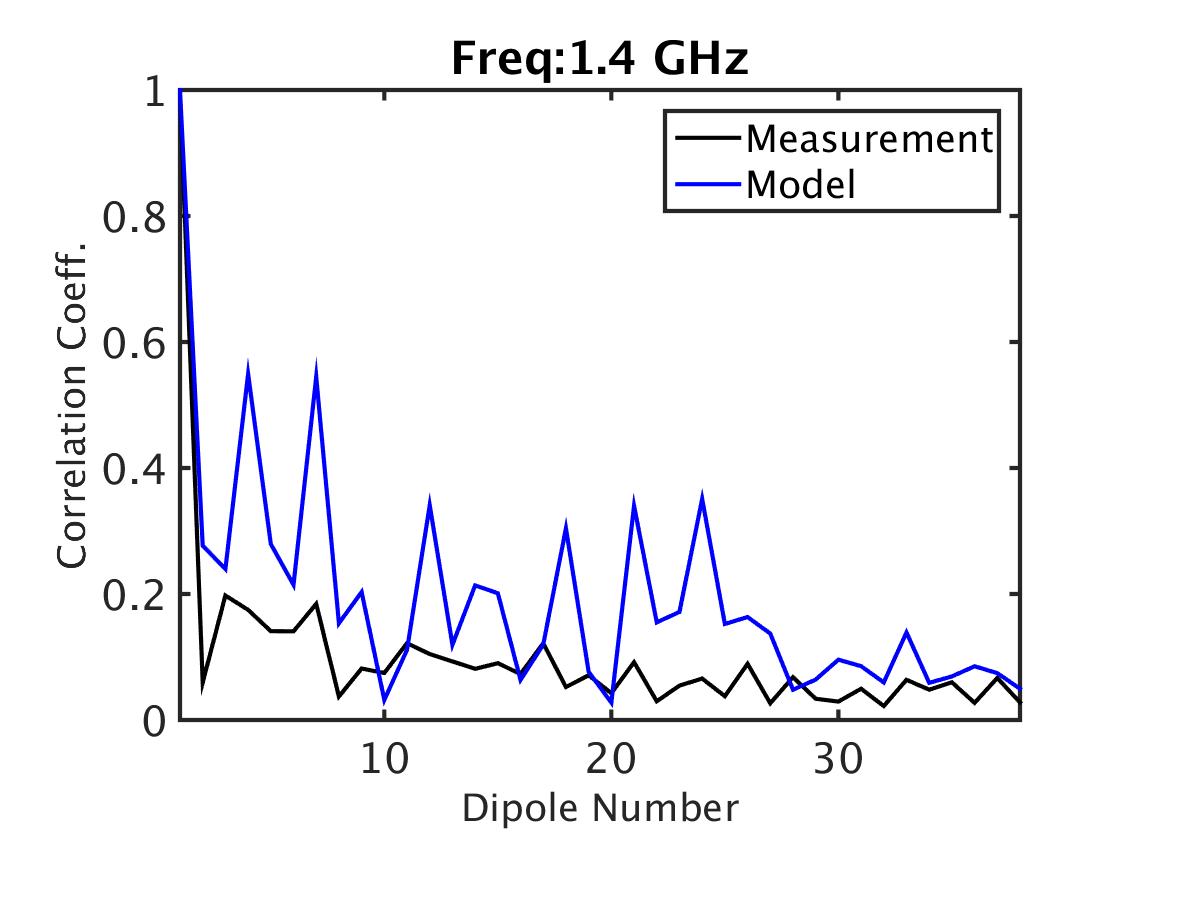} \\
\includegraphics[width=3.0in, height=3.5in, angle =0]{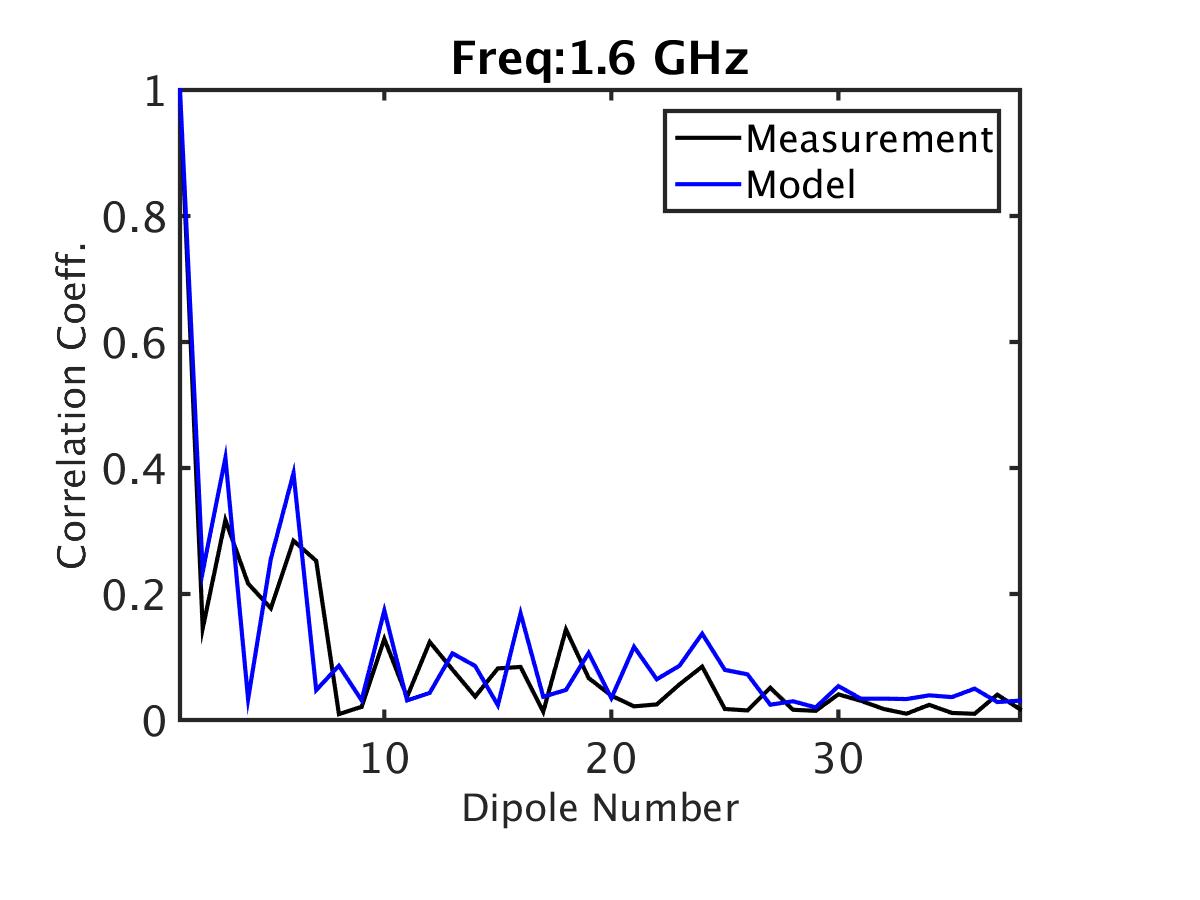} &
\includegraphics[width=3.0in, height=3.5in, angle =0]{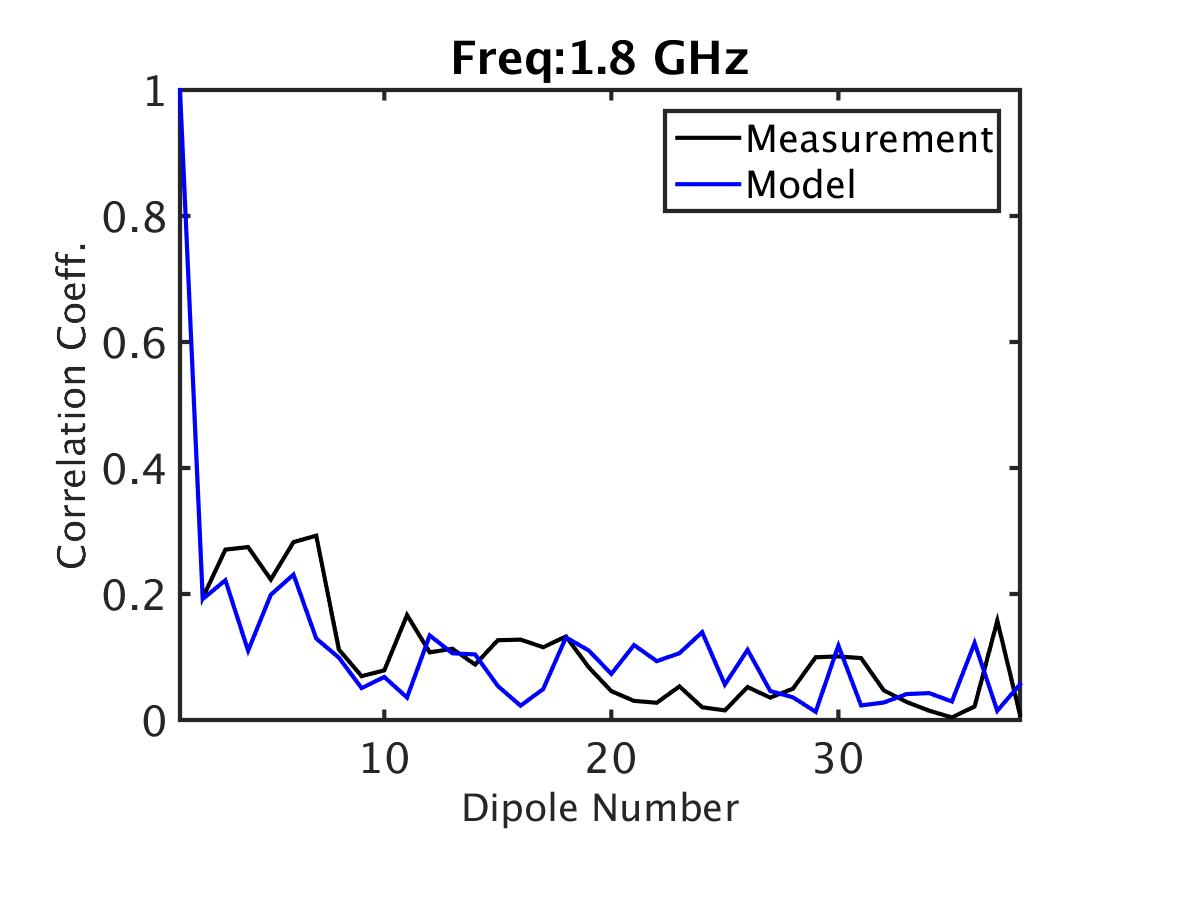} \\
\end{tabular}
\caption{
Model noise correlations (blue) and measured correlations (black)
between dipole 1 and other dipoles. Dipoles 1 to 19 correspond to X polarization
and 20 to 38 correspond to Y polarization. The frequencies at which
the correlations are obtained are marked on the plot. 
}
\label{comfig5}
\end{figure}

\begin{figure}[t]
\begin{tabular}{cc}
\includegraphics[width=3.0in, height=3.5in, angle =0]{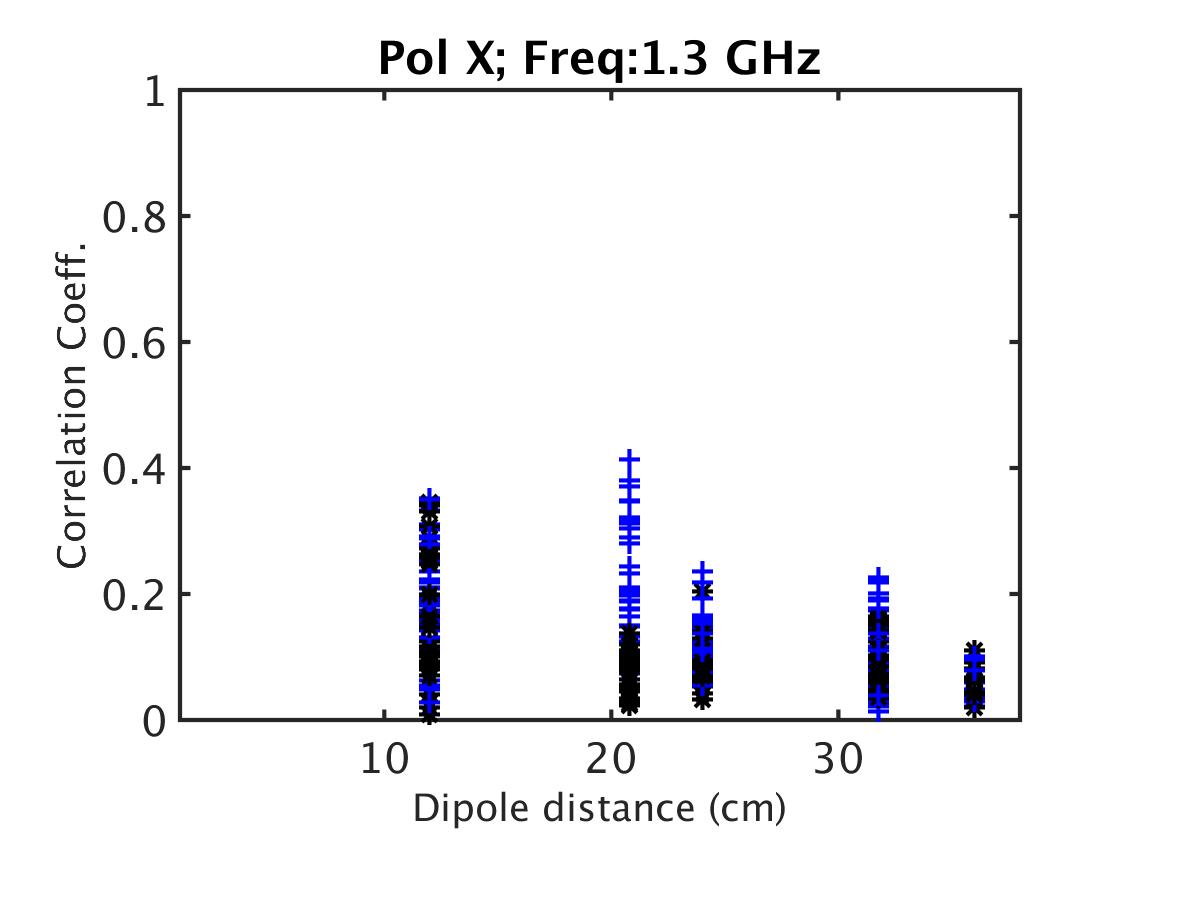} &
\includegraphics[width=3.0in, height=3.5in, angle =0]{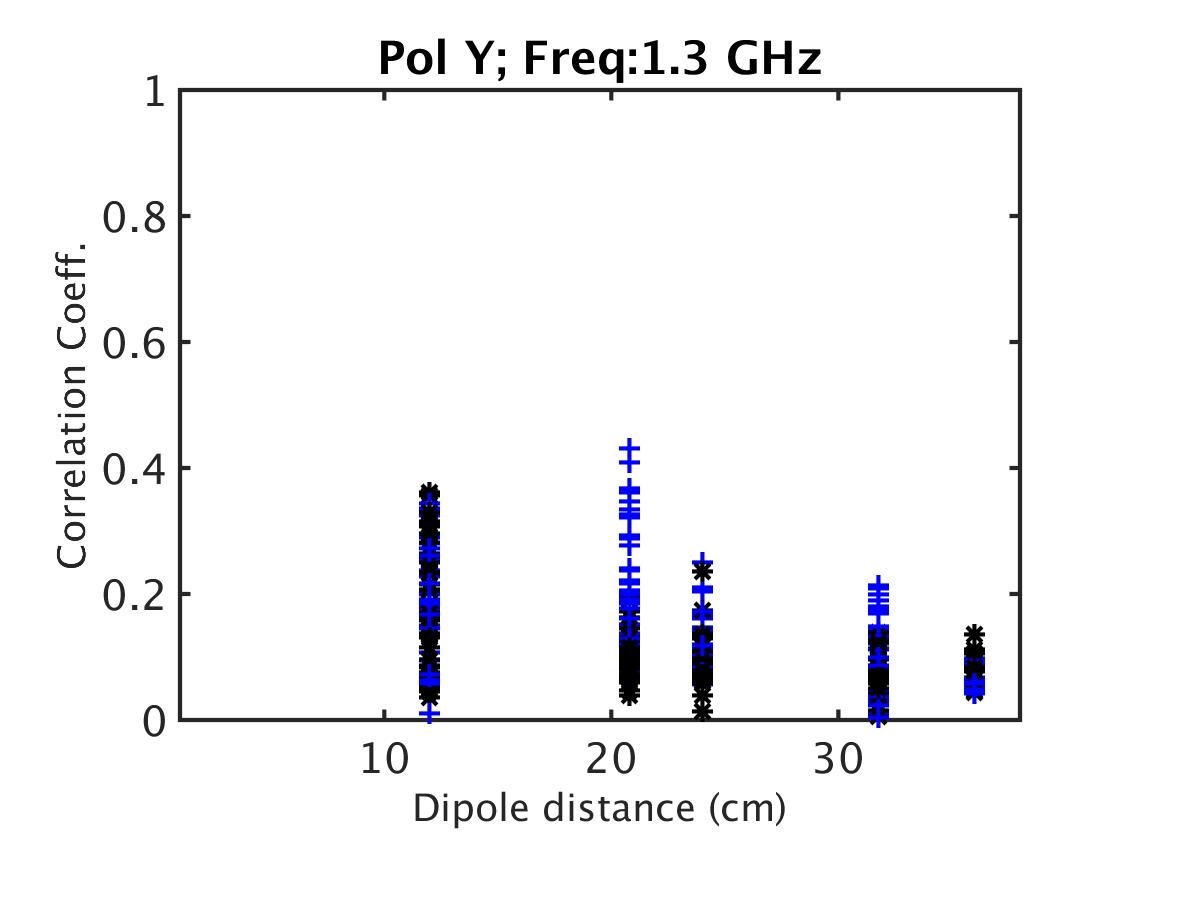} \\
\includegraphics[width=3.0in, height=3.5in, angle =0]{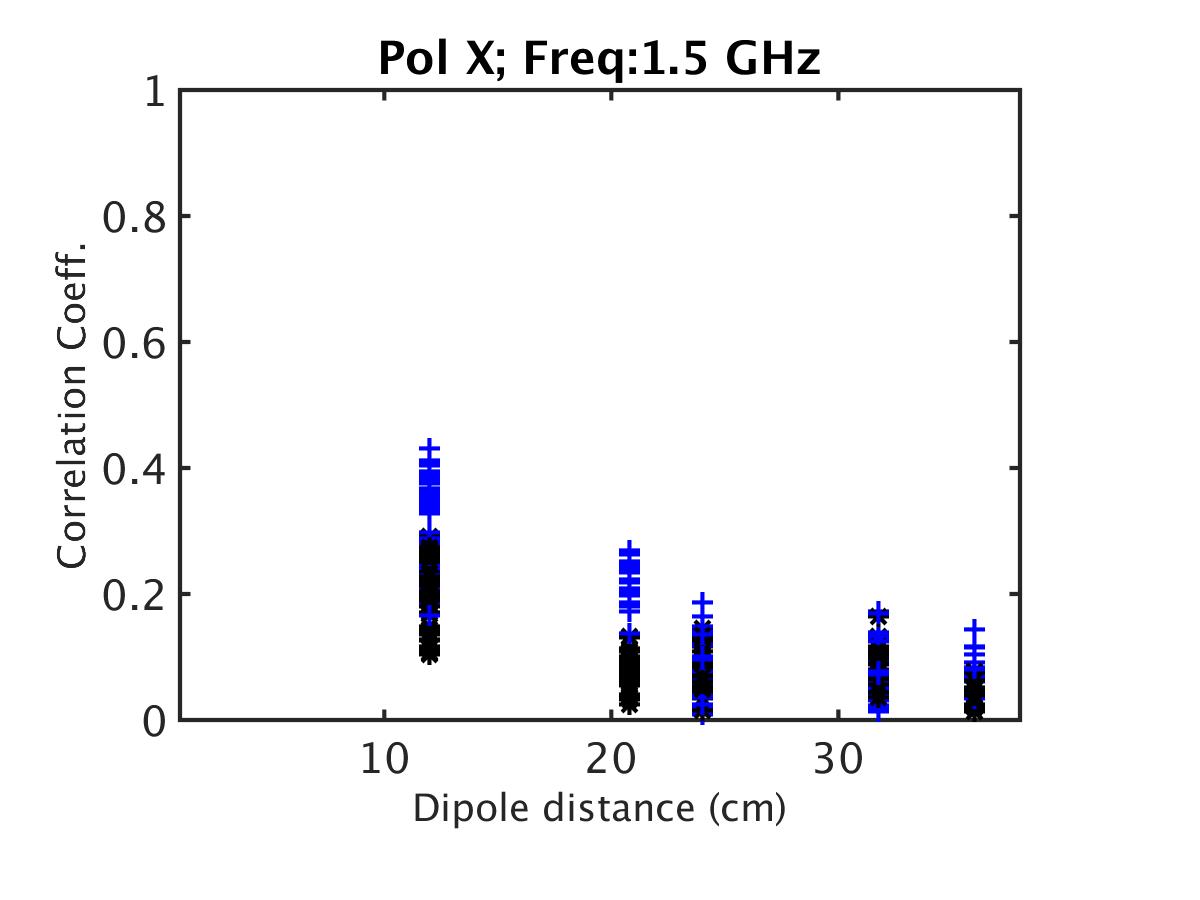} &
\includegraphics[width=3.0in, height=3.5in, angle =0]{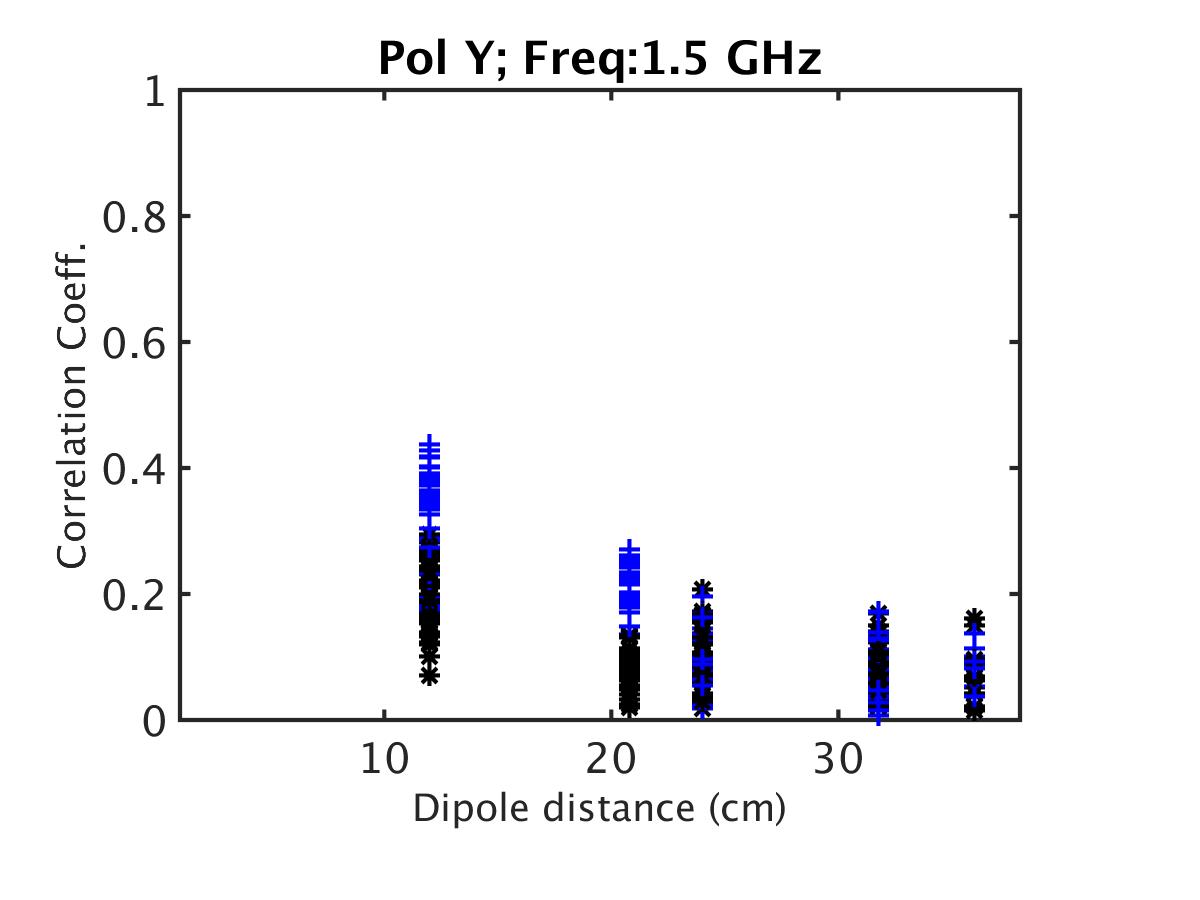} \\
\end{tabular}
\caption{
Model noise correlations (blue) and measured correlations (black)
between dipoles vs relative separation of dipoles in PAF. The correlations from X 
polarization dipole are shown on the right and those
from Y polarization dipole are shown on the left.
The frequencies at which 
the correlations are obtained are marked on the plot. 
}
\label{comfig6}
\end{figure}

\begin{figure}[t]
\begin{tabular}{cc}
\includegraphics[width=3.0in, height=3.5in, angle=0]{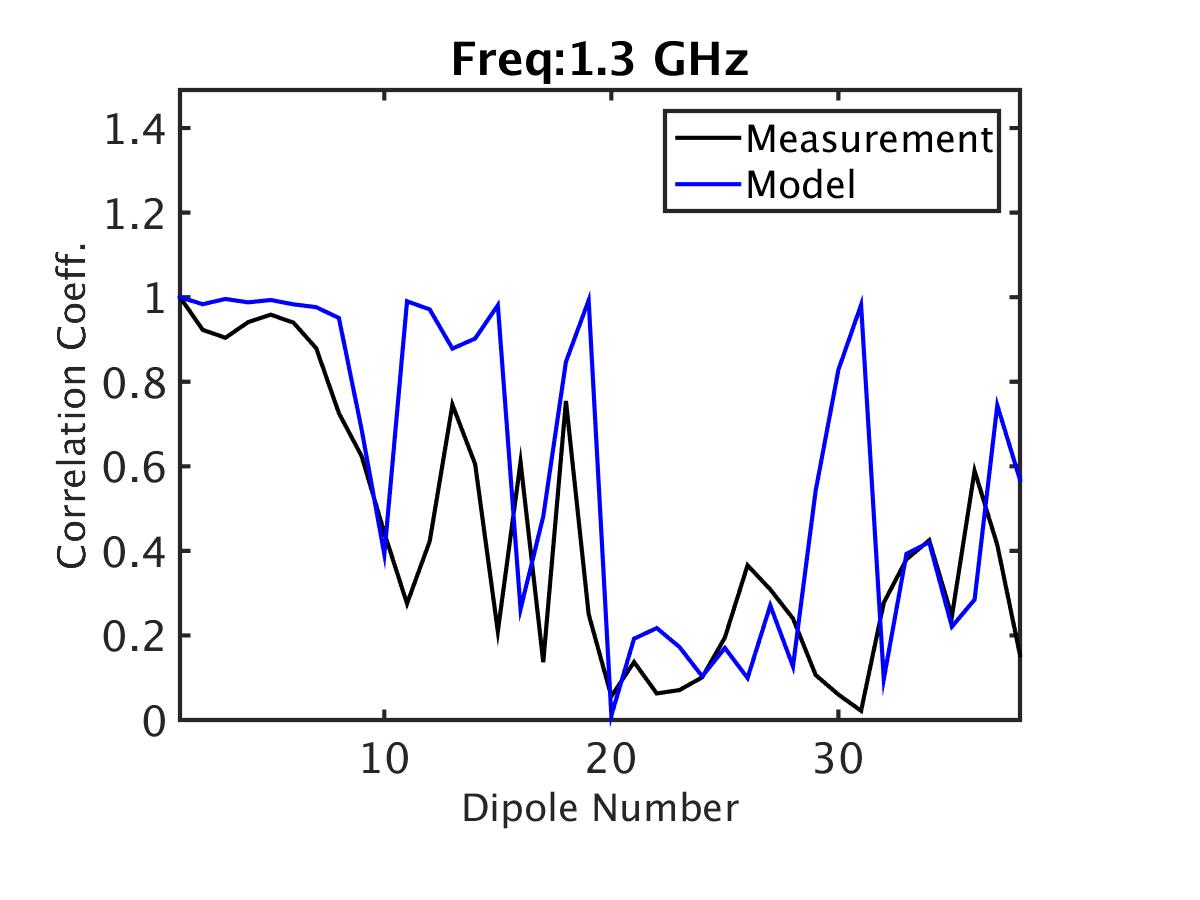} &
\includegraphics[width=3.0in, height=3.5in, angle=0]{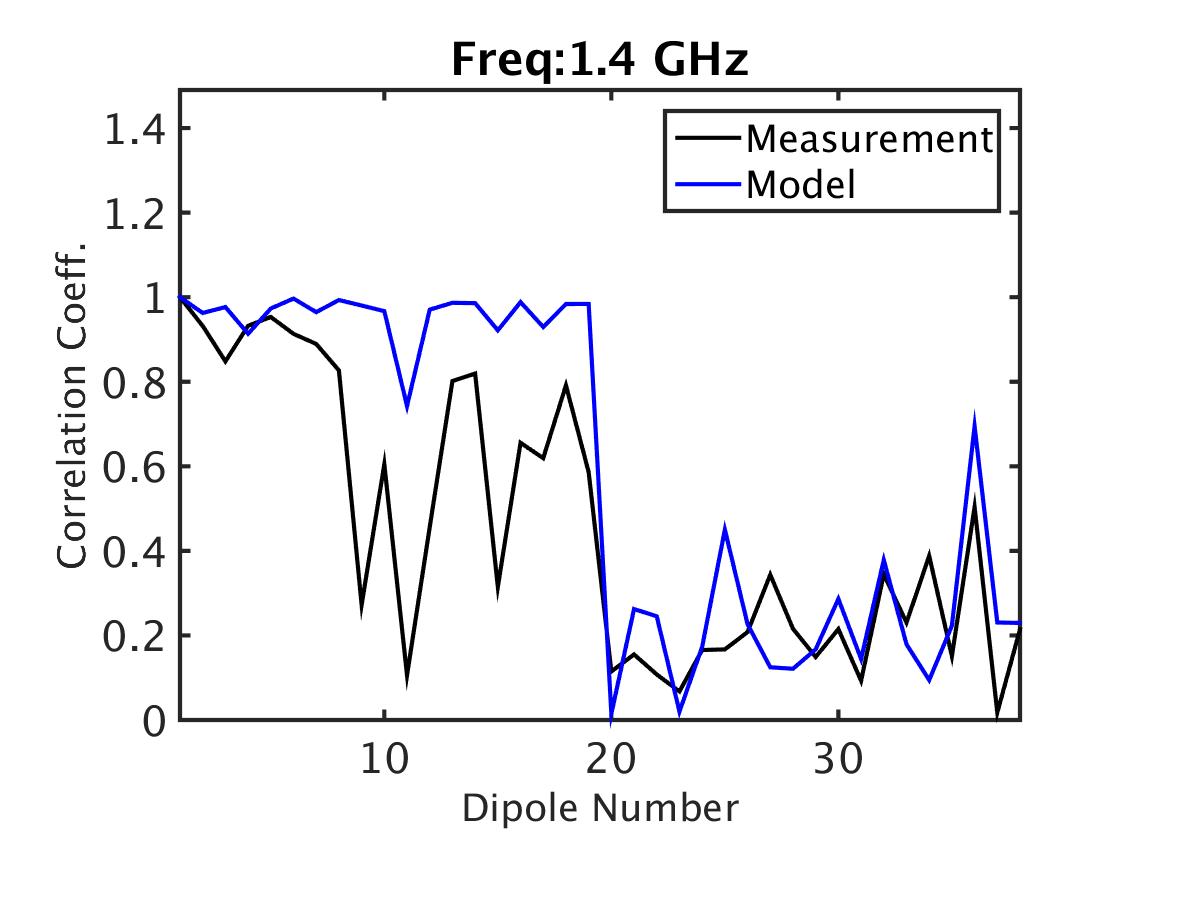} \\
\includegraphics[width=3.0in, height=3.5in, angle=0]{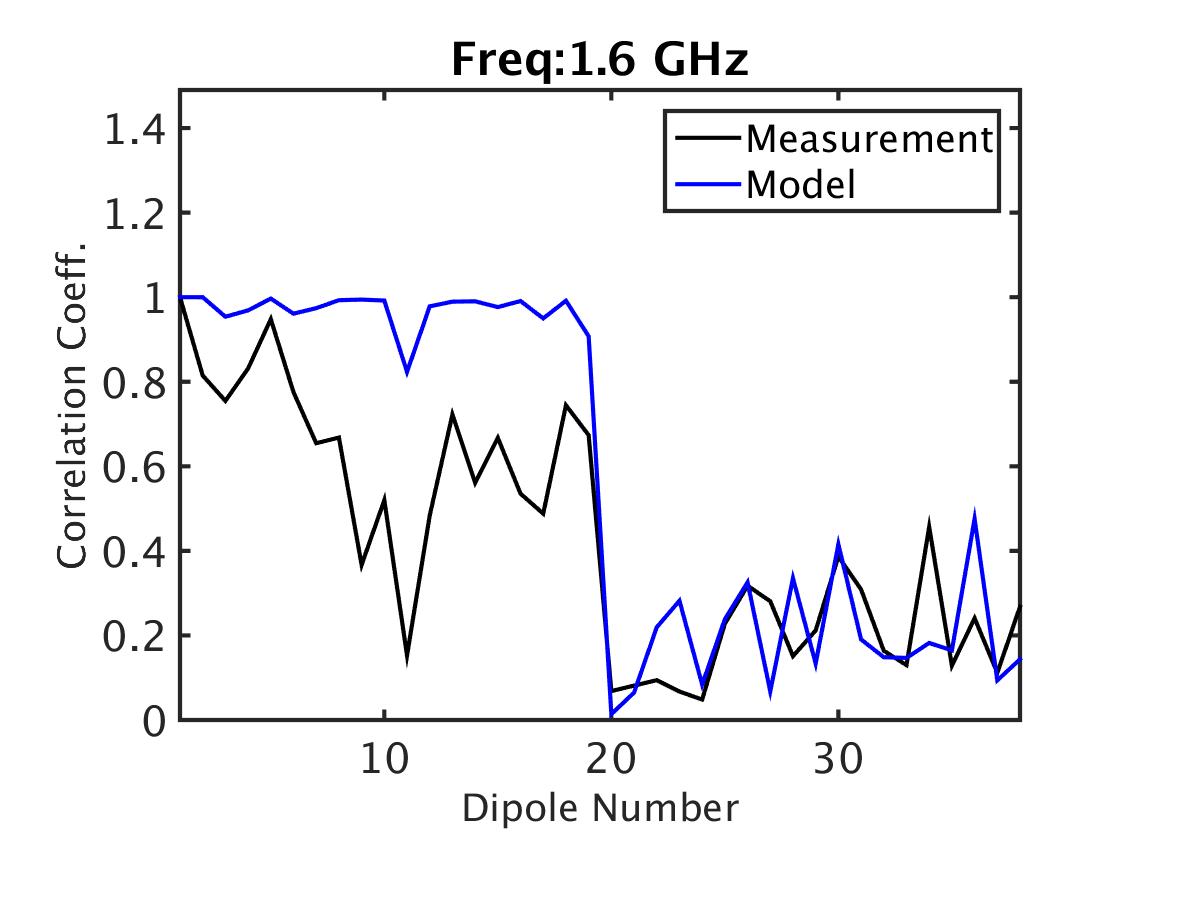} &
\includegraphics[width=3.0in, height=3.5in, angle=0]{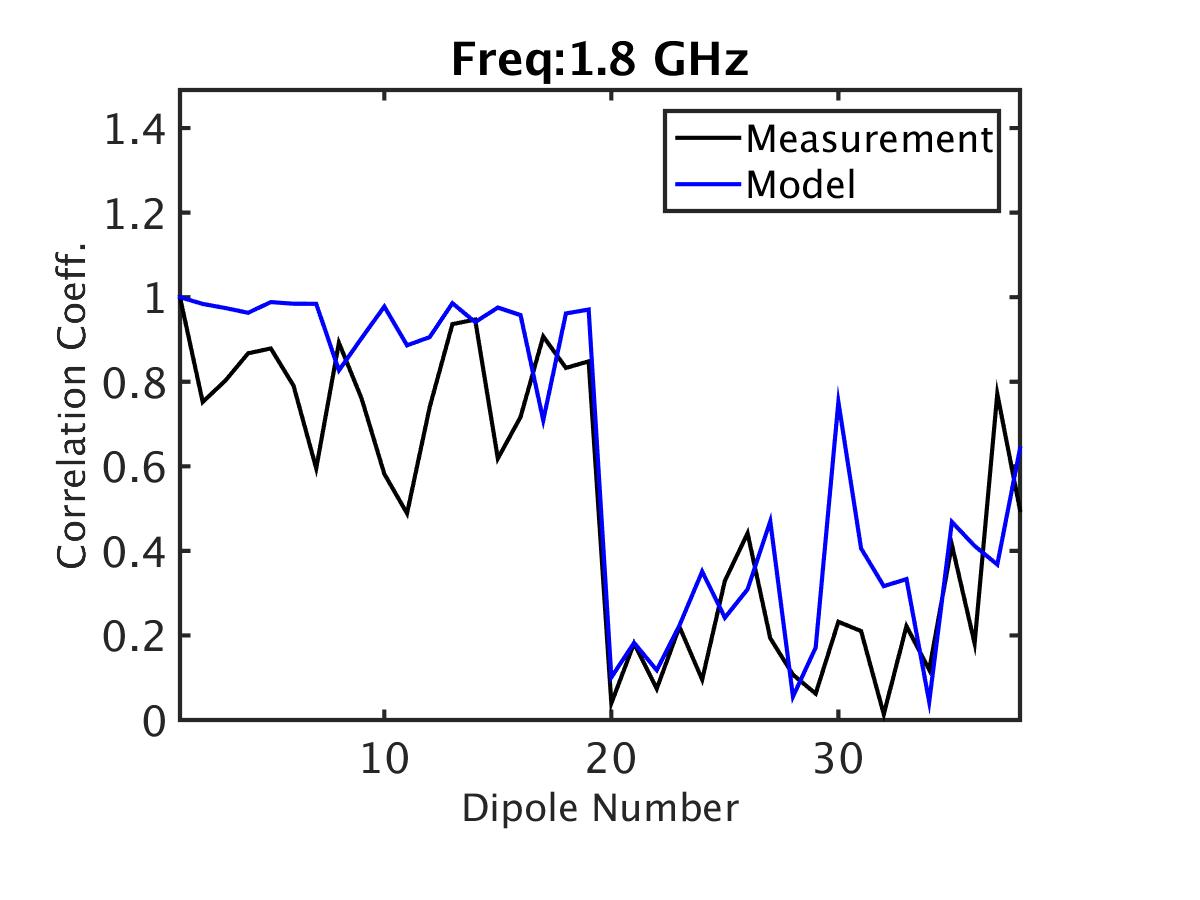} \\
\end{tabular}
\caption{
Model signal correlations (blue) and measured correlations due to source (black)
between dipole1 and other dipoles. The measured signal correlations are obtained
from the difference of on-source and off-source correlation matrices when the
Virgo A is at the boresight. Dipoles 1 to 19 correspond to X polarization
and dipoles 20 to 38 correspond to Y polarization.
}
\label{comfig7}
\end{figure}

We compare the noise correlation coefficient measured at an off-source position with
model values. Fig~\ref{comfig5} shows the noise correlation of dipole 1 with rest of
the dipoles for a set of frequencies. The modeled and measured values matched
reasonably well. Fig~\ref{comfig6} shows the noise correlation as a function of 
the relative separation of the dipole in the array. These plots show all the correlation
coefficients between dipole pairs. There are some general agreements
between modeled  and measured values.

We also compare the correlation due to signal from source between model and measurement.
Fig.~\ref{comfig7} plots the correlation of dipole 1 with rest of the dipoles for a 
set of frequencies. While the model values and measurements in general show similar trends,
the measured correlation coefficients are systematically lower than the modeled value.  
This difference in model and measurements needs further investigation.

\section*{Acknowledgment}

We acknowledge very useful discussions and suggestions from Matt Morgan
during the initial phase of the development of the PAF model. The possibility
of an analytical solution for PAF problem was pointed out to D. A. Roshi
by Stuart Hay, CSIRO. We thank Marian Pospieszalski, Anthony Kerr
for useful discussions on the noise properties of the receiver, Srikanth
for discussions on the computation of the GBT aperture field, William
Shillue for proofreading the manuscript and
William Shillue, Robert Dickman and S. K. Pan for the support
and useful discussions during the course of this work. 

\bibliographystyle{apalike}
\bibliography{ms}{}

\appendix

\section{Impedance matrix, Scattering matrix and Embedded beam pattern}
\label{A1}

A schematic of the PAF is shown in Fig.~\ref{fig1}a. The impedance matrix
$\bm Z$ of PAF relates the port voltages, $\bm V^T = [v_1, v_2, ...]$ and currents, 
$\bm I^T = [i_1, i_2, ...]$,
\be
\bm V = \bm Z \bm I.
\label{imp}
\ee
This relationship can also be written in terms of the admittance matrix $\bm Y$
\be
\bm I = \bm Y \bm V.
\label{adm}
\ee
It follows from Eq.~\ref{imp} and Eq.~\ref{adm} that $\bm Y = \bm Z^{-1}$.

Another parameter that is used to describe network is Scattering matrix $\bm S$.
This matrix relates the forward traveling wave $\bm a = [a_1, a_2, ...]$ and
the reverse traveling wave $\bm b = [b_1, b_2, ...]$ (see Fig.~\ref{fig1}a),
\be
\bm b = \bm S \bm a.
\label{Sdef}
\ee
The amplitude of the traveling waves and port voltages and currents are 
related through
\bea
\bm V & = & \sqrt{z_0} \left(\bm a + \bm b\right)  \\
\bm I & = & \frac{1}{\sqrt{z_0}} \left(\bm a - \bm b\right)  
\eea
where $z_0$ is the characteristic impedance of the transmission line.
In this report we have taken $z_0 = $ 50 $\Omega$, which is the reference impedance.
Substituting these in Eq.~\ref{imp} and keeping in mind that this equation
holds for arbitrary set of $\bm a$  we get
\be
\bm Z = z_0 \left(\bs{\mathcal{I}} + \bm S \right) \left(\bs{\mathcal{I}} - \bm S \right)^{-1}
\ee
Similarly the relationship between $\bm Y$ and $\bm S$ can be obtained as
\be
\bm Y = \frac{1}{z_0} \left(\bs{\mathcal{I}} - \bm S \right) \left(\bs{\mathcal{I}} + \bm S \right)^{-1}
\ee

In Section~\ref{recmode}, we defined the embedded beam pattern $\bs{\vec{\mathcal{E}}^e}$.
Here we present some characteristics of the embedded beam pattern and its relationship
to impedance matrix. 
\begin{enumerate}
\item
The embedded beam pattern $\vec{\mathcal{E}}^e_i$ 
is the far-field pattern when the $i^{th}$ input to the array
is excited with a harmonic signal of 1 V peak value and
all other inputs are short-circuited. For a loss less antenna,
\bea
\frac{1}{2 z_f}\int_{sphere} \vec{\mathcal{E}}^e_i \cdot \vec{\mathcal{E}}^{e*}_i \; \textrm{d}A 
           &  = & \frac{1}{2 z_f}\int_{4\pi} \vec{E}^e_i  \cdot \vec{E}^{e*}_i \; \textrm{d}\Omega  \nonumber \\
           &  =  & \frac{1}{2} i_{0_i}^2 \; \textrm{Re}\{Z_{pin_i}\}
\label{embpdis}
\eea
where $z_f$ is the free space impedance, $i_{0_i}$ is the current flowing
to port $i$ (see Fig.~\ref{fig1}b) and $Z_{pin_i}$ is the input 
impedance of port $i$ when all other ports are short circuited. The input
impedance for this case is given by
\be
Z_{pin_i} = \frac{1}{Y_{ii}}
\label{embZin}
\ee
where $Y_{ii}$ is the $i^{th}$ diagonal element of the admittance matrix ${\bm Y}$.
Thus, for 1 V excitation, $i_{0_i} = Y_{ii}$ in amps.
\item
The open circuit voltage for embedded beam excitation (i.e. $v_{0_i} = 1$ V and
0 V for all other ports) can be obtained from Eq.~\ref{parta} and Eq.~\ref{lorenz1} as
\be
v_{oc_i} = z_{pin_i} \int_{A_{free}} \left(\vec{\mathcal{E}}^e \times \vec{\mathcal{H}_r} -
          \vec{\mathcal{E}_r} \times \vec{\mathcal{H}}^e\right) \cdot \hat{n}\; \textrm{d}A.
\label{singleport}
\ee
This equation is also true for any (loss-less) single port antenna, with $z_{pin_i}$ as the input 
impedance of the antenna and $\vec{\mathcal{E}}^e$ as the beam pattern of the
antenna when excited with 1 V. 
\item
The radiated power from the PAF for an arbitrary excitation can be written in terms
of the embedded beam pattern. The radiated power $P_{rad}$ is  
\bea
P_{rad} & = & \frac{1}{2 z_f}\int_{sphere} \vec{\mathcal{E}} \cdot 
                        \vec{\mathcal{E}}^* \; \textrm{d} A \nonumber \\
        & = & \frac{1}{2 z_f}\int_{4\pi} \vec{E} \cdot \vec{E}^* \; \textrm{d} \Omega 
\label{prada1}
\eea
Using Eq.~\ref{embed}, Eq.~\ref{prada1} becomes
\bea 
P_{rad} & = & \frac{1}{2 z_f}\int_{4\pi} \bm V_0^H \bs{\vec{E}^e} \cdot \bs{\vec{E}^e}^H 
               \bm V_0 \; \textrm{d}\Omega \nonumber \\
        & = & \frac{1}{2 z_f} \bm V_0^H \left( 
             \int_{4\pi} \bs{\vec{E}^e} \cdot \bs{\vec{E}^e}^H \; \textrm{d}\Omega  
               \right) \bm V_0 \nonumber  \\
        & = & \frac{1}{2 z_f} \bm V_0^H \bm C_{Ce} \bm V_0, 
\label{prad} 
\eea
where we define
\be
\bm C_{Ce} \equiv \int_{4\pi} \bs{\vec{E}^e} \cdot \bs{\vec{E}^e}^H \; \textrm{d}\Omega,
\label{cce}
\ee
which is the correlation matrix of embedded beam patterns integrated over $4 \pi$ solid
angle.
\item
For a loss-less antenna the power dissipated at the ports should be equal to the radiated power.
Thus,
\bea
\frac{1}{2} \left(\frac{\bm V_0^H \bm I_0}{2} + \frac{\bm I_0^H \bm V_0}{2}\right)  & = & \frac{1}{2 z_f} \bm V_0^H \bm C_{Ce} \bm V_0 \nonumber \\
\frac{1}{4} \bm V_0^H \left(\bm Z^{-1} + \left(\bm Z^{-1}\right)^H \right) \bm V_0 & = &
\frac{1}{2 z_f} \bm V_0^H \bm C_{Ce} \bm V_0
\label{pdis}
\eea
Since Eq.~\ref{pdis} holds for any $\bm V_0$ it follows
\bea
\frac{1}{2}\left(\bm Z^{-1} + \left(\bm Z^{-1}\right)^H \right) & = & \frac{1}{z_f} \bm C_{Ce} \nonumber \\
& = & \frac{1}{z_f} \int_{4\pi} \bs{\vec{E}^e} \cdot \bs{\vec{E}^e}^H \; \textrm{d}\Omega  
\label{impbc}
\eea
\end{enumerate}

\section{Spectral density at Thermal Equilibrium}
\label{A2}

To express the computed power spectral density in the report
in terms of a physical temperature, it is necessary to know the spectral density
when the array is in equilibrium with a black body radiation field of temperature $T_0$
and connected with noise-less receiver system. It follows from 
the first and second law of thermodynamics, that the correlation
matrix of the open-circuit voltage at the output of antenna elements in the PAF
is \citep{Twiss1955} 
\be
\bm R_t =  \langle \bm V_{oc_t} \bm V_{oc_t}^H \rangle = 2 k_B T_0 \Big(\bm Z + \bm Z^H\Big)
\label{twiss}
\ee
By applying the appropriate weights to this voltage correlation (see for example Eq.~\ref{vout})
we can obtain the power spectral density when the PAF is at thermal equilibrium.
From the definition of noise figure\citep{kerrranda2010}, the 
conversion of any computed spectral density to a temperature is done by
first dividing the spectral density with that when the PAF is
in thermal equilibrium and then multiply by $T_0$.

\section{Voltage correlation due to a thermal radiation field}
\label{A3}

We show here that the open circuit voltage correlation when the
array is embedded in a black body radiation computed using Eq.~\ref{ocvolt}
is identical to the result given by Twiss's theorem\citep{Twiss1955}. To do that
we follow the arguments given in Section~\ref{spillmat}. 
For the case of the array embedded in a black body radiation
field, the integral in Eq.~\ref{spill5} need to be extended 
to $4\pi$ solid angle.
Thus the output voltage correlation $\bm R_t$ becomes
\bea
\bm R_t & = &\frac{4 k_B T_0}{z_f} \bm Z \left(
\int_{4\pi} \bs{\vec{E}^e}\cdot \bs{\vec{E}^e}^H \textrm{d}\Omega \right) \bm Z^H, \nonumber \\
        & = & \frac{4 k_B T_0}{z_f} \bm Z \bm C_{Ce} \bm Z^H,  
\label{thcorr}
\eea
where $\bm C_{Ce}$ is defined by Eq.~\ref{cce}. 
Substituting Eq.~\ref{impbc} in Eq.~\ref{thcorr} gives
\be
\bm R_t = 2 k_B T_0 \Big(\bm Z + \bm Z^H\Big),
\ee  
which is the voltage correlation that follows from the laws of thermodynamics.

\section{Spillover temperature}
\label{A4}

To express the spillover noise in physical temperature we
follow the prescription given in Appendix~\ref{A2}, 
\bea
T_{spill}  & = & T_0 \frac{\bm w_1^H \bm R_{spill} \bm w_1} {\bm w_1^H \bm R_{t} \bm w_1} \\
           & = & T_g 
\frac{\bm w_1^H \bm Z \left( \int_{\Omega_{spill}} \bs{\vec{E^e}}\cdot 
                     \bs{\vec{E^e}^H} \text{d}\Omega \right) \bm Z^H  \bm w_1} 
     {\bm w_1^H \bm Z \left( \int_{4\pi} \bs{\vec{E^e}}\cdot 
                      \bs{\vec{E^e}^H} \text{d}\Omega \right) \bm Z^H \bm w_1} \\
           & = & T_g 
\frac{\bm w_1^H \bm Z \bm C_{Ce1} \bm Z^H  \bm w_1}
     {\bm w_1^H \bm Z \bm C_{Ce} \bm Z^H  \bm w_1}
\label{spilltemp}
\eea
where we used Eq.~\ref{thcorr} for $\bm R_t$, Eq.~\ref{spill5} for $\bm R_{spill}$, and 
Eq.~\ref{cce} and Eq.~\ref{spill6} define $\bm C_{Ce}$ and $\bm C_{Ce1}$ respectively. 
If we consider the antenna elements in the array
are far apart such that $\bm Z$ and the embedded beam correlations become diagonal matrices,
then $T_{spill}$ due to $j^{th}$ antenna element can be obtained by considering the weight as 1 
for $j^{th}$ element and zeros for all other elements. In that case, Eq.~\ref{spilltemp} reduces to
\be
T_{spill} = T_g \frac{\int_{\Omega_{spill}} \vec{E_{j}^e}\cdot \vec{E_j^e}^* \text{d}\Omega } 
     {\int_{4\pi} \vec{E_{j}^e}\cdot \vec{E_j^e}^* \text{d}\Omega} = T_g (1-\eta_{spill}).
\ee 
Here we used the usual definition of spillover efficiency $\eta_{spill}$\citep{thomas1971}.
An alternate method to derive the result for a single port antenna is to start with
Eq.~\ref{singleport} and follow the arguments in Section~\ref{spillmat}.

\section{Antenna temperature due to a source}
\label{A5}

Following the prescription given in Appendix~\ref{A2},
the antenna temperature due to a source, $T_A$, can be written as 
\bea
T_A & = & \frac{\bm w_1^H \bm R_{signal} \bm w_1}{\bm w_1^H \bm R_t \bm w_1} T_0 \nonumber \\
    & = & \frac{S_{source}}{2 k_B} \frac{\bm w_1^H \bm Z \bm C_{Ie} \bm Z^H \bm w_1}
           {\bm w_1^H \bm Z \left(\int_{4\pi} \bs{\vec{E}^e}\cdot 
            \bs{\vec{E}^e}^H \text{d}\Omega \right) \bm Z^H \bm w_1} \nonumber \\
    & = & \frac{S_{source}}{2 k_B} \frac{\bm w_1^H \bm Z \bm C_{Ie} \bm Z^H \bm w_1}
           {\bm w_1^H \bm Z \bm C_{Ce} \bm Z^H   \bm w_1} \nonumber \\
    & = & \frac{S_{source}A_{ap}\eta_{ap}}{2 k_B}.
\label{tant}
\eea
Here we used Eq.~\ref{signal1} for the signal correlation matrix
$\bm R_{signal}$ and Eq.~\ref{thcorr} for the voltage correlation due
to a black-body field $\bm R_t$. We define the aperture efficiency as
\be
\eta_{ap} \equiv \frac{1}{A_{ap}} \; \; \frac{\bm w_1^H \bm Z \bm C_{Ie} \bm Z^H \bm w_1}
           { \bm w_1^H \bm Z \bm C_{Ce} \bm Z^H \bm w_1} 
\label{apeff}
\ee
where $A_{ap}$ is the aperture area of the telescope projected in the boresight
direction. For antenna elements
far apart and for weight 1 to $j^{th}$ element and 0 to all other elements
Eq.~\ref{apeff} reduces to the usual definition\citep{thomas1971} 
\be
\eta_{ap} = \frac{\left(
     \int_{A_{pap}} \vec{E}^e_{app,j} \textrm{d} A \right) \cdot  \left(
              \int_{A_{pap}} \vec{E}^e_{app,j} \textrm{d} A \right)^*}
     {A_{ap} \int_{4\pi} \vec{E_{j}^e}\cdot \vec{E_j^e}^* \text{d}\Omega} 
\ee

We conjecture the following relationship
\be
\bm Z \left(\bm C_{Ae} + \bm C_{ce1}\right) \bm Z^H  =  \bm Z \bm C_{ce} \bm Z^H.
\label{eff1}
\ee
where 
\be
\bm C_{Ae} \equiv \int_{A_{pap}} \bs{\vec{\mathcal{E}}^e_{app}} \cdot 
             \bs{\vec{\mathcal{E}}^e_{app}}^H \textrm{d} A.
\ee
Using Eq.~\ref{eff1}, Eq.~\ref{tant} can be written as
\be
T_A    =  \frac{S_{source} A_{ap}}{2 k_B} 
           \frac{\bm w_1^H \bm Z \bm C_{Ie} \bm Z^H \bm w_1}
                {A_{ap}\bm w_1^H \bm Z \bm C_{Ae} \bm Z^H \bm w_1}
           \left(1 - \frac{\bm w_1^H \bm Z \bm C_{ce1} \bm Z^H \bm w_1}
                     {\bm w_1^H \bm Z \bm C_{ce} \bm Z^H \bm w_1}\right).
\label{tant1}
\ee
We define the taper efficiency $\eta_t$ as
\be
\eta_t \equiv \frac{1}{A_{ap}} \;\; \frac{\bm w_1^H \bm Z \bm C_{Ie} \bm Z^H \bm w_1}
                {\bm w_1^H \bm Z \bm C_{Ae} \bm Z^H \bm w_1}.
\ee

To get the antenna temperature for a single antenna, we increase the
distance between the elements so that matrices $\bm C_{Ie}, \bm C_{Ae}, 
\bm C_{ce1}, \bm C_{ce}$ and $\bm Z$ become diagonal and use weights 1 
for one of the elements and 0 for all other elements. Eq.~\ref{tant1} then
reduces to 
\be
T_A = \frac{S_{source} A_{ap} \eta_{t} \eta_{spill}}{2 k_B}
\ee
where $\eta_{t}$ and $\eta_{spill}$ are the
taper and spillover efficiencies in the usual definition\citep{thomas1971}. 

\section{Receiver Temperature}
\label{A6}

Following the prescription in Appendix~\ref{A2}, the receiver temperature of the PAF
is
\bea
T_n & = & T_0 \frac{\bm w_1^H \bm R_{rec} \bm w_1} {\bm w_1^H \bm R_{t} \bm w_1} \nonumber \\
    & = & T_0 \; \frac{\bm w_1^H \Big(R_n\bs{\mathcal{I}} + \sqrt{R_n g_n} \;\big(\rho \bm Z + \rho^* \bm Z^H\big) + g_n \bm Z\bm Z^H \Big) \;\bm w_1} {\frac{1}{2}\bm w_1^H (\bm Z + \bm Z^H) \; \bm w_1}
\label{trec1}
\eea
where we used Eq.~\ref{tn} for $\bm R_{rec}$ and  Eq.~\ref{twiss} for $\bm R_t$. 

In radio astronomy instrumentation work\citep{marian2010} the receiver
temperature $T_n$, versus source impedance for a single antenna connected to
receiver is usually expressed as
\be
T_n = T_{min} + NT_0 \; \frac{(Z_s - Z_{opt})(Z_s - Z_{opt})^*}{\mbox{Re}\{Z_s\} \;\mbox{Re}\{Z_{opt}\}}
\label{singletn}
\ee
where $Z_s$ is the source impedance, and $Z_{opt}$ is the optimum impedance.
The minimum noise temperature, $T_{min}$, and $N$ are Lange invariants of the amplifier.
$Z_{opt}$ and the Lange invariants are related to the noise parameters through expressions
\bea
\mbox{Im}\{Z_{opt}\} & = & \frac{\rho_i \; \sqrt{R_n g_n}}{g_n} \\
\mbox{Re}\{Z_{opt}\} & = & \frac{\sqrt{R_n g_n}}{g_n}\;\sqrt{1 - \rho_i^2} \\
N & = & \mbox{Re}\{Z_{opt}\} \; g_n \\
T_{min} & = & 2T_0 \Big( N + \rho_r \; \sqrt{R_n g_n} \Big)
\eea
We show below that Eq.~\ref{trec1} 
can be written in terms of the Lange invariants as 
\be
T_n = T_{min} + N T_0 \; \frac{\bm w_1^H (\bm Z - Z_{opt} \bs{\mathcal{I}})
   (\bm Z - Z_{opt} \bs{\mathcal{I}})^H \bm w_1} {\mbox{Re}\{Z_{opt}\} \; \frac{1}{2}\bm w_1^H (\bm Z + \bm Z^H) \bm w_1}
\label{youreq}
\ee
Eq.~\ref{youreq} is a generalization of Eq.~\ref{singletn}.

Equation~\ref{trec1} can be rewritten as
\bea
&  & \frac{1}{2}\;\bm w_1^H \big(\bm Z + \bm Z^H\big) \bm w_1 \; \frac{T_n}{T_0} \nonumber \\
&  =  & \bm w_1^H \Big(R_n \bs{\mathcal{I}} + \sqrt{R_n g_n} \; \big(\rho \bm Z + \rho^* \bm Z^H\big) + g_n\bm Z\bm Z^H \Big)\bm w_1 \nonumber \\
& = & \bm w_1^H \Big(2 \; \sqrt{R_n g_n \big(1 - \rho_i^2\big)} \; \mbox{Re}\{\bm Z\} + \nonumber \\
&   & \;\;\;\;\;\;\;\; 2 \; \rho_r \sqrt{R_n g_n} \; \mbox{Re}\{\bm Z\}\Big) \; \bm w_1 + \nonumber \\
&   & \bm w_1^H \Big(\frac{R_n g_n}{g_n} \bs{\mathcal{I}} + g_n \bm Z \bm Z^H  - 2 \; \sqrt{R_n g_n \big(1 - \rho_i^2\big)} \; \mbox{Re}\{\bm Z \} \nonumber \\
&   & \;\;\;\;\;\;\;\; - 2 \; \rho_i \sqrt{R_n g_n} \; \mbox{Im}\{ \bm Z \}\Big) \; \bm w_1 \nonumber \\
& = & 2\Big(\sqrt{R_n g_n \big(1 - \rho_i^2\big)}  + \rho_r \sqrt{R_n g_n}\Big) \frac{1}{2}\bm w_1^H \big(\bm Z + \bm Z^H\big) \bm w_1 + \nonumber \\
&   & \bm w_1^H \Big(\frac{R_n g_n}{g_n} \bs{\mathcal{I}} + g_n \bm Z \bm Z^H - 2\; \sqrt{R_n g_n \big(1 - \rho_i^2\big)} \; \mbox{Re}\{\bm Z\} \nonumber \\
&  &  \;\;\;\;\;\;\;\; - 2\;\rho_i \sqrt{R_n g_n} \; \mbox{Im}\{\bm Z\}\Big)\; \bm w_1 \nonumber \\
& = & \frac{T_{min}}{T_0} \; \frac{1}{2} \; \bm w_1^H \big(\bm Z + \bm Z^H\big)\; \bm w_1 + \nonumber \\ 
&   & \bm w_1^H \Big(g_n |Z_{opt}|^2 \bs{\mathcal{I}} + g_n \bm Z \bm Z^H - 2\; g_n \; \mbox{Re}\{Z_{opt}\} \; \mbox{Re}\{\bm Z\} \nonumber \\
&   &  \;\;\;\;\;\;\;\; - 2\; g_n \; \mbox{Im}\{Z_{opt}\} \; \mbox{Im}\{\bm Z\}\Big) \; \bm w_1 \nonumber \\
& = & \frac{T_{min}}{T_0}\; \frac{1}{2}\; \bm w_1^H \big(\bm Z + \bm Z^H\big) \; \bm w_1 + \nonumber \\
&  &  \;\;\;\;\;\;\;\; g_n \bm w_1^H \big(\bm Z - Z_{opt} \bs{\mathcal{I}}\big) \big(\bm Z - Z_{opt}\bs{\mathcal{I}}\big)^H \; \bm w_1 \nonumber \\
& = & \frac{T_{min}}{T_0} \; \frac{1}{2} \; \bm w_1^H \big(\bm Z + \bm Z^H\big) \; \bm w_1 + \nonumber \\
&   & \;\;\;\;\;\;\;\; N \; \frac{\bm w_1^H \big(\bm Z - Z_{opt} \bs{\mathcal{I}}\big) \big(\bm Z - Z_{opt}\bs{\mathcal{I}}\big)^H\; \bm w_1} {\mbox{Re}\{Z_{opt}\}}
\label{youreq1}
\eea
Eq.~\ref{youreq1} can be readily rewritten as Eq.~\ref{youreq}.

\section{CST Far-field pattern, Scattering Matrix and Energy Balance}
\label{A7}

As described in Section~\ref{cstpart} the Scattering matrix and Far-field
pattern of the PAF are obtained using CST software package. The Scattering
matrix when exported in TOUCHSTONE format is normalized to a reference
impedance of 50 $\Omega$. CST computes the far-field pattern by
exciting the $i^{th}$ array element and keeping all other ports 
terminated with the transmission line impedance, which in our case is 50 $\Omega$. 
These field patterns, referred to 
as $\vec{E'}_i$, depends only on ($\theta, \phi$) and are in 
units of V. The field values are provided for a (default) 
reference distance $r$ of 1 m.   
Since all array elements are not located at the co-ordinate origin, 
a geometric phase will be present in the far-field pattern. 
The field patterns $\vec{E'}_i$ provided by CST includes this geometric 
phase factor. 

We check here whether the simulation results satisfies the energy balance.
The peak power used in CST simulation is 1 W, which corresponds to
a RMS (root mean square) power, $P_{stim}$ of 0.5 W for sinusoidal excitation. The
amplitude of traveling waves $\bm a$ when $j^{th}$ port is excited is 
\bea
a_i & =  & \sqrt{2\,P_{stim}} \quad \textrm{for} \; i=j \nonumber \\
    & =  & 0 \quad\quad\quad\quad\;\; \textrm{for}\; i \neq j
\eea
where $a_i$ are elements of vector $\bm a$. The reflected wave vector $\bm b$ is given
by Eq.~\ref{Sdef}. 
The total reflected power $P_{reflect}$, which is dissipated in the port impedance, is
\be
P_{reflect} = \frac{1}{2}\, \bm b^H \bm b = \frac{1}{2}\, \bm a^H \bm S^H \bm S \bm a
\ee
The total radiated power $P_{rad}$ is
\be
P_{rad}  = \frac{1}{2 z_f}\int_{4\pi} \vec{E'}_i \vec{E'}_i^* \; \textrm{d}\Omega 
\label{cstprad}
\ee

\begin{figure}[t]
\includegraphics[width=5.5in, height=3.5in, angle=0]{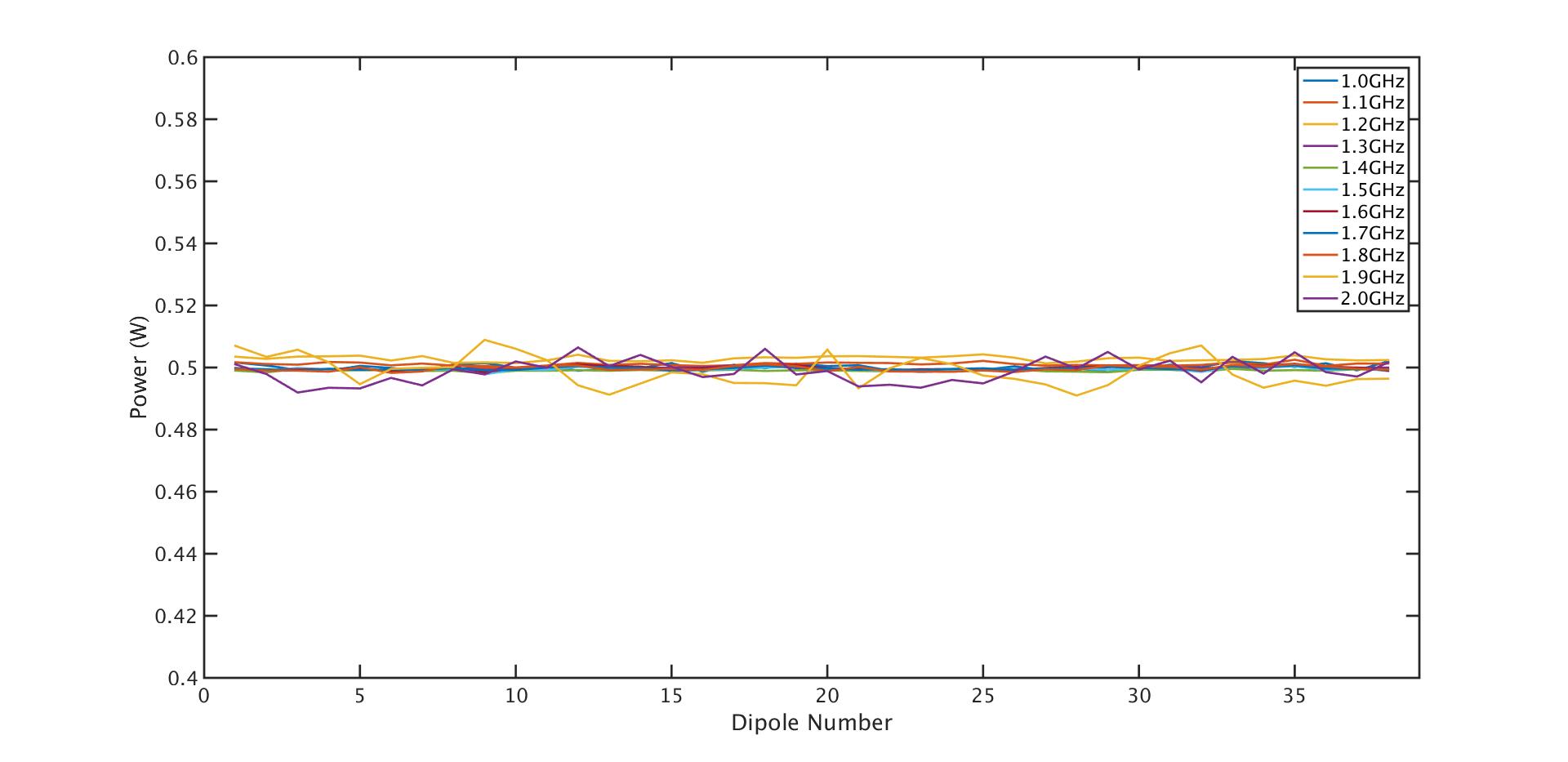}
\caption{The sum of radiated and reflected power for different dipoles in the array. These
powers are obtained using the Scattering matrix and radiation pattern provided by CST. The powers 
are computed for frequencies shown on the plot. As seen in the plot the sum of the radiated
and reflected power is about 0.5 W, which is the RMS excitation power used by CST.}
\label{figa1}
\end{figure}

A plot of $P_{rad} + P_{reflect}$ for the 38 dipoles and for frequencies in the range
1.0 to 2.0 GHz is shown in Fig.~\ref{figa1}. As seen from the plot, the calculated
total power is equal to 0.5 W. 
 
\section{Embedded beam patterns from CST far-field patterns}
\label{A8}

\begin{figure}[t]
\begin{tabular}{cc}
\includegraphics[width=3.3in, height=2.5in, angle =0]{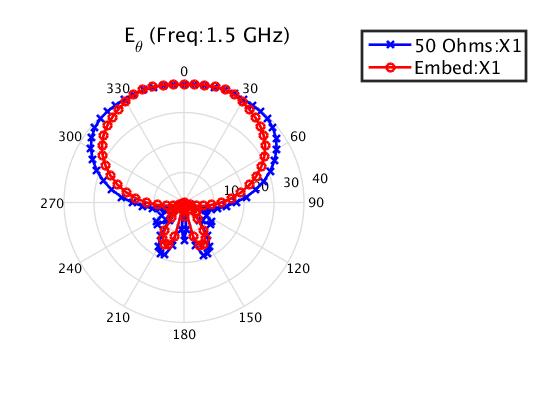} &
\includegraphics[width=3.3in, height=2.5in, angle =0]{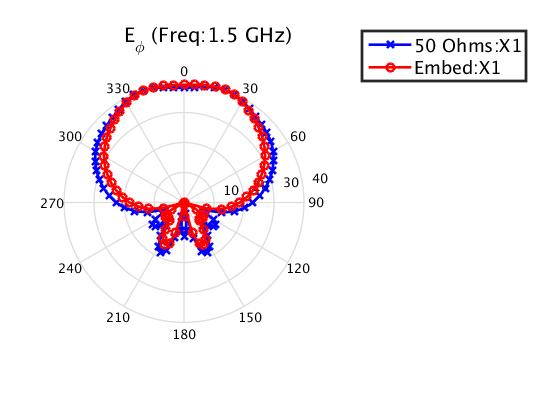} \\
\includegraphics[width=3.3in, height=2.5in, angle =0]{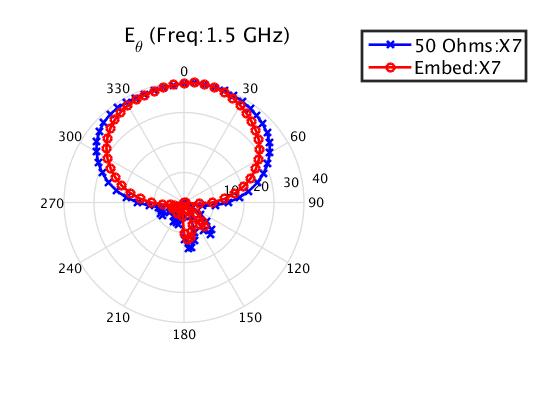} &
\includegraphics[width=3.3in, height=2.5in, angle =0]{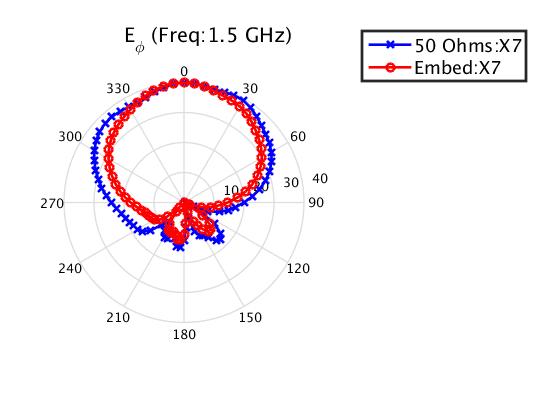} \\
\includegraphics[width=3.3in, height=2.5in, angle =0]{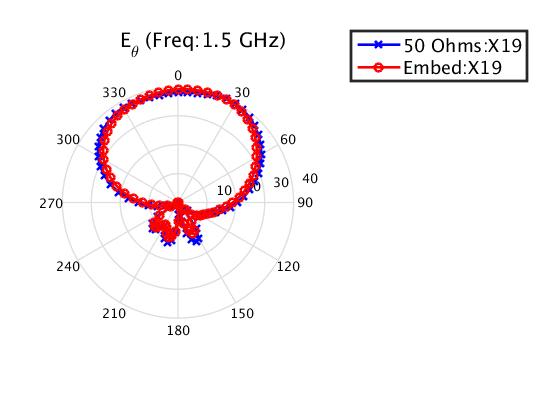} &
\includegraphics[width=3.3in, height=2.5in, angle =0]{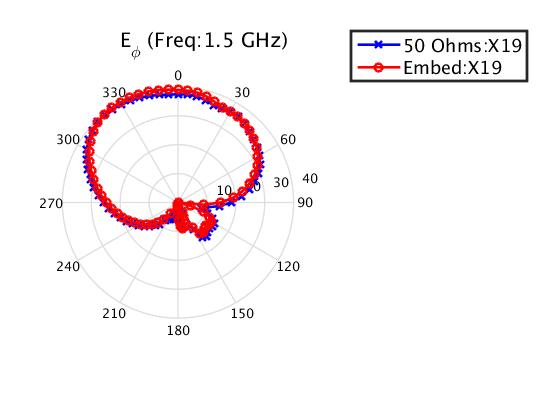}
\end{tabular}
\caption{Plots of far-field pattern from the CST software package (blue) and the
computed embedded beam patterns at 1.5 GHz (red). }
\label{figa2}
\end{figure}

CST provides far-field pattern when $j^{th}$ port is excited and all
other ports are terminated with port impedance (i.e. 50 $\Omega$).
Using Eq.~\ref{embed},
\be
\vec{E'}_j = \sum_{i=1,M} q_{ij} \; \vec{E}^e_i,
\label{cstembed}
\ee
here $q_{ij} = v_{0_i}$ are the port voltages under the excitation condition
said above. These voltages can be computed using the Scattering Matrix.
The elements of wave amplitude vector for the excitation is 
\bea
a_i & =  & \sqrt{2\,P_{stim}} \quad \textrm{for} \; i=j \nonumber \\
    & =  & 0 \quad\quad\quad\quad\;\; \textrm{for}\; i \neq j
\eea
The wave amplitude vector $\bm b$ is then
\be
\bm b = a_j \begin{bmatrix}
    S_{1j} \\
    S_{2j} \\
    \threevdots \\
    S_{jj} \\
    \threevdots \\
    S_{Mj} 
\end{bmatrix} 
,
\ee 
where $a_j$ is the $j^{th}$ element of the vector $\bm a$, 
$S_{ij}, i = 1$ to $M$ is the $j^{th}$ column of $\bm S$. The
port voltage is then
\bea
q_{ij} & = & \sqrt{z_0} (a_i + b_i) \nonumber \\
       & = & \sqrt{z_0} (1 + S_{jj}) a_j \quad \textrm{for}\;\;\; i = j \nonumber \\ 
       & = & \sqrt{z_0} S_{ij} a_j \quad\quad\quad \textrm{for}\;\;\; i \neq j 
\label{qij}
\eea 
The set of far-field patterns provided by CST along with
the port voltage can be used to obtain the embedded beam
pattern. Eq.~\ref{cstembed} for the set of far-field patterns
can be concisely written as
\be
\bs{\vec{E}^{'}} = \bm Q \; \bs{\vec{E}^e},
\ee
where the elements of $\bm Q$ are $q_{ij}$. 
This equation is valid for each $\theta, \phi$. 
Using Eq.~\ref{qij} $\bm Q$ can be written as
\be
\bm Q = \sqrt{2\, z_0\, P_{stim}} \;\;(\bs{\mathcal{I}} + \bm S).
\ee
The embedded beam patterns are
then given by
\be
\bs{\vec{E}^e} = \bm Q^{-1} \; \bs{\vec{E}^{'}},
\ee
Fig.~\ref{figa2} shows some example plots of CST far-field patterns and the computed embedded
beam patterns.

Some sanity checks on the computed embedded beam patterns are done below.
We computed the radiated power using the
embedded beam pattern (Eq.~\ref{prad}) and compared with those 
obtained using Eq.~\ref{cstprad}.
The two radiated powers have identical values.

\begin{figure}[t]
\includegraphics[width=5.5in, height=3.5in, angle=0]{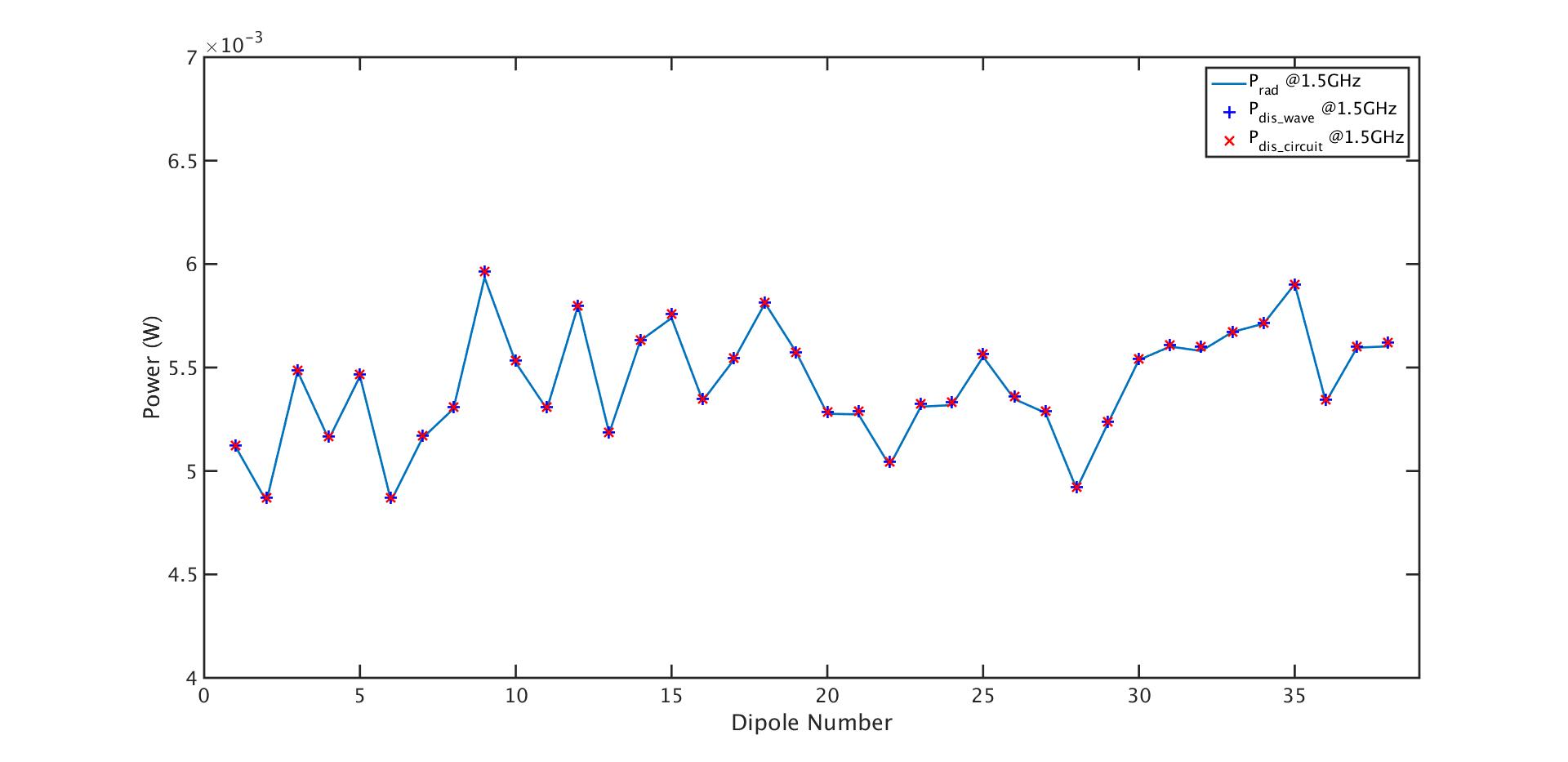}
\caption{The radiated power computed with embedded pattern is compared with
power dissipated at the excitation port. The dissipated powers are computed
using Eq.~\ref{pdis1} (marked as $P_{dis\_wave}$) and Eq.~\ref{pdis2} 
(marked as $P_{dis\_circuit}$). As
seen in the figure the radiated power is equal to the dissipated power.
}
\label{figa3}
\end{figure}

Second we check whether the embedded beam pattern satisfy the energy conservation. 
By definition of embedded beam pattern we excite the array with 1 V peak at port $j$ and short circuit 
all other ports. Thus
\bea
v_{0_i} & = & 1 \quad \textrm{for}\;\;\; i = j \\ 
    & = & 0 \quad \textrm{for}\;\;\; i \neq j.
\eea
The wave amplitudes are then 
\bea
\sqrt{z_0} (a_i + b_i) & = & 1 \quad \textrm{for}\;\;\; i = j \\
\sqrt{z_0} (a_i + b_i) & = & 0 \quad \textrm{for}\;\;\; i \neq j \\
\eea
The vector $\bm a$ can be written as
\be
\bm a = -\bm b + \frac{1}{\sqrt{z0}} \begin{bmatrix}
    0 \\
    0 \\
    \threevdots \\
    1 \\
    \threevdots \\
    0 
\end{bmatrix} 
\ee
where the non-zero element (which is 1) is located at $j^{th}$ row. Substituting 
in Eq.~\ref{Sdef} and re-arranging we get
\be
\bm b = \frac{1}{\sqrt{z0}}(\bs{\mathcal{I}} + \bm S)^{-1} \begin{bmatrix}
    S_{1j} \\
    S_{2j} \\
    \threevdots \\
    S_{jj} \\
    \threevdots \\
    S_{Mj}
\end{bmatrix}.
\ee
Power dissipated at the $j^{th}$ port is
\bea
P_{dis} & = & \frac{1}{2} (a_j a_j^* - b_j b_j^*) \\
        & = & \frac{1}{2\sqrt{z_0}} (\frac{1}{\sqrt{z_0}} - (b_j + b_j^*))
\label{pdis1}
\eea
Another method to get $P_{dis}$ is by using Eq.~\ref{embpdis} and
Eq.~\ref{embZin}; 
\be
P_{dis} = \frac{1}{2} \textrm{Re}\{Y_{jj} \}
\label{pdis2}
\ee 
The radiated power computed from embedded beam patterns are found to be equal
to $P_{dis}$ (see Fig.~\ref{figa3}).

\section{Computation of spillover integral, aperture field and aperture field integral}
\label{A8a}

\begin{figure}[t]
\includegraphics[width=5.5in, height=3.5in, angle =0]{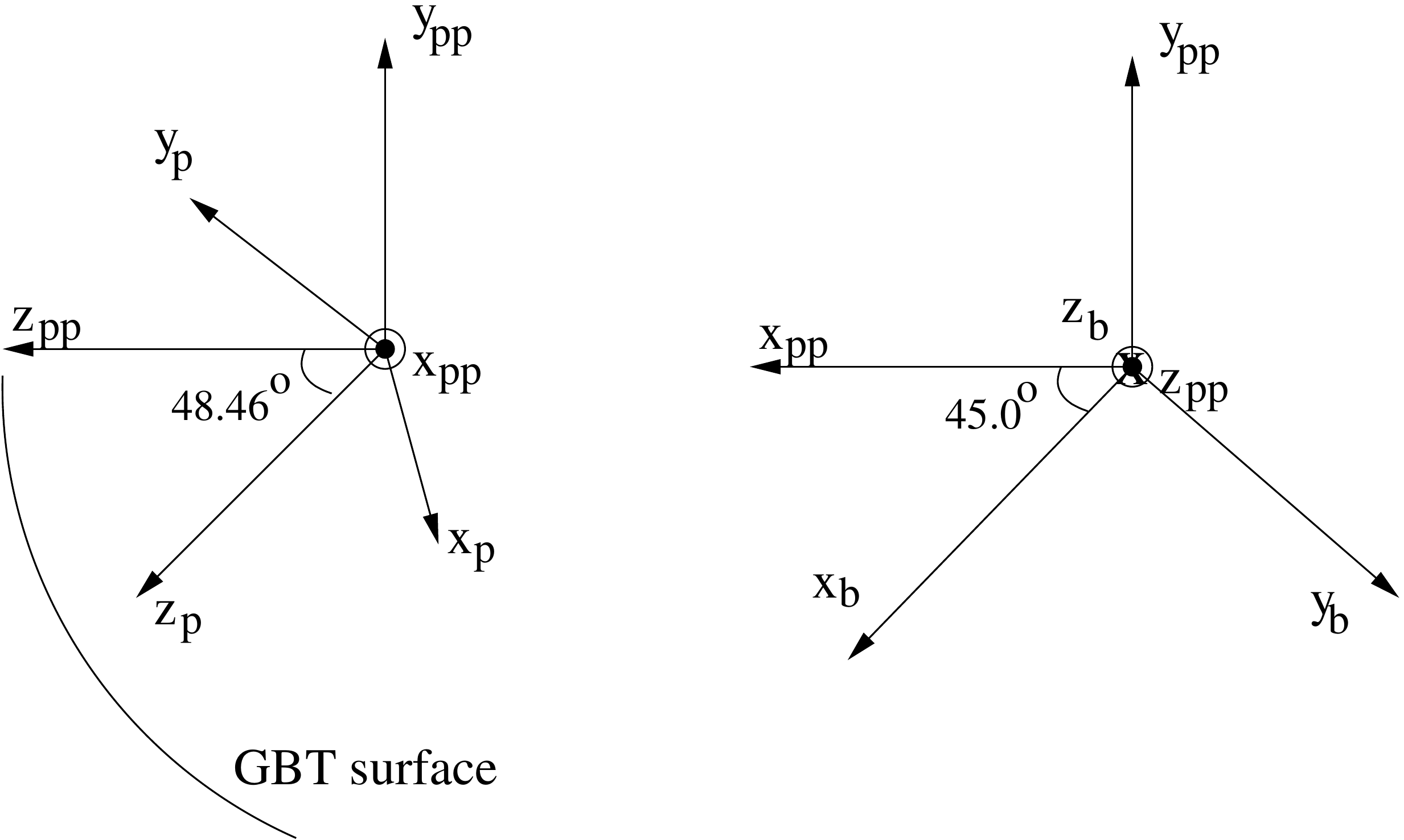} 
\caption{Coordinate system used for the computation of Eq.~\ref{spill6}
and Eq.~\ref{forsig} (see text for details).
}
\label{fig7}
\end{figure}

The integrals in Eq.~\ref{spill6} and Eq.~\ref{forsig} are evaluated using
the coordinate system described here. We also briefly describe the
computation of aperture field for the evaluation of Eq.~\ref{forsig}.

The geometry of the GBT is discussed in \citet{hallking1992,
norridsri1996}. The GBT surface is cut from a parent symmetric
paraboloid of diameter 208 m and f/D of 0.29. Here 
f = 60 m is the focal length and D = 208 m is the diameter of the 
parent paraboloid. The equation of the parent paraboloid is
\be
z_{pp} = f - \frac{x_{pp}^2 + y_{pp}2}{4f}.
\ee
The coordinate system $x_{pp}-y_{pp}-z_{pp}$ is shown in Fig.~\ref{fig7}. 
The GBT surface is
obtained by the intersection of a plane at $\sim$ 24\deg\ with the parent
paraboloid. The coordinates in m of
the intersection of the plane at $x_{pp} = 0$ are
(0, $-$4.0, 59.933) and (0, $-104$, 14.933) corresponding
to the top and bottom end of the GBT surface respectively. 

The PAF field patterns are obtained with respect to
the coordinate system $x_{p}-y_{p}-z_{p}$, which
is same as the $x-y-z$ coordinated in the CST (see
Fig.~\ref{fig5}). The $x_{p}-y_{p}$ are rotated by 45\deg
with respect to the $X-Y$ axis shown in Fig.~\ref{fig4}
as well as the $u-v$ co-ordinates of the CST.
The x and y polarization dipoles are
aligned along the $x_{p}$ and $y_{p}$ axis respectively (see Fig.~\ref{fig4}). 
The PAF is mounted on the GBT such that $z_{p}$  
passes through the projected center of the aperture plane 
on the GBT surface.  The co-ordinates of the projected center are
(0, $-$54.0, 47.85) in m. The angle between $z_{pp}$
and $z_{p}$ is 48.46\deg, which is referred to as the
feed angle $\beta$. Since the $X$ axis shown in Fig.~\ref{fig4}
is parallel to elevation axle, the orientation of the
$x_{p}-y_{p}-z_{p}$ with respect to $x_{pp}-y_{pp}-z_{pp}$
is obtained first by rotating the $x_{p}-y_{p}$ by
45\deg clockwise with respect to $x_{pp}-y_{pp}$ 
and then rotating the $z_{p}$ axis by angle $\beta$ with 
respect to $z_{pp}$. 

The boresight coordinates $x_b-y_b-z_b$ and $x_{pp}-y_{pp}-z_{pp}$
coordinates are related through the transformation
\be
\bs{PP2B} =
\begin{bmatrix} 
\textrm{cos}(\pi/4), -\textrm{sin}(\pi/4), 0 \\ 
-\textrm{sin}(\pi/4), -\textrm{cos}(\pi/4), 0 \\
0,           0,        -1 \\
\end{bmatrix}. 
\ee
The rotation of the $x_b-y_b$ coordinates by 45\deg\ will essentially
align the x and y polarization of the aperture field with these 
coordinate axes. The source coordinates
are rotated by $\theta_s, \phi_s$ with respect to the boresight coordinates.
The coordinates of the projected aperture plane in the direction of
source are obtained by defining a plane perpendicular to the unit vector
$\hat{u}_z = \textrm{cos}(\phi_s)\textrm{sin}(\theta_s) \hat{i}_b + 
             \textrm{sin}(\phi_s)\textrm{sin}(\theta_s) \hat{j}_b + 
             \textrm{cos}(\theta_s) \hat{k}_b,
$ 
where $\hat{i}_b, \hat{j}_b, \hat{k}_b$ are the unit vectors along
the $x_b, y_b, z_b$ axes.

The aperture field is obtained using Geometric optics. We also
assume that the GBT surface is a perfect conductor. The boundary
condition during reflection is that the tangential component
of the field at the surface is zero,
\be
\vec{\mathcal{E}}^{e(i)}_i - \hat{r}_n \cdot \vec{\mathcal{E}}^{e(i)}_i \hat{r}_n 
= - \vec{\mathcal{E}}^{e(r)}_i + \hat{r}_n \cdot \vec{\mathcal{E}}^{e(r)}_i \hat{r}_n,
\ee
where $\vec{\mathcal{E}}^{e(i)}_i$ and $\vec{\mathcal{E}}^{e(r)}_i$ are the
incident and reflected fields respectively and $\hat{r}_n$ is the unit
normal to the surface element where reflection is taking place. Since
the surface is assumed to be loss-less $|\vec{\mathcal{E}}^{e(i)}_i| =
|\vec{\mathcal{E}}^{e(r)}_i|$. Applying Snell's law it follows that
\be
\vec{\mathcal{E}}^{e(r)}_i  
= 2 \hat{r}_n \cdot \vec{\mathcal{E}}^{e(i)}_i \hat{r}_n - \vec{\mathcal{E}}^{e(i)}_i
\ee
The unit normal is given by
\bea
\hat{r}_n & = & - \frac{\vec{r}_{\theta} \times \vec{r}_{\phi}}
              {|\vec{r}_{\theta} \times \vec{r}_{\phi}|} \nonumber \\ 
          & = & -\textrm{cos}(\theta_{pp}/2) \hat{r}_{pp} 
              + \textrm{sin}(\theta_{pp}/2) \hat{\theta}_{pp},
\eea
where $\vec{r}_{\theta} = \frac{\partial\vec{r}_{pp}}{\partial\theta_{pp}}$,
$\vec{r}_{\phi} = \frac{\partial\vec{r}_{pp}}{\partial\phi_{pp}}$ and $(|\vec{r}_{pp}|\hat{r}_{pp}, 
\theta_{pp}\hat{\theta}_{pp}, \phi_{pp}\hat{\phi}_{pp})$ are  spherical
coordinates of the GBT surface in the $x_{pp}-y_{pp}-z_{pp}$ system.  
The incident fields are the beam patterns provided by the CST scaled by
the distance from the focus to the reflector. The reflected
field is propagated to the
source coordinate to get the aperture field, which is given by $\vec{\mathcal{E}}^{e(r)}_i 
e^{-\frac{j2\pi(2f+z^s_b)}{\lambda}}$, where $z^s_b$ is the $z$ component
of the project aperture plane in the boresight coordinates.

\section{Symbols and Notations used in this report}
\label{A9}

The bold letters represent m$\times$n, 1$\times$n and m$\times$1 matrices (example
$\bm M$). The 1$\times$n and m$\times$1 matrices are referred to as vectors. 
Transpose of a matrix is represented, for example, by $\bm M^T$ and Hermitian transpose is
represented, for example, by $\bm M^H$. Complex conjugate of variable say $v$ is 
denoted by $v^*$.  Vectors in real
space are represented with an arrow on the top of non-bold symbol (example $\vec{E}$).
An arrow on top of a bold symbol is used to represent a matrix or vector of
real space vectors (example $\bs{\vec{E}}$). For example,
\be
\bs{\vec{\mathcal{E}}^e}^T = \left[\vec{\mathcal{E}}^e_1, \vec{\mathcal{E}}^e_2, ... \right]
\nonumber
\ee
The calligraphic symbols are used
to represent {\em electromagnetic fields} which depends on the position 
vector $\vec{r}$ (example $\bs{\vec{\mathcal{E}}^e}$)  
and Latin symbols are used for {\em electromagnetic fields}  
which depends on the angular coordinates $\theta, \phi$ (example $\bs{\vec{E}^e}$).  
The operation $\bs{\vec{\mathcal{E}}^e}^T \times \bs{\mathcal{I}} \vec{\mathcal{H}_r}$
can be expanded as
\be
\left[\vec{\mathcal{E}}^e_1, \vec{\mathcal{E}}^e_2, ... \vec{\mathcal{E}}^e_M\right] \times 
\begin{bmatrix} 
1, 0, 0, ... \\
0,1,0, ... \\
\threevdots \\
\hfill .. 0,1\\
\end{bmatrix} \vec{\mathcal{H}}_r  =  
\begin{bmatrix}
\vec{\mathcal{E}}^e_1 \times \vec{\mathcal{H}}_r \\
\vec{\mathcal{E}}^e_2 \times \vec{\mathcal{H}}_r \\
\threevdots \\
\vec{\mathcal{E}}^e_M \times \vec{\mathcal{H}}_r \\
\end{bmatrix} \nonumber
\ee
Similarly the operation $\bm Z \bs{\mathcal{I}} \hat{p} \cdot \bs{\vec{\mathcal{E}^e}}$ can be
expanded as
\be
\bm Z
\begin{bmatrix} 
\hat{p}, 0, 0, ... \\
0,\hat{p},0, ... \\
\threevdots \\
\hfill .. 0,\hat{p}\\
\end{bmatrix} \cdot 
\begin{bmatrix}
\vec{\mathcal{E}}^e_1 \\
\vec{\mathcal{E}}^e_2 \\
\threevdots \\
\vec{\mathcal{E}}^e_M \\
\end{bmatrix}  =  
\begin{bmatrix}
z_{11}, z_{12}, ...,z_{1M} \\
z_{21}, z_{22}, ...,z_{2M} \\
\threevdots \\
z_{M1}, z_{M2}, ... ,z_{MM} \\ 
\end{bmatrix}
\begin{bmatrix}
\vec{\mathcal{E}}^e_1 \cdot \hat{p} \\
\vec{\mathcal{E}}^e_2 \cdot \hat{p}\\
\threevdots \\
\vec{\mathcal{E}}^e_M \cdot \hat{p}\\
\end{bmatrix}
\nonumber
\ee

\begin{tabular}{ll}
$\eta_{ap}$ & Aperture efficiency \\
$\eta_{tap}$ & Taper efficiency \\
$\eta_{spill}$ & Spillover efficiency \\
$\lambda$ & Free space wavelength of radiation \\
$\theta_s, \phi_s$ & Source position in the sky w.r.t the boresight coordinate system \\
$a_i, b_i, \bm a, \bm b$ & Traveling wave amplitudes in the transmission line \\
$\bm A$ & Matrix that transform open-circuit voltage to `loaded' voltage \\
$A$ & Enclose area in Lorentz integral (Eq.~\ref{lorentz}) \\
$A_{ap}$ & Physical area of the telescope aperture projected in the boresight direction \\
$A_{trans}$ & Cross section area in the transmission line (see Fig.~\ref{fig2}) \\
$A_{free}$ & Surface area away from the metallic surface of the PAF (see Fig.~\ref{fig2}) \\
$c$ & Velocity of light in free space \\
$D_{array}$ & Maximum physical size of the PAF \\
$\vec{\mathcal{E}}^e, \vec{E}^e$ & Embedded electric field patterns \\
$\bs{\vec{\mathcal{E}}^e}, \bs{\vec{E}^e}$ & Set of $M$ embedded electric field patterns \\
$\bs{\vec{\mathcal{E}}^e_{pap}}$ & Set of $M$ embedded electric field patterns propagated \\
 & to the aperture plane projected in the direction of the source \\
$\bs{\vec{\mathcal{E}}^e_{pap,x}}, \bs{\vec{\mathcal{E}}^e_{pap,y}}$ & x and y components of $\bs{\vec{\mathcal{E}}^e_{pap}}$ \\ 
$e_{max}$ & Maximum eigenvalue of system characteristics matrix \\
$\bm G$ & Gain matrix of the system (diagonal) \\
$\vec{\mathcal{H}}^e, \vec{H}^e$ & Embedded magnetic field patterns \\
$\bs{\vec{\mathcal{H}}^e}, \bs{\vec{H}^e}$ & Set of $M$ embedded magnetic field patterns \\
$\bs{\vec{\mathcal{H}}^e_{pap}}$ & Set of $M$ embedded magnetic field patterns propagated \\
 & to the aperture plane projected in the direction of the source \\
$i_{0_i}, \bm I_0$ & Excitation current for the PAF (source impedance = $z_0$) \\
$\bs{\mathcal{I}}$ & Identity matrix \\
$k$ & Wave vector of spherical waves of radiation pattern \\
$k_B$ & Boltzmann constant \\
$k_{inc}$ & Wave vector of the incident radiation field (plane wave and also arbitrary wave front) \\
$M$ & number of elements in the PAF \\
\end{tabular}

\begin{tabular}{ll}
$\bm M$ & Characteristics matrix of the system (i.e. PAF+telescope+receiver) \\
$\bm N$ & Open-circuit voltage correlation matrix of the total noise at the output \\
& of the antenna array \\
$\hat{p}$ & Unit vector along the polarization of an incident radiation \\
$P_{rad}$ & Power radiated by the PAF \\
$P_{reflect}$ & Power reflected from the ports of the PAF \\
$P_{stim}$ & Excitation power used by CST 1 W peak-peak \\
$\vec{r}, r$ &  Position vector and its magnitude \\
$R_n, g_n, \rho$ & Noise resistance, conductance of the low-noise amplifier and \\
& their correlation coefficient \\
$\bs{\tilde{R}}$ & Correlation of `loaded' voltages \\
$\bs{R}$ & Correlation of open-circuit voltages \\
$\bs{R_t}$ & Correlation of open-circuit voltages at the output of the antenna \\
& array when it is enclosed in a black body radiation field \\
$\bm S $ & Scattering matrix \\
$S_{source} $ & Flux density of source \\
SNR & signal-to-noise ratio \\
%\end{tabular}
%\begin{tabular}{ll}
$T_{A}$ & Antenna temperature \\
$T_{sys}$ & System temperature \\
$T_g$ & Physical temperature of the ground \\
$T_0$ & Reference temperature 290 K \\
$T_{cmb}$ & Cosmic microwave background temperature \\
$T_{sky}$ & Brightness temperature of sky background radiation at the \\
& frequency of interest \\
$v_{0_i}, \bm V_0$ & Excitation voltage for the PAF (source impedance = $z_0$) \\
$\bs{\tilde{V}}$ & `Loaded' voltage at the input of the beamformer or combiner \\
$\bs{V_{oc}}$ & Open-circuit voltage at the output of the antenna array \\
$\bm w, \bm w_1, \bm w_2$ & Weight vectors \\
$x_b,y_b,z_b$ & Boresight coordinate system \\
$x_p,y_p,z_p$ & Coordinate system on the PAF which is used to obtain \\
& the radiation pattern. In Fig.~\ref{fig5}, this is shown as the u-v-w coordinates. \\
$x,y,z$ & Coordinate system with z-axis pointing towards the source \\
$\bm Y$ & Admittance matrix of the PAF \\
$z_0$ & Reference impedance 50 $\Omega$, also characteristic impedance of \\
& the transmission line\\
$z_f$ & Free space impedance\\
$\bm Z$ & Impedance matrix of the PAF \\
$Z_{in}$ & input impedance of the low-noise amplifier \\
$\bm Z_{in}$ & input impedance (diagonal) matrix \\
$Z_{pin_i}$ & Input impedance of a port of the PAF when all other ports \\
& are short circuited \\
\end{tabular}

\end{document}